\documentclass[fleqn,usenatbib]{mnras}

\usepackage{newtxtext,newtxmath}
\usepackage[T1]{fontenc}

\DeclareRobustCommand{\VAN}[3]{#2}
\let\VANthebibliography\thebibliography
\def\thebibliography{\DeclareRobustCommand{\VAN}[3]{##3}\VANthebibliography}

\usepackage{graphicx}	% Including figure files
\usepackage{amsmath}	% Advanced maths commands
\usepackage{natbib}
\newcommand{\Lya}{{\rm Ly}$\alpha$ }
\newcommand{\Ha}{{\rm H}$\alpha$ }
\newcommand{\Hi}{H{\sc i}}
\newcommand{\fesca}{$f_{\mathrm{esc}}^{\mathrm{Ly\alpha}}$}
\newcommand{\fescc}{$f_{\mathrm{esc}}^{\mathrm{LyC}}$}
\newcommand{\nion}{$\dot{n}_{\mathrm{ion}}$}
\newcommand{\xiionUV}{$\xi_{\mathrm{ion}}^{\mathrm{UV}}$}
\newcommand{\xiionLya}{$\xi_{\mathrm{ion}}^{\mathrm{Ly\alpha}}$}
\newcommand{\xiionfesc}{$\xi\mathrm{_{ion}^{Ly\alpha}}f\mathrm{_{esc}^{LyC}}$}
\defcitealias{Begley+24}{B24} 
\defcitealias{Choustikov+24c}{C24}
\title[Ly$\alpha$ escape from LAEs \& Implication to EoR ]{Age dependence of Ly$\alpha$ escape fraction of Ly$\alpha$ emitters and their significant role in cosmic reionization}
\author[S. Shimizu et al.]{
Shunta Shimizu,$^{1}$\thanks{E-mail: s.shimizu@astron.s.u-tokyo.ac.jp}
Nobunari Kashikawa,$^{1,2}$
Satoshi Kikuta,$^{1}$
Yoshihiro Takeda,$^{1}$
Junya Arita,$^{1}$
\newauthor
Ryo Emori,$^{1}$
and Kentaro Koretomo$^{1}$
\\
$^{1}$Department of Astronomy, School of Science, The University of Tokyo, 7-3-1 Hongo, Bunkyo-ku, Tokyo 113-0033, Japan\\
$^{2}$Research Center for the Early Universe, The University of Tokyo, 7-3-1 Hongo, Bunkyo-ku, Tokyo, 113-0033, Japan\\
}

\date{Accepted XXX. Received YYY; in original form ZZZ}

\pubyear{\the\year{}}

\begin{document}
\label{firstpage}
\pagerange{\pageref{firstpage}--\pageref{lastpage}}
\maketitle

\begin{abstract}
We study the \Lya escape fraction, \fesca of \Lya emitters (LAEs) identified by Subaru/HSC narrowband imaging at $z = 2.2\mathrm{-}6.6$, using publicly available deep imaging data from HST and JWST. %By combining Subaru/HSC narrowband imaging with JWST’s deep near- and mid-infrared photometry, we identify a robust sample of LAEs and 
% We perform SED fitting for 127 LAEs 
% %at rest-wavelengths, xx um- xxum, 
% to estimate their physical properties robustly.
%As in previous studies, we 
% We confirm that two distinct LAE populations exist: young LAEs ($<100$ Myr) and old LAEs ($>100$ Myr).
We perform SED fitting for 127 LAEs at $0.4 \mathchar`- 5.0\, \mathrm{\mu m}$ to
estimate their physical properties robustly, and confirm that two distinct LAE populations exist: young LAEs (< 100 Myr) and
old LAEs (> 100 Myr).
%By comparing with SFGs, 
Young LAEs are characterized by burst-like star formation activity and low dust content, 
%exhibiting properties and 
significantly differing from Lyman-break galaxies (LBGs) at the same stellar mass, while old LAEs show similar star formation activity to  LBGs, yet with lower dust content and more compact morphology in rest-UV/optical than LBGs. 
The \fesca of LAEs is anticorrelated with stellar mass, and this correlation is found to depend on the age of LAEs, such that old LAEs show a weaker anticorrelation than young LAEs, and tend to exhibit higher \fesca than young LAEs at a given stellar mass.
%their dependencies vary with age. 
% old LAEs tend to exhibit higher \fesca than young LAEs at a given stellar mass.
This implies that \Lya photons escape more efficiently from old LAEs, possibly through the low-density channels of \Hi\ and dust created by outflows.
%This may be attributed to the more efficient formation of low-density channels in the ISM driven by outflows. 
The average \fesca of young LAEs 
%shows little to no redshift evolution and 
remains nearly constant at $\sim40$\% at $z = 2.2\mathchar`-6.6$, suggesting that the previously observed evolution of global \fesca of star-forming galaxies (SFGs) is 
%not due to the redshift evolution of \fesca itself, but rather 
due to the changes in the LAE fraction among the SFGs.
%Moreover, by estimating 
% Using either empirical or model-based conversions from \fesca to the Lyman continuum escape fraction,
Converting \fesca to Lyman continuum escape fraction using empirical relations,
%integrating the \Lya luminosity function, 
we demonstrate that LAEs alone can supply the ionizing photons necessary for reionization at $z\sim6$, causing rapid and late reionization.
%These findings show that LAEs play a important role in cosmic reionization, meaning they will remain essential for studying the early Universe.

%, although the result depends on the conversion used.
%In this model, it is also suggested that faint LAEs could play a particularly significant role in contributing to the ionizing photon budget.
\end{abstract}


%%%%%%%%%%%%%%%%%%%%%%%%%%%%%%%%%%%%%%%%%%%%%%%%%%

%%%%%%%%%%%%%%%%% BODY OF PAPER %%%%%%%%%%%%%%%%%%

\section{Introduction}
Since it was predicted in \citet{Partridge+67}, galaxies that exhibit strong \Lya emission, known as \Lya emitters (LAEs), are useful tools to reveal the distant universe.
The \Lya emission line (1216~\AA\ in the rest-frame) corresponds to the resonant $\mathrm{2p}\xrightarrow{}\mathrm{1s}$ transition of ubiquitous hydrogen atoms, and it is the brightest line in the hydrogen recombination lines.
However, the \Lya photons are resonantly scattered in the atomic gas 
and are particularly susceptible to absorption by dust in galaxies.
Consequently, the escape of \Lya photons from galaxies into the intergalactic medium (IGM) depends critically on the condition of the interstellar medium (ISM) \citep{Finkelstein+09,Herenz+25}. 
Furthermore, at $z>6$, \Lya photons are greatly absorbed by the IGM neutral hydrogen (\Hi) that has covered the entire universe since the recombination epoch (e.g. \citealt{Finkelstein+15}). %\citep{Kageura+25,Jones+25}
Recently, JWST has observed numerous LAEs \citep{Witstok+24,Protu+24} even in the reionization era, but they are thought to exist in ionized bubbles in neutral IGM.
Given that the IGM absorption at high-$z$ is significant, observations of LAEs at $z\lesssim5$, which are almost unaffected by the IGM absorption, are crucial to understand the mechanisms behind \Lya escape.

In general, LAEs are characterized as galaxies with low dust content due to their young ages (e.g. \citealt{Haro+20}) and low masses (e.g. \citealt{Khostovan+19,Santos+20}).
Similarly, in galaxies with a blue UV slope, which reflects the low amount of dust, \Lya photons are more likely to escape (e.g. \citealt{Jiang+16}).
A low \Hi\, covering fraction (e.g. \citealt{Gazagnes+20}) and metallicity (e.g. \citealt{Trainor+16}) also promote the escape of \Lya photons.
The \Lya escape fraction, \fesca, which is defined as the ratio of the escaped \Lya luminosity to that intrinsically produced, significantly increases by 2 dex from $z = 0$ to 6 \citep{Hayes+11}, in line with the evolution of both dust and \Hi\, covering fraction.
The escape mechanism of \Lya photons in galaxies is thought to further depend strongly on the detailed ISM physical condition, through a clumpy ISM \citep{Finkelstein+09} or vigorous outflows \citep{Trainor+15}.
The outflows in LAEs are thought to originate from their compact nature and extremely high star formation surface densities \citep{Afonso+18,Pucha+22,Kim+25}. 
These outflows may facilitate the creation of low-density channels in their ISM that aid the escape of \Lya photons \citep{Verhamme+15}.
\citet{Marchi+19} 
found that galaxies with higher ISM outflow velocities exhibit smaller \Lya velocity shifts, which implies
a lower \Hi\, column density based on the shell model of \citet{Verhamme+06}, suggesting that \Lya photons can escape more easily. 
These findings support the idea that the physical conditions of the ISM, particularly those driven by outflows, play a crucial role in regulating \Lya escape.

While LAEs are generally thought of as young, less massive star-forming galaxies (SFGs), 
some LAEs have been found to show \Lya emission despite their old stellar ages 
(e.g. \citealt{Lai+08,Finkelstein+09,Ono+10,Rosani+20,Iani+24,Firestone+25}). 
As suggested in \citet{Iani+24}, the difference in the spectral energy distribution (SED) between young ($\mathrm{age}<100\,\mathrm{Myr}$) and old ($\mathrm{age}>100\,\mathrm{Myr}$) LAEs is most noticeable at wavelengths longer than the Balmer break, which is important for determining the age of the galaxy.
Therefore, JWST observations capable of deeply tracing long wavelengths up to $20\mu$m are particularly important for high-$z$ LAEs with faint continuum fluxes. 
\citet{Iani+24} reported that, at $z\sim3$, approximately 29\% of the LAEs have old ages, and the fraction of old LAEs decreases at higher redshifts \citep{Nilsson+09}.
It is speculated that these old LAEs are sufficiently evolved galaxies in the process of experiencing rejuvenation. 
Given that young and old LAEs may have undergone distinct star formation histories \citep{Shimizu+10, Firestone+25}, it is conceivable that the mechanisms facilitating the escape of \Lya photons differ between them \citep{Herenz+25}.

The intense \Lya radiation is also widely recognized as a proxy of Lyman-continuum (LyC) leakage (e.g. \citealt{Verhamme+17,Marchi+18,Steidel+18,Pahl+21}), which are expected to efficiently reionize the universe.
It is easy to imagine that LyC photons can escape from galaxies through low-density ISM channels \citep{Gazagnes+20,Erik+25} in the same way as \Lya photons, so it is not surprising that LAEs could make an important contribution to cosmic reionization.
As many studies have suggested, a positive correlation exists between the escape of \Lya photons and LyC photons (e.g. \citealt{Verhamme+15,Dijkstra+16,Izotov+20,Begley+24,choustikov+24a,Choustikov+24c,Izotov+24,Le_Reste+25}).
The main driver of cosmic reionization is still under debate, but two primary scenarios have been proposed for driving reionization: 
one in which numerous faint galaxies in UV luminosity are the dominant ionizing sources (e.g. \citealt{Finkelstein+19,Lin+24}), and the other in which relatively bright galaxies play a more significant role (e.g. \citealt{Naidu+20, Matthee+22}).
Although there has been much debate,
it is also important to re-examine the issue from the perspective of \Lya radiation brightness, which could be more closely correlated with LyC radiation.
Indeed, \citet{Matthee+22} argued that LAEs could have driven late and rapid reionization at $z\approx6\mathchar`-9$, based on the reasonable assumption that half of the bright LAEs with \Lya luminosities exceeding $10^{42.2}\, \mathrm{erg\,s^{-1}}$ are LyC leakers.
A comprehensive assessment of LAEs, from their intrinsic properties and escape fractions to their contribution to the ionizing photon budget, is essential for unraveling the reionization history of the Universe.

In this study, we derive the stellar population properties of LAEs at $z=2\mathchar`-6$ by applying the SED fitting to ultra-deep HST and JWST data sets at $0.4 \mathchar`- 5.0\, \mathrm{\mu m}$, to examine the dependence of \fesca on these properties.
Understanding the physics of \fesca in the post-reionization universe will help interpret the origin of the \Lya emission observed at the epoch of reionization.
Building on this analysis, we further quantify the ionizing photon budget from LAEs during the epoch of reionization.
In contrast to previous studies, which often rely on simplified assumptions or indirect estimates of the escape fraction of ionizing photons, this work takes an observational approach by modelling the relationship between \Lya luminosity and the number of ionizing photons escaping into the IGM. 
This allows a more quantitative understanding of the role of LAEs in driving cosmic reionization.
In this study, we utilize LAE samples identified through narrowband (NB) imaging, a method that enables a robust measurement of the total \Lya flux.
It may be more reliable to use the JWST spectroscopically confirmed LAEs, but due to the extended nature of \Lya emission (e.g. \citealt{Momose+14,Kikuta+23_halo}), the NIRSpec/micro-shutter assembly can suffer from slit losses and has been known to miss \Lya flux altogether (e.g. \citealt{Bhagwat+24,Jiang+24}).

The structure of this paper is as follows. 
Section~\ref{sec:data} describes the LAE catalogue selected using the Subaru NB filter and the JWST imaging catalogue. 
In Section~\ref{sec:sample_method}, we explain the matching of these two datasets to select LAEs, the methods used to characterize their properties (SED fitting, size measurement, and \fesca calculation), 
and the selection of Lyman-break galaxies (LBGs) for comparison.
Section~\ref{sec:result} presents the main results.
First, we compare the properties of LAEs and LBGs (Section~\ref{sec:LAE_LBG}), and then discuss how the characteristics of LAEs vary with their age (Section~\ref{diff_young_old}). 
Next, we examine the continuum sizes of LAEs (Section~\ref{sec:LAE_size}) and 
their age dependence of the mass–\fesca relation, and further investigate the redshift evolution of the \fesca of LAEs (Section~\ref{sec:Lya_escape}).
Finally, we evaluate the contribution of LAEs to cosmic reionization (Section~\ref{sec:reionization}) and the evolution of IGM neutral fraction (Section~\ref{sec:reionization_history}).
Throughout this paper, a flat $\Lambda$CDM cosmology is assumed with $H_0=70\,\mathrm{km\,s^{-1}\,Mpc^{-1}}$, $\Omega_\mathrm{m}=0.3$, and $\Omega_\Lambda=0.7$. All magnitudes in this paper refer to AB magnitude \citep{Oke}.

\section{Data} \label{sec:data}
\subsection{Subaru/HSC NB imaging} \label{Data_HSC}
We use the LAE sample obtained from the Hyper Suprime-Cam Subaru Strategic Program (HSC-SSP; \citealt{Aihara+18a, Aihara+18b, Aihara+19, Aihara+22}) S21A internal release, one of the most comprehensive NB imaging surveys for studying LAEs across a wide range of redshifts.
Using the HSC-SSP full-depth NB and broadband (BB) imaging data, \citet{Kikuta+23} incorporated deep imaging data from the CHORUS  (Cosmic HydrOgen Reionization Unveiled with Subaru) programme \citep{Inoue+20}. 
This combined dataset enabled the identification of 20,567 LAEs over 25 $\mathrm{deg^2}$ at redshifts $z=2.2, 3.3,4.9, 5.7, 6.6,7.0$, and 7.3, corresponding to NB387, NB527, NB718, NB816, NB921, NB973, and NB1010 filters, respectively.
These observations spanned multiple fields, including COSMOS
and Subaru XMM-Newton Deep Survey field (SXDS). 
LAEs were selected based on the NB excess relative to BB, and their selection criteria roughly correspond to $>10\mathchar`-20$~\AA~ in the rest-frame \Lya equivalent width (EW$_{\mathrm{Ly\alpha}}$), depending on the transmission curves \citep{Ono+21}. 
Please refer to \citet{Kikuta+23} for the details of the NB observations, the data reduction, and LAE selection.
In this study, as detailed in Section~\ref{sec:JWST}, we use parts of the COSMOS and SXDS fields where JWST data is available.
These regions correspond to the ultradeep (UD) fields of the HSC-SSP, where the deepest observations have been conducted.
Within HSC-SSP, these fields are referred to as UD-COSMOS and UD-SXDS, while in JWST observations, they are designated as PRIMER COSMOS and PRIMER UDS, as part of the Public Release IMaging for Extragalactic Research (PRIMER; \citealt{Dunlop+21}) programme.
 All the NB imaging data mentioned above are available in the PRIMER COSMOS field, while in the PRIMER UDS field, only NB816, NB921, and NB1010 imaging data are available. 
 The $5\sigma$ depths for each NB image in the UD regions are summarized in Table~\ref{tab:limitmag}. 
 To calculate the limiting $5\sigma$ magnitude, we first make segmentation maps for the images in each filter using SExtractor \citep{sextractor}. 
 The detected sources are then masked, and $2.0\arcsec$ apertures are randomly scattered on the background to estimate its standard deviation.

\subsection{JWST/NIRCam and MIRI imaging} \label{sec:JWST}
We use the JWST imaging data and photometric catalogues in DAWN JWST Archive (DJA)\footnote{\url{https://dawn-cph.github.io/dja/index.html}}.
DJA is a repository of public JWST data, and we use their \texttt{v7} imaging data and photometric catalogues of PRIMER COSMOS and UDS.
In this programme, the imaging observations were performed over a very wide wavelength range using NIRCam's F090W, F115W, F150W, F200W, F277W, F356W, F410M, and F444W filters, and MIRI's F770W and F1800W filters. 
In addition, deep imaging was also carried out in these regions by HST CANDELS survey \citep{CANDELS}, which, together with the JWST, covers a wavelength range of $0.4 \mathchar`- 18\, \mathrm{\mu m}$.
The JWST and HST images were reduced with \texttt{grizli}\footnote{\url{https://github.com/gbrammer/grizli}} \citep{Brammer+23} and \texttt{msaexp}\footnote{\url{https://github.com/gbrammer/msaexp}} \citep{msaexp}, and reconstructed with a pixel scale of $0.04\arcsec$ using \texttt{drizzle }\citep{drizzle}.
For source detection, the composite images created by optimally weighting the noise maps of NIRCam’s long-wavelength filters (F277W, F356W, and F444W) were used. 
Based on that source detection, DJA/\texttt{grizli} catalogues contain the photometries of the HST and JWST filters shown in Table~\ref{tab:limitmag}.
Please refer to \citet{Valentino+23} for the details of the data reduction and source detection.
The $5\sigma$ depths, measured using $0.7\arcsec$ aperture, and the full width at half maximum (FWHM) of the point spread functions (PSFs) for the HST and JWST filters are also summarized in Table~\ref{tab:limitmag}. 
The method of PSF measurement is described in Section~\ref{PSF_mesurement}. 
 
\begin{table}
\centering
\caption{The $5\sigma$ depths and FWHMs of the PSF for each filter used in this paper. For Subaru/HSC, the $5\sigma$ depth measurements are made using a $2.0\arcsec$ aperture, and for HST and JWST, a $0.7\arcsec$ aperture is used. The FWHM of the PSF is measured only for the filters used in the size analysis of this paper.}
\begin{tabular}{llccc}

Instrument & Filter & COSMOS & UDS & PSF [$\arcsec$] \\ \hline
Subaru/HSC & NB387  & 25.70  & --  & 0.964 \\
 & NB527  & 26.82  & --  & 0.788 \\
& NB718  & 26.45  & --  &  -- \\
 & NB816  & 26.21  & 26.18 & -- \\
& NB921  & 26.17  & 26.15 & \\
 & HSC \textit{g} &  27.51      &  --      & 0.745 \\
 & HSC \textit{r} &  27.18     &   --     & 0.808 \\ \hline
HST/ACS & F435W  & 28.05     & 26.97 & -- \\
& F475W  & 27.16     & --  & --\\ 
 & F606W  & 27.60 &  27.63 & 0.094\\
 & F814W  & 27.55 &  27.65 & 0.095\\ 
HST/WFC3-IR & F105W & 27.52 & 27.52 & --\\
 & F125W & 27.30 &  27.64 & --\\
 & F140W & 27.22 & 27.18 & --\\ 
 & F160W & 27.33 & 27.47 & --\\ \hline
JWST/NIRCam & F090W &  27.78& 27.43 & --\\
 & F115W & 27.52 & 27.44 & 0.054\\
 & F150W & 27.73 & 27.73 & 0.066\\
 & F200W & 28.32 & 27.85 & 0.075\\
 & F277W & 28.40 & 28.25 & 0.118\\
 & F356W & 28.72 & 28.72 & 0.132\\
 & F410M & 28.21 & 27.61 & --\\
 & F444W & 28.13 & 27.97 & 0.158\\
JWST/MIRI & F770W & 26.55 & 26.39 & --\\
 & F1800W & 24.51 & 24.49 & -- \\ \hline

Area [arcmin$^2$]   &  & $\sim$190      &   $\sim$230 
\end{tabular}
\label{tab:limitmag}
\end{table}

\section{Sample and Method} \label{sec:sample_method}
\subsection{LAE sample selection} \label{LAE selection}
We select LAEs from \citet{Kikuta+23} in regions that overlap with the PRIMER COSMOS and UDS fields. 
Unfortunately, there are no $z=7.0$ and 7.3 LAE samples at PRIMER COSMOS and UDS fields.
We identify initial samples of 54, 154, 27, 37, and 18 LAEs at $z=$ 2.2, 3.3, 4.9, 5.7, and 6.6, respectively, totaling 290 LAEs. 
Next, we perform a coordinate matching of these LAEs with those in the DJA/\texttt{grizli} photometric catalogues, using a matching radius of $0.5\arcsec$. 
As a result, 185 LAEs are matched to a single object, 35 are matched to multiple objects, and the remaining 70 have no corresponding objects. 
A single object is defined as a case where only one counterpart of an LAE, presumably the host galaxy’s continuum component, is found within a $0.5\arcsec$ radius from the \Lya emission peak. 
Conversely, if two or more objects are found within this radius, the LAE is classified as a multiple object.
Even if it is a single object in the HSC NB image,  
there are cases where it corresponds to multiple objects that JWST could resolve with its high spatial resolution.
Moreover, as suggested by \citet{Matthee+21}, LAEs may have clumpy structures, which could lead to a single object being split into multiple components. 
Possible explanations for the LAEs with no corresponding objects are that they are continuum-faint galaxies with extremely strong \Lya emission, or fluorescent sources \citep{Cantalupo+12,Marino+18}.
It is important to note, however, that as previously mentioned, the DJA/grizli photometric catalogues are constructed based on the detections in NIRCam’s long-wavelength filters. 
Therefore, a "single object" here refers to an object identified as a single source in these filters. 
In some cases, higher-resolution images in shorter-wavelength filters reveal that these sources are in fact composed of multiple components. 
Such cases are excluded from the single object sample through subsequent visual inspections.
Similarly, the LAEs without counterparts may simply lack detections in the NIRCam long-wavelength filters.
However, since the composite images of these long-wavelength filters are significantly deeper than those of the shorter-wavelength filters, sources detected only in the latter are typically extremely faint and are detected in only a limited number of bands. 
Such sources generally fall below the minimum number of photometric bands required for reliable SED fitting, as described below, and are therefore excluded from the final sample.

In this study, we only focus on objects matched with a single counterpart to reliably characterize the nature of the host galaxy emitting \Lya photons. 
Multiple objects, including clumpy galaxies, are excluded from this sample because there is a considerable uncertainty in the SED fitting depending on which components are selected as the NB counterparts.
Furthermore, even in clumpy galaxies, there would be significant ambiguity when distributing the NB flux of a single object among multiple counterparts.
For the 185 LAEs matched with a single counterpart, to ensure the reliability of the SED fitting performed in Section~\ref{SED_fitting}, 
we restrict our sample to LAEs detected above the $2\sigma$ limit in at least eight bands.
All of our samples have been confirmed to have detections with $>3\sigma$ significance in five or more bands.
In addition, to ensure a reliable estimation of the UV $\beta$ slope, sources detected in fewer than two bands within the UV wavelength range ($1500 \mathchar`- 3000$~\AA) are also excluded from the sample.
We exclude AGNs, as they are thought to have a different origin for \Lya photons.
To remove AGNs, as done by \citet{Iani+24}, we employ two approaches: (1) cross-matching with X-ray and radio catalogues and (2) imposing an upper limit on \Lya luminosity. 
In the first approach, we use publicly available X-ray and radio catalogues. 
For PRIMER COSMOS, we use the X-ray point source catalogues from the Chandra Observatory \citep{chandra-cosmos} and XMM-Newton \citep{XMM-cosmos}, and the $3 \mathrm{GHz}$ (10 cm) radio source catalogue from the Very Large Array (VLA) \citep{VLA-cosmos}. 
Similarly, for PRIMER UDS, we use the X-ray point source catalogues from Chandra \citep{chandra-uds} and XMM-Newton \citep{XMM-uds}, and the $1.4 \mathrm{GHz}$ radio source catalogue from the VLA \citep{VLA-uds}. 
As a result, we exclude one object at $z\sim2.2$ in PRIMER-COSMOS. 
In the second approach, we exclude objects with $L_{\mathrm{Ly\alpha}}$ greater than $10^{43} \, \mathrm{erg\,s^{-1}}$, where 
the nature of LAEs shifts from being dominated by star formation to being AGN-dominated ($\sim60\%$ at $L_{\mathrm{Ly\alpha}}>10^{43}\, \mathrm{erg\,s^{-1}}$, \citealt{Sobral+18b}). 
While this luminosity threshold may also exclude high-mass SFGs with high \Lya luminosities 
without AGNs, distinguishing them from genuine AGNs based solely on photometry data is challenging. 
Therefore, we uniformly applied this cut.
Through this approach, five LAEs at $z=3.33$, four at $z=4.90$, three at $z=5.72$, and one at $z=6.58$ are excluded. 
The method for calculating $L_{\mathrm{Ly\alpha}}$ is described in Section~\ref{sec:Lya_escape}. 
Finally, we conduct a visual inspection to exclude objects significantly affected by cosmic rays or nearby bright stars, or that appear as multiple objects in other wavelength images.
Finally, a total of 127 LAEs are selected.

\subsection{Estimates of physical properties} 
\subsubsection{SED fitting} \label{SED_fitting}
To estimate the physical properties of LAEs from the multi-wavelength photometric data, we perform the SED fitting using CIGALE \citep{Boquein+19}. 
To combine the ground-based NB photometries with the space-based photometries from HST and JWST while minimizing flux loss due to PSF differences, we use \texttt{MAG\_AUTO} of SExtractor for the NB data. 
For the broadband photometries on HST and JWST, we also employ the total magnitudes from the DJA/\texttt{grizli} photometric catalogues. 
These total magnitudes are derived by applying an aperture correction \citep{Valentino+23} in the detection filter, based on a $0.5\arcsec$ aperture photometry and Kron aperture photometry \citep{Kron}.
As shown in Table~\ref{tab:limitmag}, since the FWHMs of PSF in the NB images are considerably larger than those of HST and JWST, the NB photometries may potentially include fluxes from nearby sources. 
However, as noted in Section~\ref{LAE selection}, we use LAE samples that have no nearby sources within $0.5\arcsec$ and are not resolved into multiple components in the JWST images, thereby minimising the possibility of contamination.

\begin{figure*}
	\includegraphics[width=\textwidth]{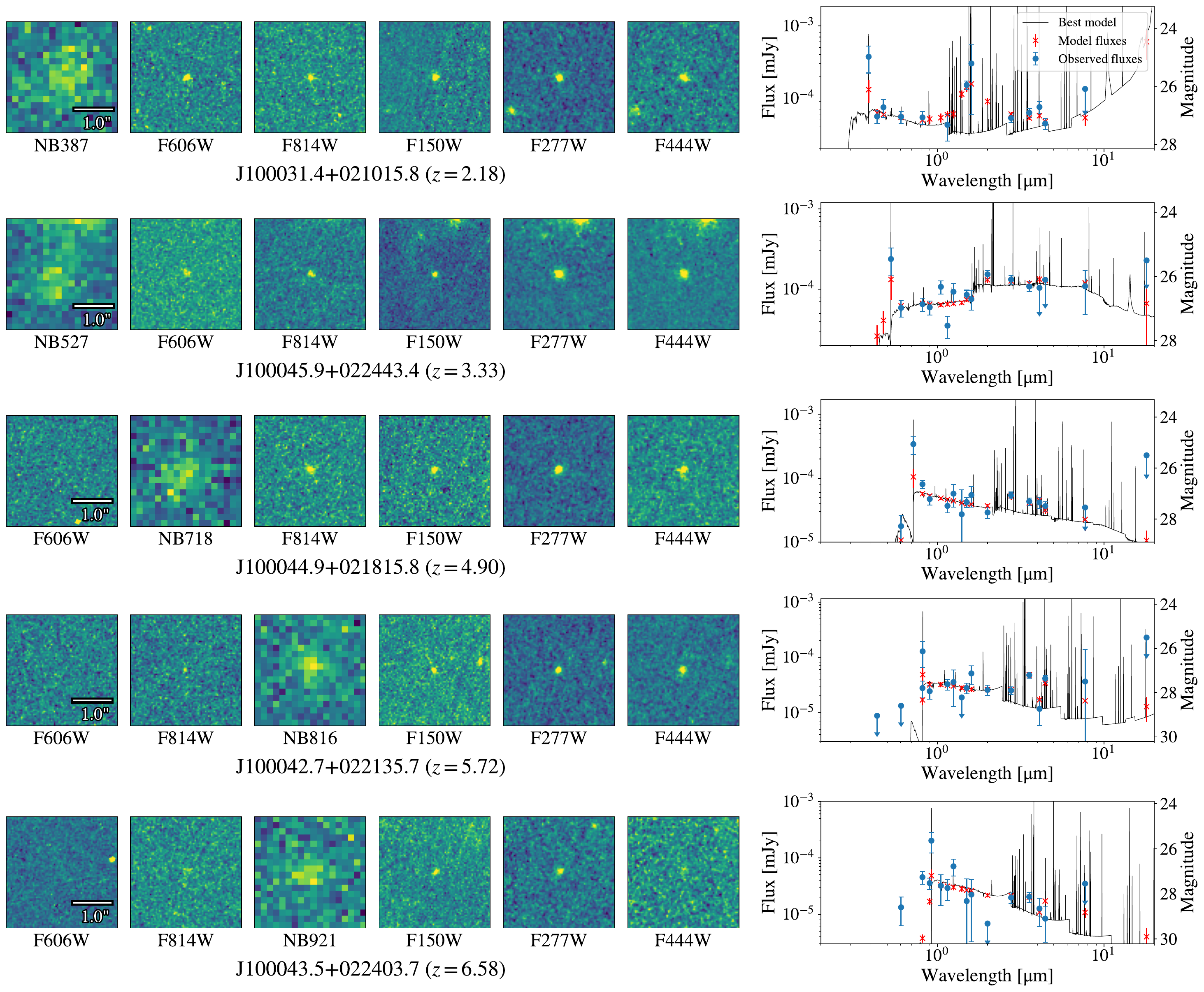}
    \caption{Examples of cutout images and SED fitting results for LAEs at each redshift. The left-hand panels show $3.0\arcsec \times 3.0\arcsec $ cutout images, with the filter name indicated below each panel. The right-hand panel shows the SED of the LAE: the blue points represent the observed fluxes, the black line indicates the best-fitting SED model, and the red points mark the model-predicted fluxes in each filter.}
    \label{fig:SED}
\end{figure*}

CIGALE provides SED libraries containing various models of star formation histories (SFHs), stellar populations, and dust attenuations, which are used to fit the model SEDs to the observed photometries. 
Bayesian methods are employed to estimate the physical properties and their errors. 
In this study, we assume a Chabrier initial mass function (IMF, \citealt{Chabrier+03}) for the single stellar population (SSP) model \citep{BruChar+03}. 
The potential effects on our results from using different modeling assumptions, such as binary stellar populations (e.g. \citealt{Eldridge+17}), a top-heavy IMF, or a different high-mass cutoff for the IMF, are discussed in Section~\ref{sec:reionization_history}.
For the SFH, we adopt a $\tau$-model, where star formation rate, $\mathrm{SFR}(t) \propto \exp{\left(-\frac{t}{\tau}\right)}$, and parameterize their stellar age on a logarithmic scale from 2 Myr up to the cosmological age at the corresponding redshift. 
The stellar age we use in this study represents the time elapsed since the onset of the first star formation in the assumed SFH.
In this age definition, it should be noted that the age uncertainty increases for galaxies that had a burst of star formation a long time ago but then experienced a more recent burst.
We take this uncertainty into account in the subsequent analysis.
The metallicity is considered over the range $Z = 0.02Z_{\odot} \mathchar`- Z_{\odot}$. 
We use the attenuation law by \citet{Calzetti+00}, with E(B-V) ranging from 0 to 1.0. 
The redshift of LAE is fixed to the redshift of the \Lya emission line at the central wavelength of the corresponding NB filter.
To ensure the accuracy of the SED fitting, magnitudes fainter than the $2\sigma$ depth in each filter are treated as upper limits. 
We note that our sample includes photometric data at wavelengths longer than the Balmer break at all redshifts. 
This increases the likelihood that the degeneracy of ages and dust amounts can be resolved, as noted in \citet{Iani+24}. 
After performing the SED fitting, 
we assess the goodness of fit using \texttt{reduced\_chi\_square} ($\chi^2_{\texttt{reduced}}$) of CIGALE, which quantifies how well the model matches the observed data.
The LAEs with $\chi^2_{\texttt{reduced}}$ greater than five are excluded from the sample due to their large uncertainty in the SED fitting. 
In practice, however, all objects satisfied the $\chi^2_{\texttt{reduced}}$ criterion, and no sources were removed based on this threshold.
Figure \ref{fig:SED} shows examples of the SED fitting results at each redshift.
To check the validity of physical quantities such as SFR, stellar age, and stellar mass output by CIGALE, we compare our results with those of the SED fitting carried out in DJA/grizli photometric catalogue using EAZY \citep{EAZY} for all objects in the catalogue, and confirm that there are no significant discrepancies.
We also utilize the \texttt{mock\_flag} function in CIGALE to assess the reliability of our fitting procedure.
This function automatically generates mock catalogs by adding observational errors to model SEDs based on the best-fit parameters, and then perform SED fitting on these mocks again. 
The physical quantities recovered through this process show no significant difference from the input values, confirming that our fitting method robustly retrieves these parameters.
The posterior distributions of the physical parameters obtained with CIGALE are derived using a method that explores the entire model grid, allowing the likelihood for each parameter value to be determined. 
Using these likelihoods, we further confirm that their age and E(B-V) are not degenerate in the SED fitting performed with the current dataset. 
The number of the final samples of LAEs is summarized in Table~\ref{tab:LAE_samples}.
\begin{table}
	\centering
	\caption{Summary of the number of LAE samples.}
	\label{tab:LAE_samples}
	\begin{tabular}{lccr} % four columns, alignment for each
		\hline
		NB filter & Area & Redshift & Number\\
		\hline
		NB387 & COSMOS & 2.18 & 20\\
		NB527 & COSMOS & 3.33 & 69\\
            NB718 & COSMOS & 4.90 & 13\\
            NB816 & COSMOS \& UDS & 5.71 & 20\\
            NB921 & COSMOS \& UDS & 6.57 & 5\\
            \hline
            Total       &       &   &  127\\

	\end{tabular}
\end{table}

\subsubsection{SFR and UV magnitude}
Although SFR is estimated through the SED fitting, we use the SFR evaluated from the UV luminosity to enable a fair comparison with other studies.
To calculate $\mathrm{SFR_{UV}}$ for each LAE, the model flux at 1500\,\AA\, obtained from the SED fitting is converted to apparent magnitude ($m_{1500})$. 
The absolute magnitude ($M_{1500}$) is then derived by:
\begin{equation} \label{eq:Mag}
    M_{1500}=m_{1500}-5\log{\left(\frac{D_L(z)}{10\mathrm{pc}}\right)}+2.5\log(1+z),
\end{equation}
where $D_L$ is the luminosity distance. 
Next, we apply a dust correction to obtain the intrinsic UV absolute magnitudes, using the E(B-V) values derived from the SED fitting and the Calzetti law \citep{Calzetti+00}. 
The corrected absolute magnitude is then converted into the monochromatic luminosity at 1500\,\AA\, ($L_{1500}$). Following \citet{Kennicutt+98}, the $\mathrm{SFR_{UV}}$ is calculated as:
\begin{equation}
    \mathrm{SFR_{UV}} =0.63\times \mathrm{C_{UV}}\,L_{1500}\, ,
\end{equation}
where $\mathrm{C_{UV}}=1.4\times 10^{-28} \,\mathrm{M_{\odot}\, yr^{-1}\, erg^{-1}\, s\, Hz}$, which is the conversion factor assuming a Salpeter IMF \citep{Salpeter}. 
To maintain consistency with the Chabrier IMF that we use in the SED fitting in this study, we multiply $\mathrm{SFR_{UV}}$ by 0.63
\citep{Madau+14}.

\subsubsection{Ly$\alpha$ escape fraction}
We evaluate the \fesca, which indicates the fraction of \Lya photons produced within the galaxy that escape into the IGM.
The relative intensity of the \Lya and \Ha emission lines due to hydrogen recombination depends on the temperature and electron density of the gas cloud \citep{Hummer}. 
Here, similar to other studies (e.g. \citealt{Lin+24}), we assume Case B recombination in a gas cloud with a temperature of $T\sim10^4\,\mathrm{K}$ and an electron density of $n_e\sim350\,\mathrm{cm^{-3}}$. 
In this case, the intrinsic \Lya/\Ha ratio is 8.7, and \fesca can be expressed as follows:
\begin{equation}\label{eq:f_esc}
f_{\mathrm{esc}}^{\mathrm{Ly\alpha}}=\frac{F_{\mathrm{Ly\alpha,obs}}}{8.7F_{\mathrm{H\alpha,int}}},
\end{equation}
where $F_{\mathrm{Ly\alpha,obs}}$ is the observed \Lya flux and $F_{\mathrm{H\alpha,int}}$ is the intrinsically produced \Ha flux. 
The \Lya flux, $F_{\mathrm{Ly\alpha}}$, is calculated using the NB magnitude, $m_{\mathrm{NB}}$,  and the continuum flux at 1216\AA\,, $F_{1216}$, derived from the SED fitting, using a formula similar to that of \citet{Shibuya+18}: 
\begin{equation} \label{eq:shibuya}
    48.6+m_{\mathrm{NB}}=-2.5\log_{10}\frac{\int^{\infty}_{0}[F_{1216}+F_{\mathrm{Ly\alpha}}\delta(\nu-\nu_{\mathrm{Ly\alpha}})]T_{\mathrm{NB}}\mathrm{d\nu}}{\int^{\infty}_{0}T_{\mathrm{NB}}\mathrm{d\nu}},
\end{equation}
where $T_{\mathrm{NB}}$ is the transmission curve of the NB filter, and $\nu_{\mathrm{Ly\alpha}}$ is the observed frequency of the \Lya line.
Here we assume $\nu_{\mathrm{Ly\alpha}}$ is the central frequency of the $T_{\mathrm{NB}}$, 
and the \Lya line is approximated as a $\delta$-function.
For the continuum at wavelengths shorter than the \Lya line, the IGM absorption is accounted for using the optical depth calculated from the model of \citet{Madau+1995}. 
In addition to the continuum, it is well established that \Lya emission is increasingly affected by IGM absorption at higher redshifts.
In this study, 
IGM absorption of \Lya emission lines is also taken into account, assuming that 
the intrinsic line profile is symmetric with respect to its peak, and that the blue half of the line is absorbed by the IGM. 
We assume that the \Lya line width is sufficiently narrow such that the variation in IGM optical depth across the line can be neglected. 
Based on this method, we apply redshift-dependent corrections to the observed \Lya fluxes, with attenuation factors computed to be 0.91 at $z=2.2$, 0.79 at $z=3.3$, 0.59 at $z=4.9$, 0.54 at $z=5.7$, and 0.51 at $z=6.6$.
Next, the \Ha flux for each LAE is estimated from the SED fitting results.
The \Ha flux reflects star formation activity on short timescales, approximately $5\,\mathrm{Myr}$ (e.g. \citealt{Clarke+24}), and is often used to calculate SFR based on \citet{Kennicutt+98} (e.g. \citealt{Lin+24,Clarke+24}). 
However, since spectroscopic data for the \Ha line are not available for all sample, we adopt an inverse approach: estimating the \Ha luminosity from the SFR.
The UV luminosity and its derived $\mathrm{SFR_{UV}}$ trace star formation over a longer timescale of about $100\,\mathrm{Myr}$. Consequently, $\mathrm{SFR_{UV}}$ is not a suitable indicator for estimating \Ha flux. 
In contrast, the $\mathrm{SFR_{SED}}$, derived from SED fitting, represents the current SFR within the star formation history model that best reproduces the full set of photometric data by self-consistently modelling both star formation activity and dust properties across all wavelengths. 
We, therefore, consider $\mathrm{SFR_{SED}}$ to be a more direct representation of the star formation responsible for \Ha emission within our model compared to $\mathrm{SFR_{UV}}$.
The approach of estimating \Ha luminosity from the SFR obtained via SED fitting and subsequently using this to calculate the \fesca has also been employed in \citet{Goovaerts+24b}. 
Specifically, the \Ha luminosity is calculated using the following equation, based on the method of \citet{Kennicutt+98} adapted for \citet{Chabrier+03} IMF:
\begin{equation} \label{eq:SFR_Ha}
    \log L_{\mathrm{H\alpha,int}}\,[\mathrm{erg/s}]=\log \mathrm{SFR_{SED}}\,\mathrm{[M_{\odot} yr^{-1}]}+41.35.
\end{equation}
To validate this estimate of the \Ha luminosity, we calculate the \Ha luminosity using the broadband photometric measurements that include the \Ha emission line and the continuum estimates derived from the SED fitting, following the approach described in \citet{Begley+24}. 
These show general agreement with each other, and the details of this comparison are provided in Appendix~\ref{sec:Ha_check}.
Thus, we conclude that the estimate of the \Ha luminosity is robust and use it in equation~\ref{eq:f_esc} to calculate the \fesca.

\subsection{LBG sample selection}
We select SFGs based on their continuum properties from the DJA/\texttt{grizli} photometric catalogues to provide a comparative sample to the emission-line selected LAEs.
In this study, we refer to this continuum-selected sample as LBGs, following the convention in previous studies (e.g., \citealt{Haro+20,Iani+24}), although our specific selection relies on the SED fitting rather than traditional color-dropout criteria. 
We first select objects whose photometric redshift (photo-z) estimated by EAZY is within a redshift range of $\pm$0.1 around the redshifts of our LAE sample.
To ensure that the photo-z is appropriately constrained, we apply additional selection criteria.
Following \citet{Valentino+23}, we require $\chi^2/N_{\mathrm{filt}}\leqq8$, where $N_{\mathrm{filt}}$ is the number of filters used for each galaxy, and also constrain
the redshift probability distribution function, $p(z)$, as $p(z)_{\mathrm{peak}} > 0.5$ and $(z_{84} - z_{16})/(2z_{50}) < 0.3$, where $z_{i}$ indicates the redshift at the $i$th percentile of $p(z)$.
Next, we restrict the sample using the same criteria for the available number of bands as described in Section~\ref{LAE selection}. 
We perform SED fitting using exactly the same parameters as those applied to the LAEs in Section~\ref{SED_fitting}.
The $\chi^2_{\texttt{reduced}}$ criterion is set more stringently for LBGs, which have larger redshift uncertainties than LAEs, excluding objects with $\chi^2_{\texttt{reduced}}$ above the 84th percentile, %of the SFG $\chi^2_{\texttt{reduced}}$ distribution,
corresponding to a threshold of  $\chi^2_{\texttt{reduced}} >2$. 
Following the method described in \citet{Pacifici+16}, we apply a specific star formation rate (sSFR) cut at each redshift to identify quiescent galaxies, but found that none of the LBGs in our LBG sample meet this criterion at any redshift.
Although some LBGs lie below the star-forming main sequence (MS) in the stellar mass–SFR plane (Fig.~\ref{fig:MS}), the number of such galaxies is negligible, and even if they are quiescent, their presence does not significantly affect our subsequent analysis.
The LBG sample, which is basically selected by its photo-z, also inevitably includes some LAEs. 
We remove apparent LAEs, which are located within $0.5\arcsec$ from the LAEs in our catalogues, regardless of whether they are single or multiple counterparts.
The final LBG samples used in this study consist of 1,112 galaxies at $z \sim 2.2$, 819 at $z \sim 3.3$, 465 at $z \sim 4.9$, 300 at $z \sim 5.7$, and 505 at $z \sim 6.6$.

\citet{Goovaerts+23} reported that the LAE fraction ($X_{\mathrm{LAE}}$, the number ratio of \Lya emitting galaxies to all SFGs) with EW$_{\mathrm{Ly\alpha}}>$ 25\,\AA\, is approximately 20\% at $-22<\mathrm{M_{1500}}<-18$ at $z\sim3$, and increases to 40\% towards $z\sim5$. 
Therefore, when comparing LAE and LBG samples, we should be aware of the possibility of LAE contamination
, especially at high-$z$. 

\subsection{Size measurement} \label{sec:size_mesurement}
\subsubsection{PSF measurement} \label{PSF_mesurement}
When quantifying the continuum and \Lya spatial extent of LAEs, it is crucial to account for the PSF broadening. 
Here, we describe the method to measure the PSF size in the images for each filter.
We impose the following criteria outlined in \citet{Ito+24} on the objects, which are detected using SExtractor in each filter, to construct the PSF model:
(1) To ensure reliable morphological information, objects must have a signal-to-noise ratio S/N $>20$ within a $0.5\arcsec$ aperture for HST and JWST, or within a $1.5\arcsec$ aperture for Subaru/HSC, and must not be saturated.
(2) The \texttt{ellipticity}, expressed as (a-b)/(a+b), where a and b are the semi-major and semi-minor axes, respectively, must be less than 0.1.
(3) The objects must be at least 25 pixels away from other detected sources or the image edges to ensure accurate measurement of their sizes.%to surely measure their sizes.
(4) When plotting magnitude versus \texttt{flux\_radius}, the objects must lie within $1\sigma$ of the mean sequence to select only stars that can be regarded as point sources.

The selected objects are cut out into 50 $\times$ 50 pixels, and the PSF is constructed by stacking these cutouts for each band using the \texttt{ePSF} module in \texttt{photutils} v2.0.2 \citep{photutils}. 
The FWHM of the constructed PSF is measured, with the results summarized in Table~\ref{tab:limitmag}. 
Additionally, the half-light radius of the constructed PSF, which is used in Section~\ref{sec:measure_cont}, is measured by SExtractor.

\subsubsection{Measurement of continuum size of LAEs} \label{sec:measure_cont}
We measure the size of LAEs in the rest-UV and the rest-optical continuum.
While the distribution of UV emission traces regions of active star formation, that of optical emission reflects the overall extent of the galaxy, including older stellar populations (e.g. \citealt{Allen+24}). 
Therefore, the rest-optical size is crucial for understanding the size of the LAEs.
For the rest-UV (optical) continuum, we adopt wavelengths around 2000\AA\,(6000\AA) across all redshifts. Specifically, we use the F606W (F200W) filter at $z=2.2$, F814W (F277W) at $z=3.3$, F115W (F356W) at $z=4.9$, F150W (F444W) at both $z=5.7$ and $z=6.6$.

First, we cut out each object at each HST and JWST filter with a wide area of $75\times75$ pixels ($0.04\arcsec/\mathrm{pixel}$).
At the same time, for the same band image, we extract 1,000 background images free from any objects and add them to the LAE images respectively, following the approach by \citet{Zhu+24}, to estimate size errors due to background fluctuations.
Using SExtractor, we then measure the half-light radius for each LAE image with 1,000 different background images, and calculate the average value using 3$\sigma$ clipping and the standard deviation.
For each redshift, LAEs without corresponding images in either the UV or optical bands designated above are excluded from the size measurements. 
In addition, some objects are extremely faint 
and do not meet the detection criteria of SExtractor (e.g. \texttt{DETECT\_MINAREA} and \texttt{DETECT\_THRESH}).
In the case of non-detection in certain filters, their sizes are set to an upper limit corresponding to the half-light radius of the PSF.
After measuring the half-light radius of LAEs for each filter, we correct the PSF broadening using the following equation, as described in \citet{Zhu+24}:
\begin{equation}
    r_{\mathrm{50,int}}=\sqrt{\mathrm{r_{50,obs}^2}-r_{\mathrm{50,PSF}}^2}
\end{equation}
where $r_{\mathrm{50,int}}$ represents the corrected half-light radius, $r_{\mathrm{50,obs}}$ is the observed half-light radius measured from the images, and $r_{\mathrm{50,PSF}}$ is the half-light radius of the PSF. 
For objects with sizes set to the upper limit (corresponding to the half-light radius of the PSF), this correction is not applied.

\section{Results \& Discussion} \label{sec:result}
\subsection{Comparison of LAE and LBG Properties} \label{sec:LAE_LBG}
\begin{figure}
    \centering
    \includegraphics[width=\columnwidth]{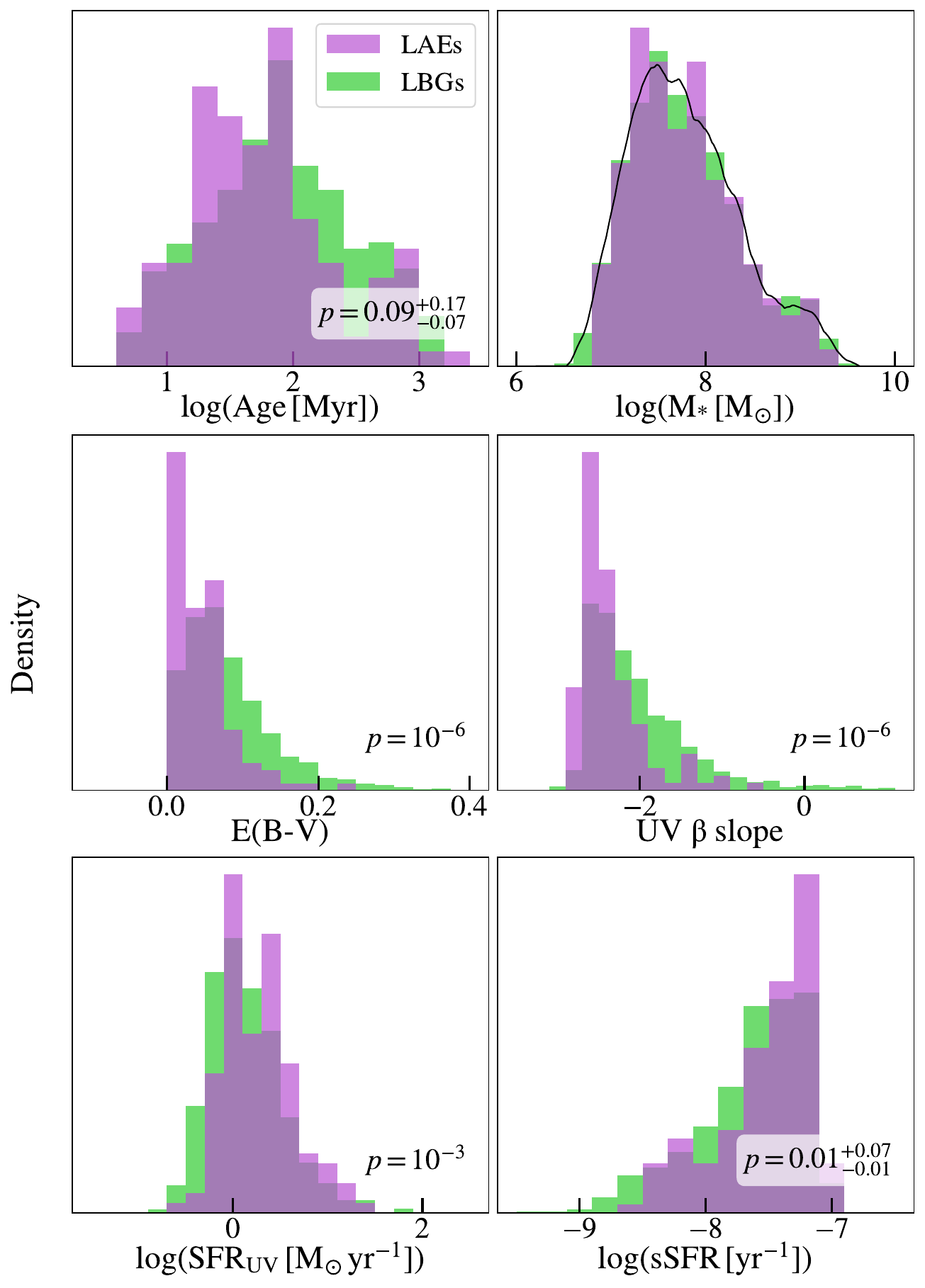} 
    \caption{Comparison of physical properties (stellar mass, age, E(B-V), UV slope $\beta$, $\mathrm{SFR_{UV}}$, and sSFR) between the LAE sample (purple) and the LBG sample (green). To mitigate biases due to differences in stellar mass distributions, the LBGs are resampled to match the stellar mass distribution of the LAEs. As one example, the PDF of the LAE stellar masses used for the resampling is overplotted as a black line in the stellar mass panel (top right). The p-values from the KS test for the two samples are shown in each panel. For clarity, p-values significantly below 0.05 are shown in exponential notation without their associated errors.}
    \label{fig:ALL_LAE_hist}
\end{figure}
Following the method of \citet{Iani+24}, we mitigate the biases arising from stellar mass difference by matching the mass distribution of LBGs to that of LAEs, in order to enable a fair comparison of their physical properties.
Specifically, for both LAE and LBG samples, we estimate the probability density functions (PDFs) of the stellar mass distributions using kernel density estimation (KDE). 
The PDF for the LAE sample is overplotted as a solid black curve on the mass distribution in Fig.~\ref{fig:ALL_LAE_hist}. 
For each LBG, we then calculate a selection probability as the ratio of the LAE PDF to the LBG PDF at the given stellar mass. 
LBGs are randomly resampled according to this selection probability.  
To quantitatively assess whether the observed distributions of these physical quantities shown in Fig.~\ref{fig:ALL_LAE_hist} originate from distinct parent distributions, we also conduct a two-sample Kolmogorov-Smirnov (KS) test.
We adopt a p-value threshold of 0.05 to signify that two samples are drawn from different distributions. 
LBG resampling is repeated 100 times, and for each repetition, the parameters are perturbed
assuming a normal distribution based on their errors, resulting in a total of 10,000 KS tests.
\begin{figure}
    \centering
    \includegraphics[width=\columnwidth]{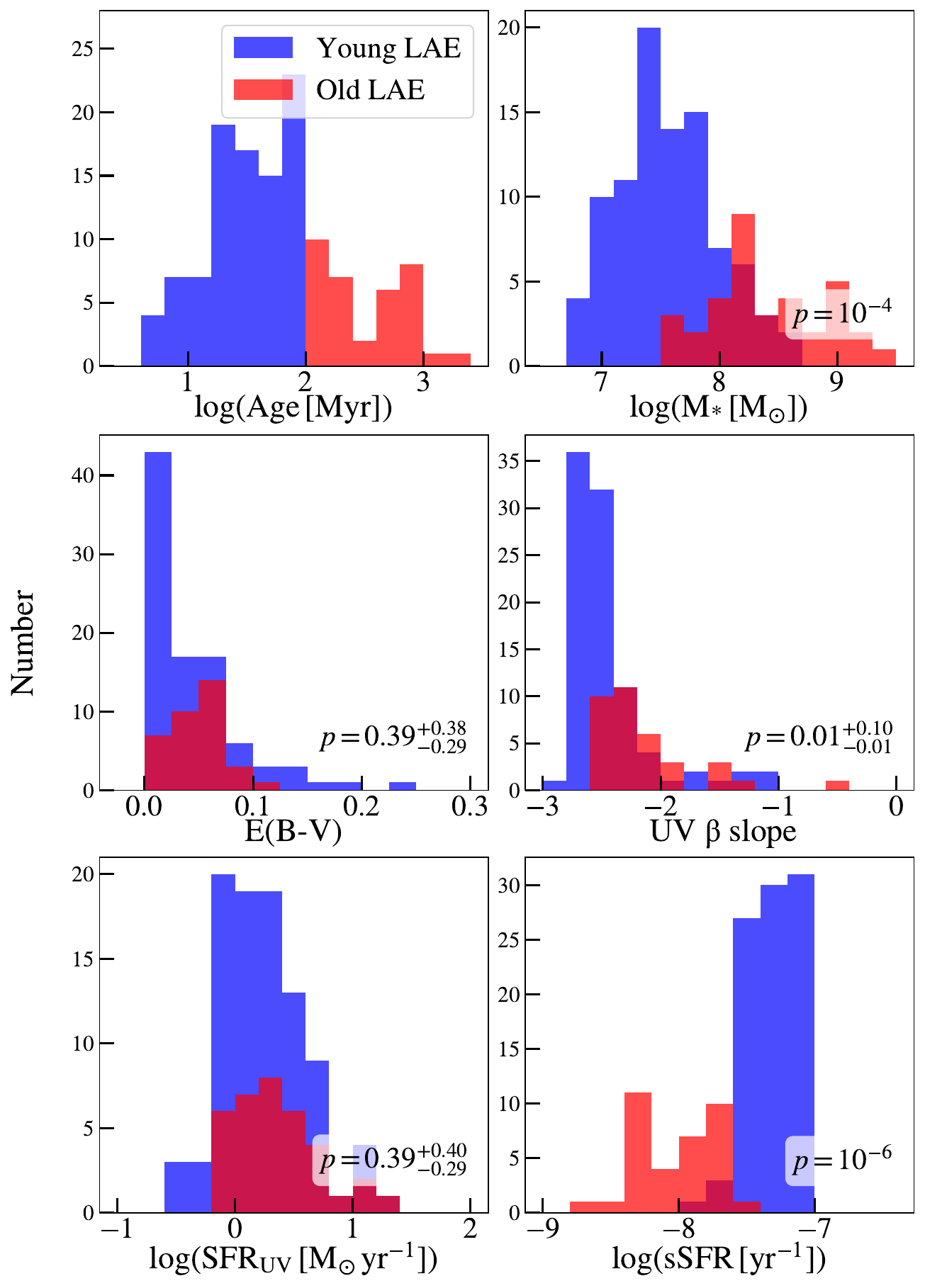}
    \caption{Comparison of physical properties (stellar mass, age, E(B-V), UV slope $\beta$, $\mathrm{SFR_{UV}}$, and sSFR) between the young LAE sample (blue) and the old LAE sample (red). The p-values from the KS test for the two samples are shown in each panel.}
    \label{fig:YOhist}
\end{figure}
\begin{figure*}
    \centering
    \includegraphics[width=\textwidth]{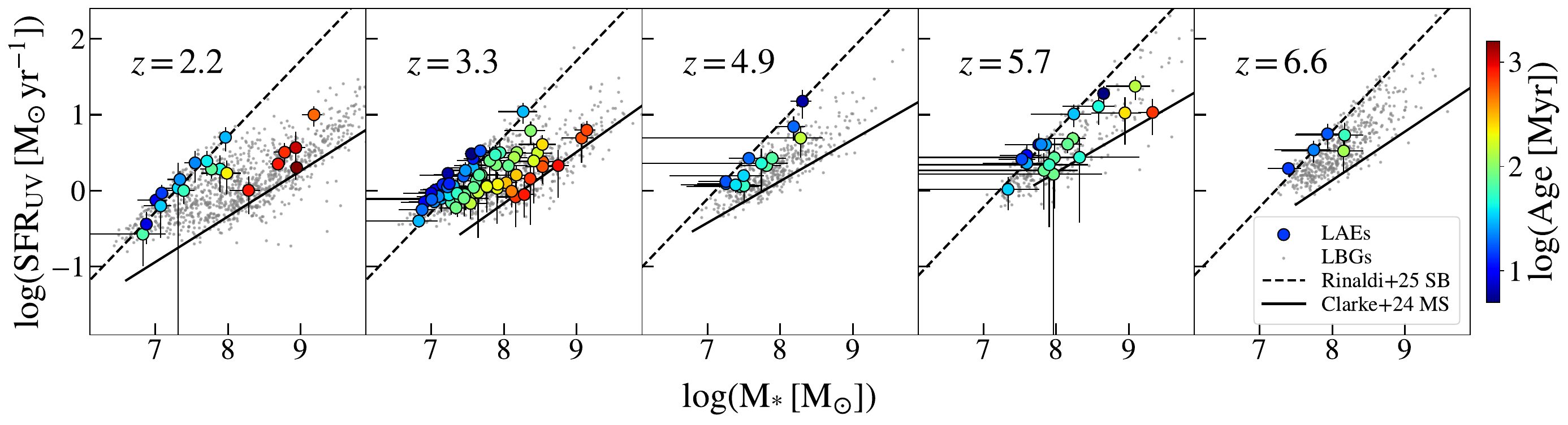}
    \caption{The relationship between $\mathrm{SFR_{UV}}$ and stellar mass for LAEs at each redshift. The colour of each point indicates the age of the LAE, with redder colours corresponding to older ages and bluer colours to younger ages. The LBG samples at each redshift are shown as grey points. The black dashed line represents the SB reported by \citet{Rinaldi+24}, while the black solid line indicates the MS at each redshift as reported by \citet{Clarke+24}.}
    \label{fig:MS}
\end{figure*}

Fig.~\ref{fig:ALL_LAE_hist} shows the comparison of the physical properties between LAEs and LBGs along with p-values.
The histograms are normalized since the sample sizes of LAEs and LBGs differ.
The distributions and p-values of E(B-V) and UV $\beta$ slope suggest that LAEs contain less dust than LBGs.
The trends are consistent with the previous studies (e.g. \citealt{Gawiser+06,Santos+20,Iani+24}).
Additionally, the distribution and p-value of the sSFR, defined as $\mathrm{SFR_{UV}}$ divided by stellar mass, reveals that LAEs are skewed towards higher sSFR than LBGs.
This indicates that LAEs are likely undergoing star formation more efficiently.
Focusing on the ages, the distribution for LAEs is slightly skewed towards younger ages compared to LBGs. 
However, as \citet{Haro+20} and \citet{Rosani+20} demonstrated, some LAEs with ages exceeding 100 Myr are also present. 
We divide our LAE sample into two groups: those younger than 100 Myr (young LAEs) and those older than 100 Myr (old LAEs) as performed by \citet{Iani+24}.
These old LAEs constitute $34^{+2}_{-4}$\% 
of our total LAE sample, with the uncertainty derived from the errors of age.
At $z\lesssim4.9$, the fraction of old LAEs tends to decrease with redshift, as $40^{+5}_{-5}$\% at $z=2.2$, $36^{+4}_{-6}$\% at $z=3.3$, $15^{+15}_{-7}$\% at $z=4.9$.
This can be naturally understood, as higher redshifts correspond to younger cosmic ages, where younger galaxies are expected to dominate. 
On the other hand, the fraction of old LAE is $30^{+5}_{-10}$\% at $z=5.7$, and $20^{+20}_{-20}$\% at $z=6.6$.
This increase at $z\gtrsim4.9$ is observed \citep{Iani+24}, though it might be due to a selection effect: old LAEs tend to be brighter in optical/near-IR, making them easier to detect. 
In our current sample, the $z=6.6$ subset includes only five LAEs, of which just one has an age exceeding 100 Myr. Due to the limited sample size, it is challenging to draw robust conclusions regarding this trend in our analysis.

\subsection{Differences in the properties of young LAEs and old LAEs} \label{diff_young_old}
\subsubsection{Comparison of young LAEs and old LAEs} \label{comp_young_old}
We compare the physical properties of 92 young LAEs and 35 old LAEs.
Fig.~\ref{fig:YOhist} presents histograms of the two populations across several key parameters: age, stellar mass, E(B-V), $\beta$, $\mathrm{SFR_{UV}}$, and sSFR. 
The results reveal that old LAEs are more massive and exhibit lower sSFR than young LAEs, consistent with \citet{Lai+08} and \citet{Rosani+20}.
The KS test is performed iteratively, 10,000 times in total, comprising 100 resampling iterations of the young and old LAE populations according to their age errors, combined with 100 iterations where the parameters are perturbed assuming a normal distribution based on their errors.
The p-values of these results are
shown in each panel of Fig.~\ref{fig:YOhist}.

The results indicate that young and old LAEs exhibit distinct distributions in terms of stellar mass, $\beta$, and sSFR, whereas no clear differences are observed in their E(B-V) and $\mathrm{SFR_{UV}}$ distributions. 
The discrepancy in $\beta$ despite the similarity in dust content suggests a difference in the stellar population compositions between young and old LAEs, as is evident from their definitions.
Furthermore, despite the larger stellar masses of old LAEs, they exhibit SFRs comparable to those of young LAEs, consequently leading to the distinct distribution in sSFR.
It should be noted that these results remain unaffected even when the threshold for young and old LAEs is changed from 100 Myr to 75 Myr or 125 Myr.

Fig.~\ref{fig:MS} shows 
the relationship between stellar mass and $\mathrm{SFR_{UV}}$ of our LAE sample at each redshift. 
In this figure, coloured points represent LAEs, and overall, they are found close to the starburst sequence (SB) \citep{Rinaldi+24}.
However, as the age increases, LAEs deviate from the SB and instead move closer to the MS \citep{Clarke+24}. 
This indicates that young LAEs are near the SB, while old LAEs exhibit SFRs similar to MS galaxies.
The trend is also found by \citet{Iani+24}, suggesting that young LAEs are undergoing an initial starburst phase, while old LAEs may be once-evolved galaxies. 
According to the speculation of \citet{Iani+24}, after some event, these old LAEs may have undergone rejuvenation, leading to the new star formation and \Lya photon emission. 
While our comparison of young and old LAEs largely corroborates the findings of \citet{Iani+24}, a discrepancy can be found in the distributions of E(B-V) and $\beta$. 
Specifically, we find that $\beta$ exhibits distinct distributions between the two populations, whereas E(B-V) does not, contrasting with \citet{Iani+24}, who reported the dust content of old LAEs was significantly lower than that of young LAEs.  
Moreover, other studies focusing on old LAEs \citep{Pentericci+09,Nilsson+09}, suggested that evolved, older LAEs tend to be more massive and dusty.
Thus, the dust content in old LAEs has been the subject of considerable debate.
These discrepancies can be explained by the scenario proposed by \citet{Gazagnes+20}, in which \Lya photons escape from galaxies through low-density channels of \Hi\ and dust created, for example, by outflows.
Whether old LAEs appear dust-rich or dust-poor may depend on the covering fraction of the low-density channel, i.e., the relative size of the low-density channels compared to the overall size of the galaxy. 
Therefore, when comparing with young LAEs, it becomes difficult to assess whether old LAEs are intrinsically dustier or less dusty, as the observed dust content is heavily influenced by the geometry of the ISM and viewing angle.
We examine the geometry of the ISM in old LAEs in more detail in Section~\ref{escape_oldLAE}.

\subsubsection{Comparison among young LAE, old LAE and LBGs} \label{sec:YOLBG}
\begin{figure}
    \centering
    \includegraphics[width=\columnwidth]{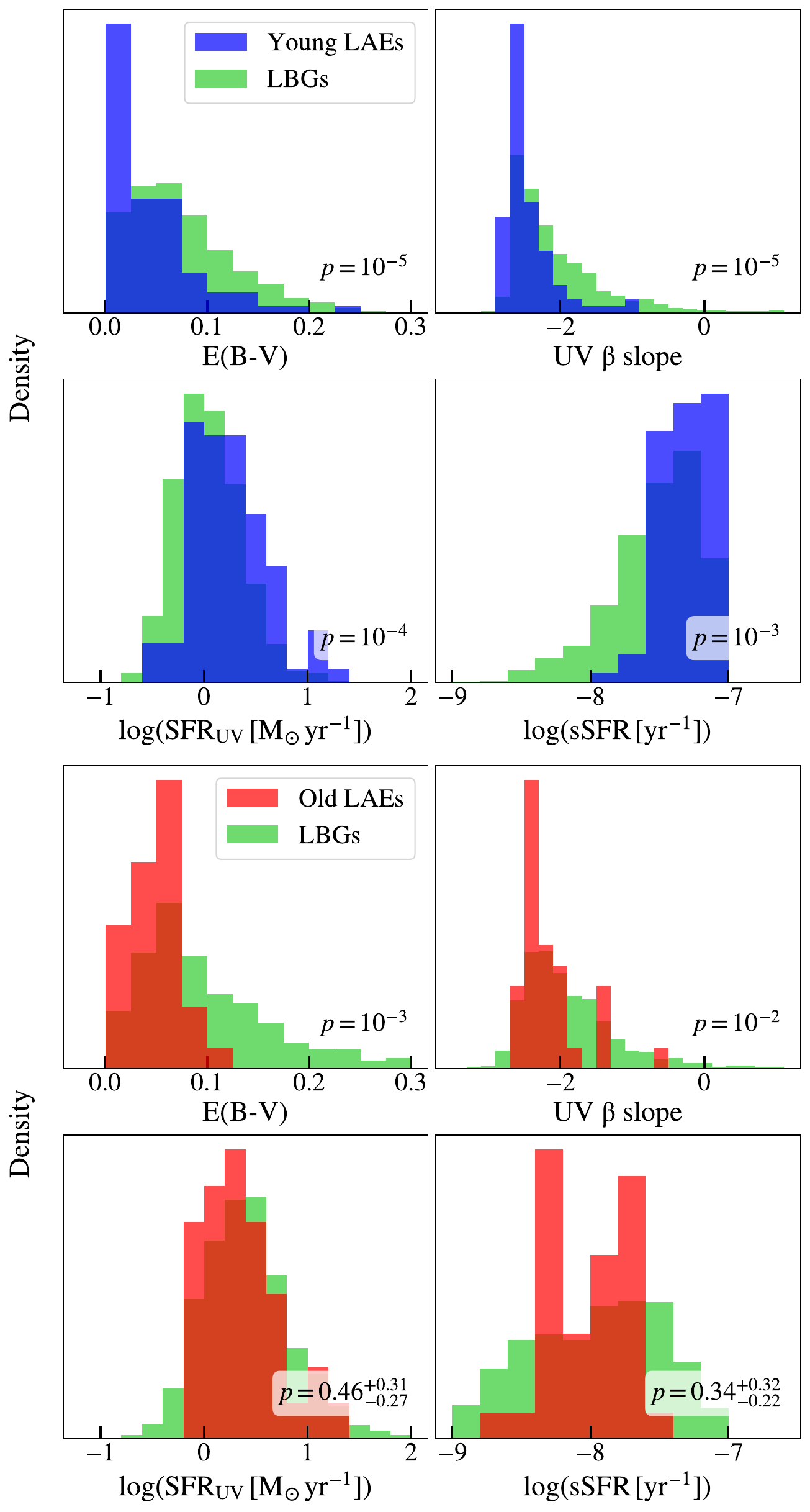}
    \caption{ Comparison of physical properties (E(B-V), UV slope $\beta$,  $\mathrm{SFR_{UV}}$, and sSFR) between young LAEs (blue) and LBGs (green) in the top four panels, and between old LAEs (red) and LBGs in the bottom four panels. In each case, the LBG sample is resampled to match the stellar mass distribution of the corresponding LAE subsample. The p-values from the KS tests are shown in each panel.}
    \label{fig:YOhist_LBG}
\end{figure}

Previous studies (e.g. \citealt{Haro+20,Iani+24}) have compared the properties of the LAE population as a whole with those of LBGs.
However, given the potential dichotomy within the LAE population based on age, 
such a comparison alone is insufficient to reveal the nature of LAEs. 
Therefore, we undertake a separate comparative analysis, examining the properties of old LAEs and young LAEs individually with those of LBGs. 
In this analysis, we resample LBGs so that their stellar mass distributions match those of young and old LAEs individually, mirroring the procedure performed in Section~\ref{comp_young_old}. 
Fig.~\ref{fig:YOhist_LBG} shows the comparisons of physical properties and p-values derived from the KS tests.
For the p-value calculation, we employ 10,000 iterations, comprising 100 iterations of LBG resampling and, for each resampled LBG, 100 iterations to account for the errors associated with parameters.
We note that the p-value remains unchanged even when we include a resampling step for the young and old LAEs based on the errors in their ages.
Focusing first on the young LAEs, we find all p-values are smaller than 0.05 in the four top panels of Fig.~\ref{fig:YOhist_LBG}, suggesting that the distributions of E(B-V), $\beta$, SFR, and sSFR of young LAEs are all distinct from those of LBGs with comparable stellar masses. 
The results suggest that, compared to LBGs of similar stellar mass, young LAEs are less dusty and undergo more burst-like star formation \citep{Kim+25}.
In contrast, old LAEs exhibit distinct distributions only in E(B-V) and $\beta$ compared to LBGs of similar stellar mass. 
This suggests that while old LAEs may have star formation properties similar to those of LBGs, as also seen in Fig.~\ref{fig:MS}, they are characterised by significantly lower dust content.
This raises the possibility that we observe these old LAEs through a preferential line of sight with low dust and \Hi\, column density in the ISM, which facilitate \Lya escape.
\citet{Iani+24} reported no significant differences between LAEs as a whole and LBGs except for E(B-V) and $\beta$.
However, our results show that significant differences in star formation properties emerge when focusing solely on young LAEs, even though our sample possesses a similar fraction of young and old LAEs to \citet{Iani+24}.
This finding is consistent with \citet{Gawiser+06} who compared LBGs specifically with LAEs exhibiting EW$_{\mathrm{Ly\alpha}}>150$\AA, which almost corresponds to young LAEs in our sample.

In summary, the above analysis suggests that young LAEs may have distinct properties in terms of star formation and dust content compared to LBGs even with similar stellar mass. 
This suggests that young LAEs are not simply the low-mass end population of LBGs, but may be a distinct population with active star formation or low dust content even among low-mass galaxies.
On the other hand, old LAEs have similar characteristics to LBGs, but may be observed as LAEs simply due to a favorable line of sight with low dust, which facilitates \Lya escape.
How the \Lya emission escapes from such young and old LAEs is discussed in detail in Section~\ref{escape_oldLAE}.

\subsection{Size-mass relation of LAEs} \label{sec:LAE_size}

\begin{figure*}
    \centering
    \includegraphics[width=\textwidth]{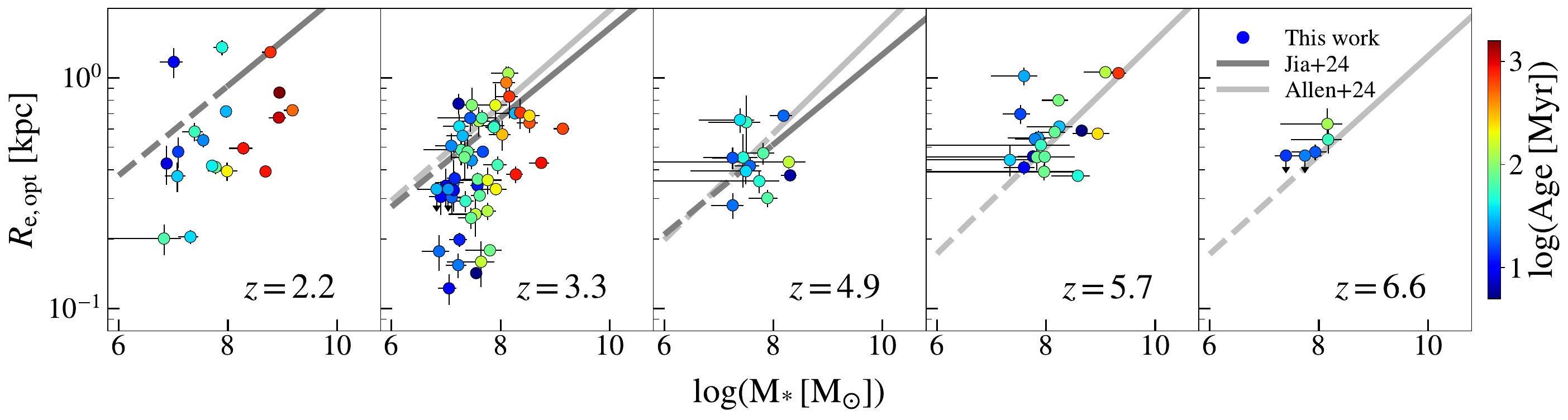}
    \caption{The relation between the size at rest-optical (0.6 $\mathrm{\mu m}$) and stellar mass of LAEs at each redshift. Those for which only the upper limit of size,  corresponding to the PSF half-light radius for objects unresolved in the continuum images, has been determined are indicated by downward-pointing arrows. The colour of each point indicates the age of the LAE, with redder colours representing older ages and bluer colours indicating younger ages. The dark and light grey lines represent extrapolations of the size-mass relations for SFGs \citep{Allen+24,Jia+24} into the low-mass regime ($\log(\mathrm{M_{*}\,[M_{\odot}])} \lesssim 8$).
}
    \label{fig:size_mass}
\end{figure*}

Fig.~\ref{fig:size_mass} presents the relation between the half-light radius measured at rest-optical wavelength and stellar mass of LAEs at each redshift. 
The black and grey lines represent the extrapolated size-mass relations of SFGs (\citealt{Jia+24}; \citealt{Allen+24}) toward the low-mass regime. 
We find that LAEs at $z=2.2$ and $z=3.3$ tend to lie systematically below the size-mass relation, indicating that they are more compact than SFGs.
This is consistent with the conclusion of \citet{Liu+24}, who investigated the rest-optical size of LAEs at $z=3.1$. 
On the other hand, at $z\gtrsim4.9$, we find that the sizes of LAEs lie on the typical stellar size-mass relation of SFGs.
A similar trend is also found when measuring LAE sizes in the rest-UV, which is consistent with \citet{Afonso+18}.
They reported that LAEs at $z<2$ are typically $2 \mathchar`- 4$ times smaller than SFGs in the rest-UV, while at $z>5$, their sizes become comparable to SFGs, suggesting that at higher-z, compact galaxies such as LAEs may represent the typical size of SFGs.
In this study, we demonstrate that the trend of LAE sizes approaching those of SFGs towards higher redshifts holds not only for rest-UV sizes, which are strongly influenced by star formation activity, such as inside-out growth \citep{Matharu+24}, but also for rest-optical sizes, which reflect the overall stellar extent of the galaxy. 
This finding reinforces the idea that compact LAEs represent a typical population of SFGs at high-$z$.
This is consistent with the observed increase in the $X_{\mathrm{LAE}}$ towards $z\sim6$ \citep{Kusakabe+20,Goovaerts+23}.
However, it should be noted that the convergence of LAE sizes with those of typical SFGs at $z\gtrsim4.9$ contrasts with their position on the stellar mass-SFR plane shown in Fig.~\ref{fig:MS}, where LAEs tend to lie in the SB rather than on the MS, even at $z\gtrsim4.9$.
This suggests that, although low-mass, compact galaxies such as LAEs become more prevalent at higher-$z$, LAEs may represent a subset of these galaxies that exhibit particularly intense star formation activity, as indicated in Section~\ref{sec:YOLBG}.

As discussed in Section~\ref{comp_young_old}, it is expected that old LAEs, as more evolved systems, would have larger sizes in rest-optical than those of young LAEs, and that their sizes would be comparable to those of typical SFGs.
However, Fig.~\ref{fig:size_mass} does not show a clear trend in which older LAEs approach the size-mass relation of SFGs more closely.
This is inconsistent with the finding that old LAEs have characteristics similar to LBGs, except for dust content, as seen in Section~\ref{sec:YOLBG}.
Despite exhibiting SFRs consistent with the MS (Fig.~\ref{fig:MS}), 
their compactness results in high star formation surface densities, which may promote efficient outflows and enhance \Lya escape \citep{Davies+19,Kim+25}.
This would allow them to be observed as LAEs. 
To fully test this hypothesis, however, a larger sample of old LAEs is needed.

\subsection{Ly$\alpha$ escape fraction of LAEs} \label{sec:Lya_escape}

\subsubsection{Age-related differences in Ly$\alpha$ escape} \label{escape_oldLAE}
\begin{figure}
    \centering
    \includegraphics[width=\columnwidth]{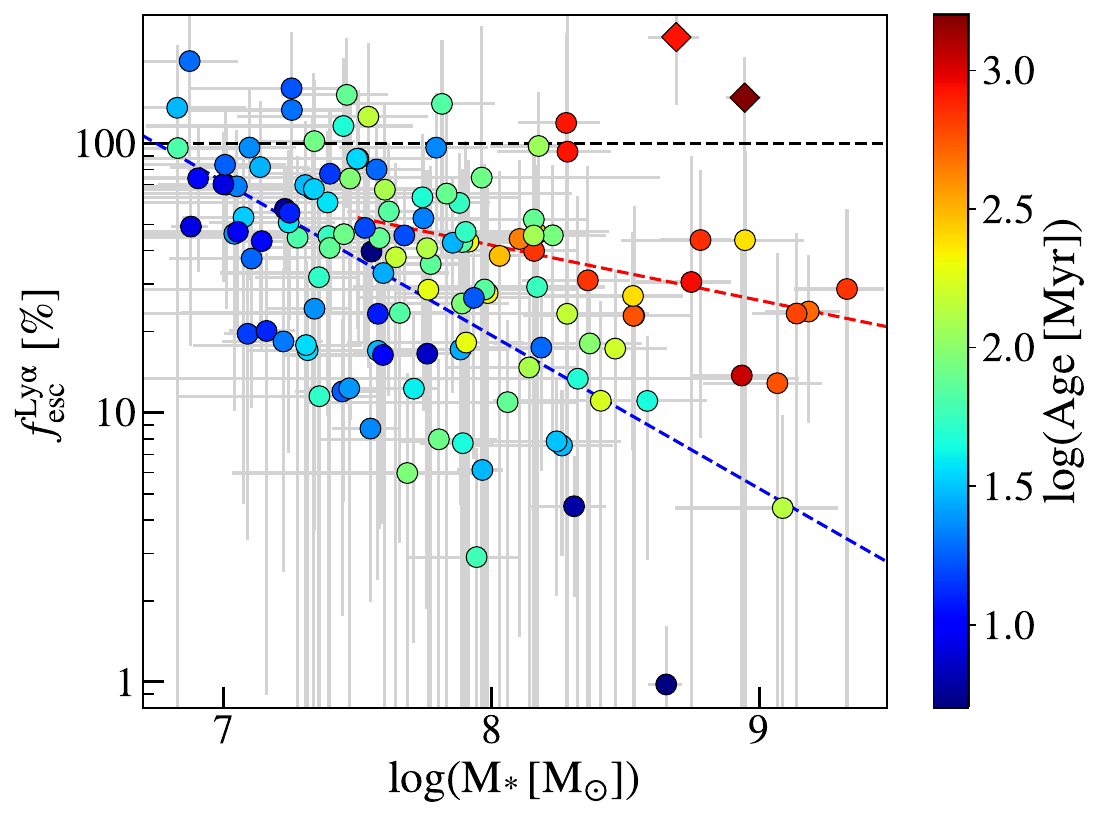}
    \caption{The \fesca of LAEs as a function of stellar mass. The colour indicates the age of the LAE, with redder colours representing older ages and bluer colours indicating younger ages. The black horizontal dashed line indicates a \fesca of 100\%. The blue and red dashed lines show linear fits to the young LAEs ($<100$ Myr) and old LAEs ($>100$ Myr), respectively. The diamonds indicate the two old LAEs with extended \Lya haloes: J100040.0+022215.9 and J100029.2+021007.0 which are discussed in Appendix~\ref{app:halo}.}
    \label{fig:f_esc_mass}
\end{figure}
Fig.~\ref{fig:f_esc_mass} presents the relationship between the stellar mass and the \fesca for the LAE sample. 
The p-value of the Spearman rank correlation is $p\sim10^{-6}$, and in general, a trend of decreasing \fesca with increasing stellar mass is observed, consistent with the previous findings (e.g. \citealt{Oyarzn+17,Weiss+21,Lin+24}). 
However, when this relationship is examined separately for young and old LAEs, distinct dependencies emerge. 
The relationship forms two sequences, one with young LAEs (blue dashed line) and the other with old LAEs (red dashed line), and at a given stellar mass, old LAEs tend to exhibit higher \fesca than young LAEs. 
The p-value
for this relation is significantly small for young LAEs ($p\sim10^{-7}$), whereas for old LAEs it is considerably larger ($p\sim0.08$), which suggests that the \fesca dependence on stellar mass is weaker in old LAEs.
This result is unaffected by resampling the young and old LAE based on the errors in their ages.
This observed trend can be interpreted through the following scenario: young LAEs represent galaxies in the early stages of star formation with low stellar masses and correspondingly low dust content, allowing \Lya photons to escape.
However, as their age and stellar mass increase, the associated increase in dust content makes \Lya escape more difficult, resulting in a steep decline in the \fesca with stellar mass among young LAEs.
Subsequently, enough dust accumulated to prevent \Lya photons from being emitted, and if left as they are, they would no longer be LAEs.
However, if outflows in an evolved galaxy preferentially clear low-density channels in the ISM along a specific direction, the galaxy might be observed as an old LAE when viewed from that line of sight. 
As a result, old LAEs may exhibit relatively high \fesca even at higher stellar masses.
In this case, the efficiency of \Lya photon escape in old LAEs would depend less on stellar mass itself, and more on the orientation of cleared channels along the line of sight. 
This naturally explains the weaker dependence of the \fesca on stellar mass observed in the old LAEs, compared to the young LAEs.
In this scenario, old LAEs do not necessarily require a rejuvenation event to become visible as LAEs.
Instead, they could simply be evolved galaxies that have maintained sufficient star formation activity to cause outflows.
In this case, old LAEs would be observed to have lower dust content along the line of sight where the outflow has blown away the dust.
This aligns well with our finding that old LAEs have SFRs similar to LBGs but with lower dust content.
Furthermore, as discussed in Section~\ref{sec:LAE_size}, 
the higher star formation rate surface densities of old LAEs compared to SFGs may indicate that outflows are more efficiently driven (e.g. \citealt{Davies+19}) in old LAEs, which further supports the scenario.

\subsubsection{Evolution of the Ly$\alpha$ escape fraction of LAE} \label{sec:f_esc_evo}
The evolution of the global \fesca of SFGs at a given epoch has been statistically studied using the UV luminosity function (UV LF) and \Lya luminosity function (\Lya LF) (e.g. \citealt{Hayes+11,Konno+16,Goovaerts+24b}). 
These studies have reported that the global \fesca increases at higher redshifts. 
The global \fesca has also been observationally determined by measuring the \Lya luminosity of \Ha emitters as representatives of SFGs at a given epoch (e.g. \citealt{Matthee+16,Lin+24}), and the results generally agree with those derived from the LFs. 
In this study, we focus on the evolution of the \fesca specific to LAEs.
To understand the contribution of LAEs to reionization,  we consider only \fesca of young LAEs, because the fraction of old LAEs during the reionization era is low.

\begin{figure}
    \centering
    \includegraphics[width=\linewidth]{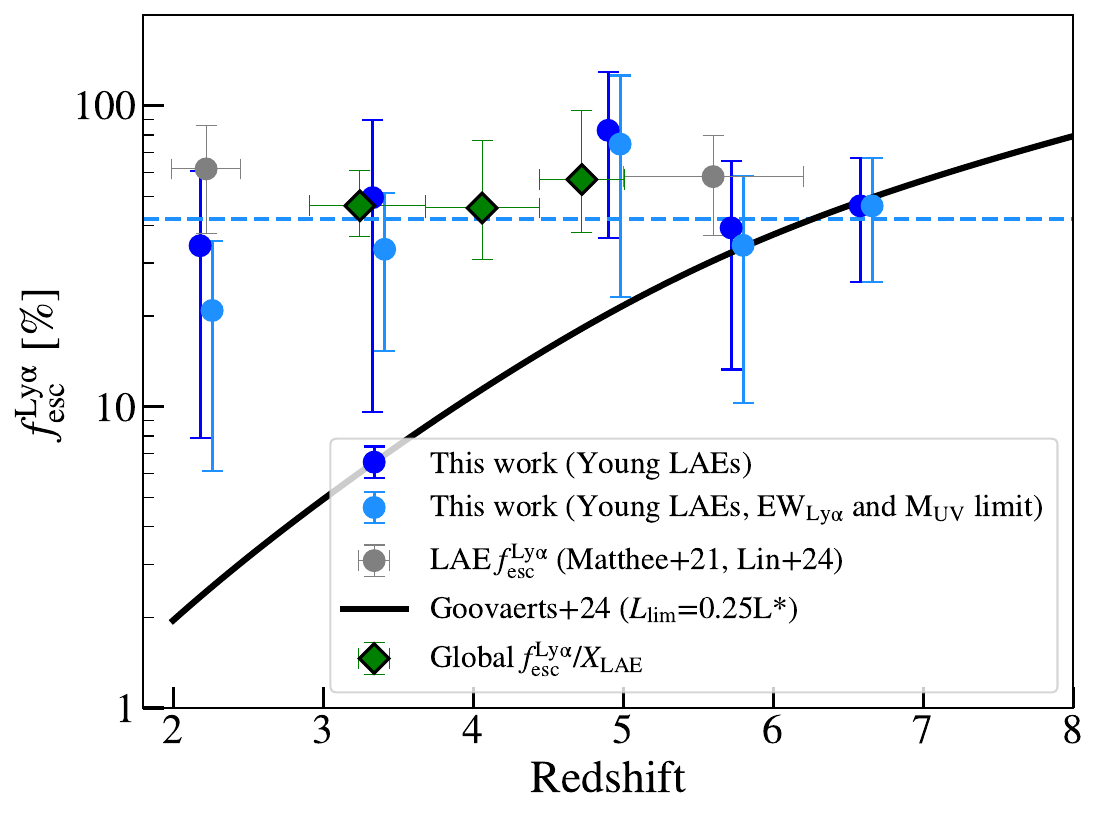}
    \caption{The evolution of \fesca. The blue points represent the average IGM-corrected \fesca derived from the full young LAE sample at each redshift, while the light blue points show the \fesca limited to EW$_{\mathrm{Ly\alpha}}>$ 25\,\AA\,, and $-20.25<\mathrm{M_{UV}}<-18.75 $. Similarly, the grey points present the IGM-corrected \fesca for LAEs selected with the same criteria in EW$_{\mathrm{Ly\alpha}}$ and $\mathrm{M_{UV}}$ for \citet{Matthee+21} ($z\sim2.2$) and \citet{Lin+24} ($z=5\mathchar`-6$). The black curve shows the evolution of the global \fesca for SFGs \citep{Goovaerts+24b} with an integration limit of $0.25 L*$ for comparison. The green diamond points indicate the values obtained by dividing the global \fesca by the $X_{\mathrm{LAE}}$ \citep{Kusakabe+20}. For visual clarity, the light blue points are slightly offset to the right, and the green point at $z\sim3$ is slightly offset to the left.}
    \label{fig:f_esc_evo}
\end{figure}

Fig.~\ref{fig:f_esc_evo} presents the average \fesca of young LAEs (blue points) at each redshift.
It is almost constant over redshifts, contrasting with that seen in the global \fesca (black line) of the general SFG population \citep{Goovaerts+24b}, which increases significantly toward higher redshifts. 
To make a fair comparison, we apply the same UV luminosity cut as our LAE sample, $\mathrm{M_{UV}} < -18.75$, which corresponds to $L_{\mathrm{lim}}=0.25L*$, where $L*$ represents the characteristic luminosity of the LF, to the data from \citet{Goovaerts+24b}, along with the same IGM correction.
It should be noted that the completeness of our LAE sample, which is selected by NB excess, in terms of $\mathrm{M_{UV}}$ is uncertain and varies across redshift bins.
We examine the completeness based on the NB limiting magnitudes and find that, up to $z=4.9$, the sample is almost complete for LAEs with EW$_{\mathrm{Ly\alpha}}>25$\AA\, and $\mathrm{M_{UV}}<-18.75$.
We restrict the samples to EW$_{\mathrm{Ly\alpha}}>25$\AA\ and $-20.25<\mathrm{M_{UV}}<-18.75$, where our sample is reasonably complete, and this is the same condition as the $X_{\mathrm{LAE}}$ sample used for comparing results later.
The same cut is applied for the samples at $z=5.7$ and $6.6$, where the completeness in $\mathrm{M_{UV}}$ is not fully guaranteed.
The light blue points in Fig.~\ref{fig:f_esc_evo} correspond to the average \fesca of young LAEs within this EW$_{\mathrm{Ly\alpha}}$ and $\mathrm{M_{UV}}$ range. 
As previously noted by \citet{Goovaerts+24b}, the average \fesca of LAEs remains roughly constant at $\sim40$\%, indicating little to no redshift evolution.
This result is in good agreement with previous studies (grey points, \citealt{Matthee+21} at $z\sim2.2$ and \citealt{Lin+24} at $z=5 \mathchar`- 6$), when similar selection criteria in EW$_{\mathrm{Ly\alpha}}$ and $\mathrm{M_{UV}}$ are applied and IGM correction is taken into account.
The lack of notable redshift evolution of \fesca over $z=2 \mathchar`- 7$ in LAEs is also seen in their average \Lya profile \citep{Hayes+21}, 
UV slope $\beta$ \citep{Santos+20}, and UV size \citep{Afonso+18}.
However, it should be noted that our small sample size of young LAEs may obscure subtle evolutionary trends due to intrinsic scatter within the population.

Furthermore, the green diamond points in Fig.~\ref{fig:f_esc_evo} indicate the \fesca of only LAEs, which is obtained by dividing the global \fesca by the $X_{\mathrm{LAE}}$ \citep{Kusakabe+20}.
The points closely match the average \fesca of our LAE sample and remain nearly constant with redshift. 
The result suggests that the observed evolution of global \fesca may not be due to the redshift evolution of \fesca itself, but rather to changes in the $X_{\mathrm{LAE}}$ within the SFG population. 
This supports the scenario in which LAEs consistently represent an early phase of galaxy formation, exhibiting similar properties across cosmic time.

\subsection{Contribution of LAEs to reionization} \label{sec:reionization}
In order to quantify the ionizing photon contribution of LAEs to reionization, we evaluate \nion, which represents the number of ionizing photons emitted per unit time per unit comoving volume. 
Following the method used in \citet{Matthee+22,Goovaerts+24b}, we calculate \nion\ using:
\begin{equation} \label{eq:nion_Lya}
    \dot{n}_{\mathrm{ion}}= \xi\mathrm{_{ion}^{Ly\alpha}}\, f\mathrm{_{esc}^{LyC}}\,\rho_{\mathrm{Ly\alpha}},
\end{equation}
where \xiionLya denotes LyC production rate per \Lya luminosity, \fescc is LyC escape fraction, and $\rho_{\mathrm{Ly\alpha}}$ represents the \Lya luminosity density.
The \xiionLya is expressed using the LyC production rate, $N\mathrm{(H^0)}$, as follows:
\begin{equation}
    \xi\mathrm{_{ion}^{Ly\alpha}} = \frac{N\mathrm{(H^0)}}{L_{\mathrm{Ly\alpha}}},
\end{equation}
where
\begin{equation}
    N\mathrm{(H^0)}=\frac{L_{\mathrm{H\alpha}}}{1.36\times 10^{-12}(1-f\mathrm{_{esc}^{LyC}})}.
\end{equation}
Here, $L_{\mathrm{H\alpha}}$ is the dust-corrected \Ha luminosity, which traces the recombination radiation from ionized hydrogen and is directly related to the number of ionizing photons produced by young, massive stars.
Using equation~\eqref{eq:f_esc}, \xiionfesc can be transformed as follows:

\begin{align*} \label{eq:xion_fescc}
    \xi\mathrm{_{ion}^{Ly\alpha}}\,f\mathrm{_{esc}^{LyC}} & = \frac{L_{\mathrm{H\alpha}}}{L_{\mathrm{Ly\alpha}}} \frac{f\mathrm{_{esc}^{LyC}}}{1.36\times 10^{-12}(1-f\mathrm{_{esc}^{LyC}})} \\
   & = \frac{f\mathrm{_{esc}^{LyC}}}{1.18\times 10^{-11}f\mathrm{_{esc}^{Ly\alpha}}(1-f\mathrm{_{esc}^{LyC}}) }.
\stepcounter{equation}\tag{\theequation}
\end{align*}
We note that our definition of \xiionLya has different units from the commonly used \xiionUV because its denominator is the observed \Lya luminosity (in units of $\mathrm{erg\,s^{-1}}$) rather than the UV luminosity density (in $\mathrm{erg\,s^{-1}\,Hz^{-1}}$).

To estimate the \fescc for LAEs, we utilize the two relationships between \fescc and \fesca established in \citet{Begley+24} (hereafter \citetalias{Begley+24}) based on observational data and \citet{Choustikov+24c} (\citetalias{Choustikov+24c}) derived from theoretical modelling.
\citetalias{Begley+24} estimated the \fesca for 152 SFGs at $z\sim4\mathchar`-5$ by combining \Lya line flux from their spectra and \Ha line flux from broadband IRAC 3.6 $\mathrm{\mu m}$ flux excess. 
They inferred \fescc from the EW of low-ionization interstellar absorption lines ($W_{\mathrm{LIS}}$), which trace the neutral gas covering fraction.
The $W_{\mathrm{LIS}}$-\fescc\ relation \citep{Saldana+22a} they used was calibrated using direct, reliable \fescc\ measurements of low-z galaxies in the LzLCS survey \citep{Flury+22}. 
\citetalias{Begley+24} empirically derived a relationship: $f\mathrm{_{esc}^{LyC}}\simeq 0.15^{+0.06}_{-0.04}\,f\mathrm{_{esc}^{Ly\alpha}}$. 
In contrast, \citetalias{Choustikov+24c} employed cosmological hydrodynamical simulations with mock spectral and photometric data to investigate correlations between \fescc and various physical properties, including \fesca. 
Their findings indicate that the \fesca is a reliable tracer of LyC leakage in galaxies with low \Hi\, column densities and minimal UV dust attenuation. 
By averaging over the viewing angle of galaxies along different lines of sight, they derived a relationship: $\log_{10} (f\mathrm{_{esc}^{LyC}}) =1.753 \log_{10} (f\mathrm{_{esc}^{Ly\alpha}}) -0.4683$ at $Z\sim0.03Z_{\odot}$, a metallicity at which their model was shown to best reproduce observations of low-z LyC leakers (e.g. \citealt{Flury+22,Izotov+24}).

\begin{figure}
    \centering
    \includegraphics[width=\linewidth]{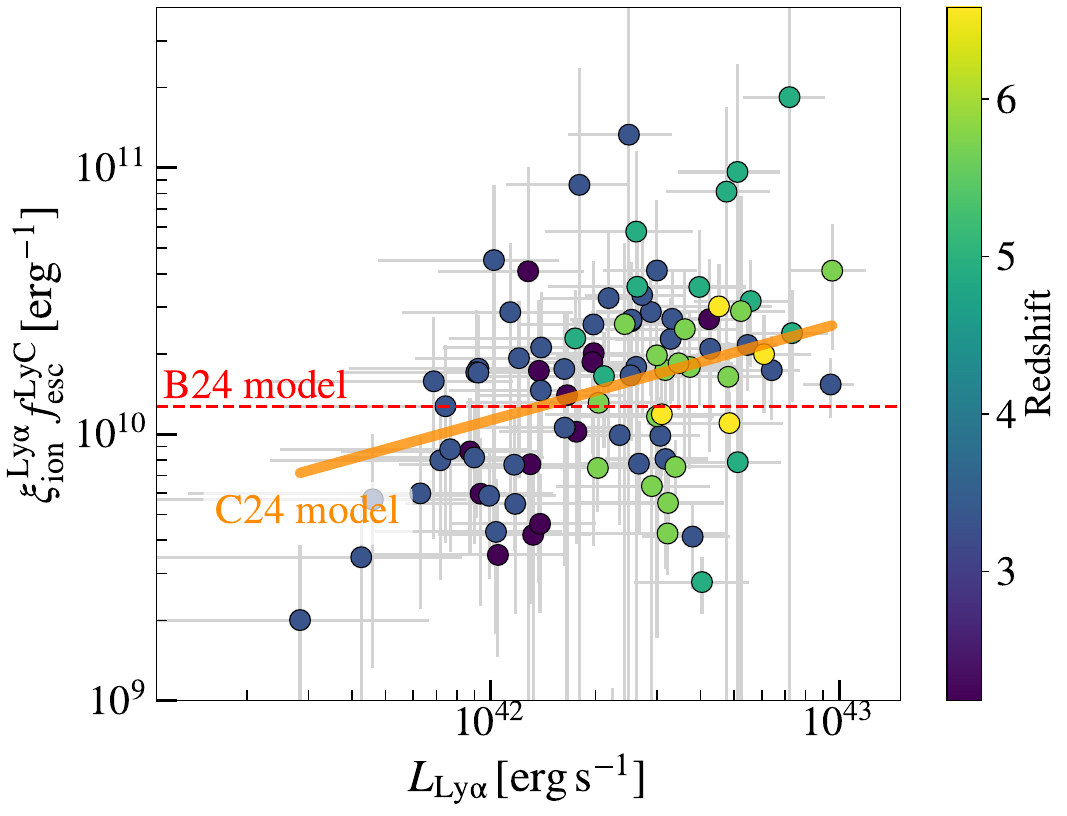}
    \caption{Dependence of \xiionfesc on \Lya luminosity for young LAEs in the \citetalias{Choustikov+24c} model. 
    The \xiionfesc predicted by the \citetalias{Choustikov+24c} model for young LAEs are colored according to redshift. 
    The fitted relation for the \citetalias{Choustikov+24c} model is overplotted as an orange line. 
    The red dashed line indicates the constant \xiionfesc based on the \citetalias{Begley+24} model for a reference.
    }
    \label{fig:xifesc}
\end{figure}

Using these relations and equation~\eqref{eq:xion_fescc}, we can calculate \xiionfesc, which represents the effective number of photons that ionize the IGM per unit \Lya luminosity.
In the \citetalias{Begley+24} model, \fescc is proportional to \fesca, 
and the average \fescc is low ($\sim 5\%$), 
resulting in only minimal variation ($\sim 0.02\,\mathrm{dex}$) in \xiionfesc. 
Therefore, the \xiionfesc in the \citetalias{Begley+24} model is treated as a constant value of $10^{10.1}\, \mathrm{erg^{-1}}$ in our subsequent analysis.
On the other hand, Fig.~\ref{fig:xifesc} plots the relationship between \Lya luminosity and \xiionfesc derived using the \citetalias{Choustikov+24c} model.
This is shown for young LAEs, in order to focus on the ionizing photon emissions from LAEs in the reionization era, when there are few old LAEs. 
While, due to the observational bias, the low \Lya luminosity regime is mostly dominated by samples at $z=2.2$ and 3.3, 
there is no difference in trend with redshift. 
To examine the correlation between \Lya luminosity and \xiionfesc for the \citetalias{Choustikov+24c} model, we perform Spearman’s rank correlation test across all young LAEs. 
The result yields a small p-value ($p\sim10^{-4}$), indicating a statistically robust correlation.
Although the scatter of this relationship is large, likely due to factors such as the viewing angle \citepalias{Choustikov+24c}, it should be useful for investigating the average photon budget of the universe.
To calculate \nion\ based on equation~\eqref{eq:nion_Lya},
we perform a linear fit to the relationship between \Lya luminosity and \xiionfesc for the \citetalias{Choustikov+24c} model as shown in Fig.~\ref{fig:xifesc}.
Regarding the $\rho_{\mathrm{Ly\alpha}}$,
we use the \Lya LF derived by \citet{Thai+23}, fitted with the traditional Schechter function \citep{Schechter}.
We utilize the \Lya LF averaged over the redshift range $2.9 < z < 6.7$, because previous studies \citep{Herenz+19,Thai+23,Umeda+25a} have shown that the redshift evolution of the \Lya LF is minimal at $z < 6$, where IGM absorption is negligible.
Although we use LF averaged up to $z=6.7$, the effect of including $z>6$ is considered to be small because it is almost identical to the averaged \Lya LF at $2.2 < z < 4$ shown in \citet{Thai+23}.
Since \citet{Thai+23} derived the \Lya LF down to $10^{39}\,\mathrm{erg\,s^{-1}}$ by observing LAEs in lensed regions, we calculate \nion\ by integrating $\xi\mathrm{_{ion}^{Ly\alpha}} f\mathrm{_{esc}^{LyC}} \rho_{\mathrm{Ly\alpha}}$, expressed as a function of \Lya luminosity, over the range $10^{39}$ to $10^{43}\,\mathrm{erg\,s^{-1}}$ for both \citetalias{Begley+24} and \citetalias{Choustikov+24c} models.
For the \citetalias{Choustikov+24c} model, we assume that the \Lya luminosity dependence of \xiionfesc can be extrapolated down to this lower limit.
However, it is important to note that the \Lya LF from \citep{Thai+23} has large uncertainties, particularly at the faint end, due to significant completeness corrections.

\begin{figure}
    \centering
    \includegraphics[width=\linewidth]{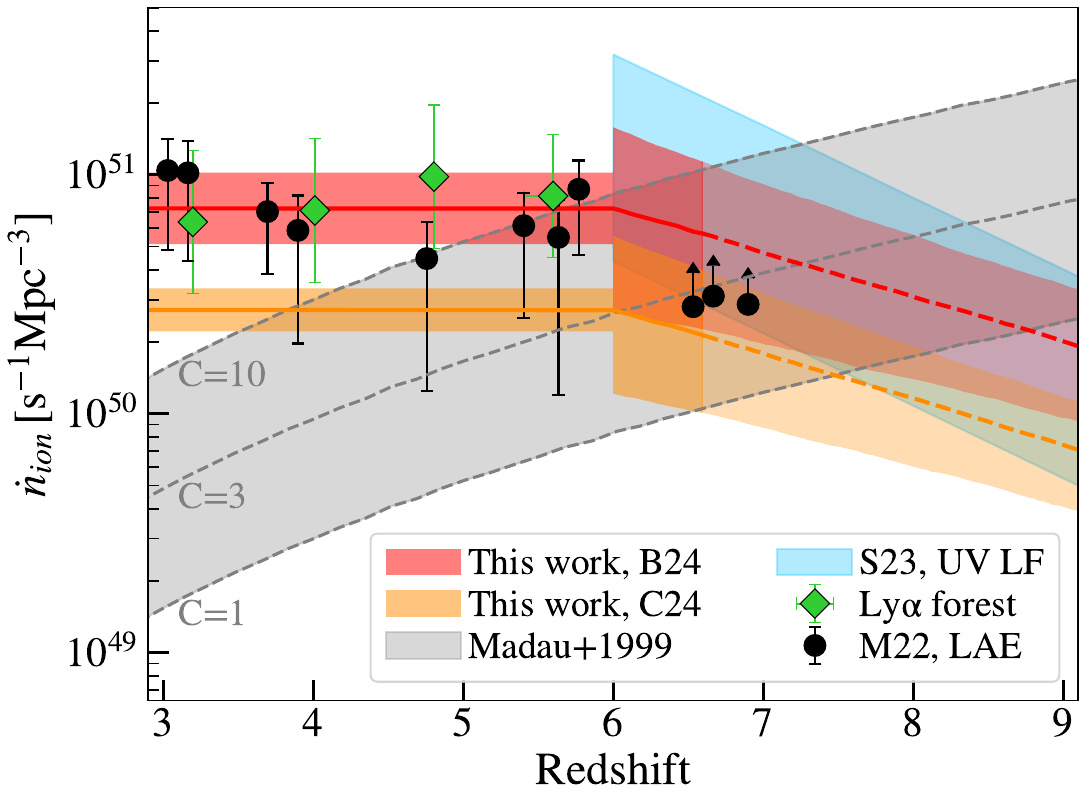}
    \caption{Redshift evolution of \nion\ contributed by LAEs. The red (orange)-shaded region indicates the evolution of \nion\ estimated using the \fescc based on the model of \citetalias{Begley+24} (\citetalias{Choustikov+24c}), which remains constant up to $z=6$. Beyond $z=6$, \nion\ declines following the UV LF of SFGs \citep{Harikane+25}, with its uncertainty reflecting both the errors from $z<6$, including those of \xiionLya, \fesca, the conversion between \fesca and \fescc, and the uncertainties in the UV LF. We also show independent \nion\ estimates from: measurements from the \Lya forest on quasar spectra (green points; \citealt{Becker+13,Davies+24}), contributions from only bright LAEs (black points; \citealt{Matthee+22}), and estimates assuming an escape fraction of 10\%–20\% for all SFGs based on their UV luminosities (cyan-shaded region; \citealt{Simmonds+23}). The grey-shaded region from \citet{Madau+1999} shows the number of ionizing photons required to maintain the ionized state of the IGM at each redshift.}
    \label{fig:nion}
\end{figure}

As a result, we obtain $\dot{n}_{\mathrm{ion}}= 10^{50.86\pm0.15} \mathrm{s^{-1}\,Mpc^{-3}}$ for the \citetalias{Begley+24} model and $\dot{n}_{\mathrm{ion}}= 10^{50.50\pm0.09} \mathrm{s^{-1}\,Mpc^{-3}}$ for the \citetalias{Choustikov+24c} model for the number of ionizing photons emitted by LAEs.
The lower limit of the integration down to $10^{39}\,\mathrm{erg\,s^{-1}}$ is the same as in \citet{Goovaerts+24b}, but even when the \Lya LF from \citet{Thai+23} is extrapolated and integrated down to a lower (higher) luminosity limit of $10^{38}$ ($10^{40}$) $\mathrm{erg\,s^{-1}}$, the resulting value of \nion\ changes by only 0.14 (0.17) dex for the \citetalias{Begley+24} model and by merely 0.02 (0.04) dex for the \citetalias{Choustikov+24c} model.
Since neither \xiionfesc nor the \Lya LF exhibits redshift evolution over $z<6$, \nion\ should be considered to be constant within this interval. 
However, at $z > 6$, where IGM absorption is significant, the evolution of the intrinsic \Lya LF is unknown.
Therefore, we assume the shape of the \Lya LF remains unchanged and that the evolution of the normalization factor ($\phi^*$) of intrinsic \Lya LF 
at $z>6$ follows the decline in the $\phi^*$ of UV LF 
of SFGs \citep{Harikane+25}, which traces the UV LF evolution up to $z\sim14$ using spectroscopically confirmed samples.
This approach is motivated by several observational trends. 
As shown in Fig.~\ref{fig:f_esc_evo}, the global \fesca of SFGs is known to increase significantly towards higher redshifts (e.g. \citealt{Goovaerts+24b}), suggesting that a larger fraction of the SFG population becomes visible as LAEs. 
In addition, galaxies at these high redshifts are expected to be generally characterized by low dust content and vigorous star formation (e.g. \citealt{Roberts-Borsani+24}), which is consistent with the properties of young LAEs as shown in Section~\ref{sec:YOLBG}.
This is further supported by our analysis in Section~\ref{sec:LAE_size}, which shows that the rest-optical sizes of LAEs converge with those of the general SFG population at $z\gtrsim4.9$.
These observational trends suggest that LAEs become increasingly representative of the typical galaxy population at the epoch of reionization. 
While these lines of evidence do not strictly imply that the $X_{\mathrm{LAE}}$ reaches 100\%, they provide strong support for approximating the evolution of $\phi^*$ of the intrinsic \Lya LF with that of the UV LF.
It should be noted that whether we change this hypothetical turning point at $z=6$ to $z=5.7$ or $z=6.7$, does not affect the subsequent discussion regarding Fig.~\ref{fig:nion}. 
Furthermore, this choice has little to no impact on the evolution of the IGM neutral fraction in Section~\ref{sec:reionization_history}.

The resulting evolution is shown in Fig.~\ref{fig:nion}, where the red-shaded region corresponds to \nion\ based on the \citetalias{Begley+24} model, while the orange region represents \nion\ based on the \citetalias{Choustikov+24c} model.
We here assume the dependence of \xiionfesc on \Lya luminosity beyond $z>6.6$ is the same as that at $z<6.6$.
No evolution in $\dot{n}_{\mathrm{ion}}$ at $z<6$ is consistent with the independent observational constraints of $\dot{n}_{\mathrm{ion}}$ at $z=3 \mathchar`- 6$ estimated by the \Lya forest on quasar spectra \citep{Becker+13,Davies+24}, as shown by the green points in Fig.~\ref{fig:nion}, which is in excellent agreement with \nion\ obtained using the \citetalias{Begley+24} model.
The grey shaded area represents the amount of ionizing photons required to maintain the ionisation of hydrogen, according to the model by \citet{Madau+1999}. 
A larger clumping factor, C, indicates a higher number of recombinations occurring in the IGM, leading to more ionizing photons to maintain the ionized state. 
Comparing our results with this region, we find that under both \citetalias{Begley+24} and \citetalias{Choustikov+24c} models, LAEs alone can supply the ionizing photon budget required to fully ionize the universe at $z\sim6$.
The light blue area represents the \nion\ estimated from \citet{Simmonds+23}, focusing on the $\mathrm{M_{UV}}$ of SFGs and integrated down to $\mathrm{M_{UV}} = -16$. 
The upper and lower bounds correspond to cases where the \fescc is assumed to be 20\%, 10\%, respectively.
Their result is in good agreement with our result, despite the completely different assumptions of the \fescc. 
This demonstrates that consistent results are obtained even when \nion\ is estimated from different perspectives: UV luminosity and \Lya luminosity. 
The results are also in good agreement with \citet{Matthee+22}, who evaluate the ionizing emissivity of LAEs by assuming that half of the very bright ($L_{\mathrm{Ly\alpha}}>10^{42.2}\, \mathrm{erg\, s^{-1}}$) LAEs are LyC leakers with a fixed \fescc and \xiionLya, based on a detailed analysis of the \Lya profile at $z\sim2$ \citep{Naidu+22}.
The difference is that they assigned a fixed \fescc of 50\% exclusively to the bright LAEs,
whereas we assume it to be a function of the \fesca.
These studies imply that LAEs play important roles in reionization, even if different assumptions are made in the calculation of \nion.

\begin{figure}
    \centering
    \includegraphics[width=\linewidth]{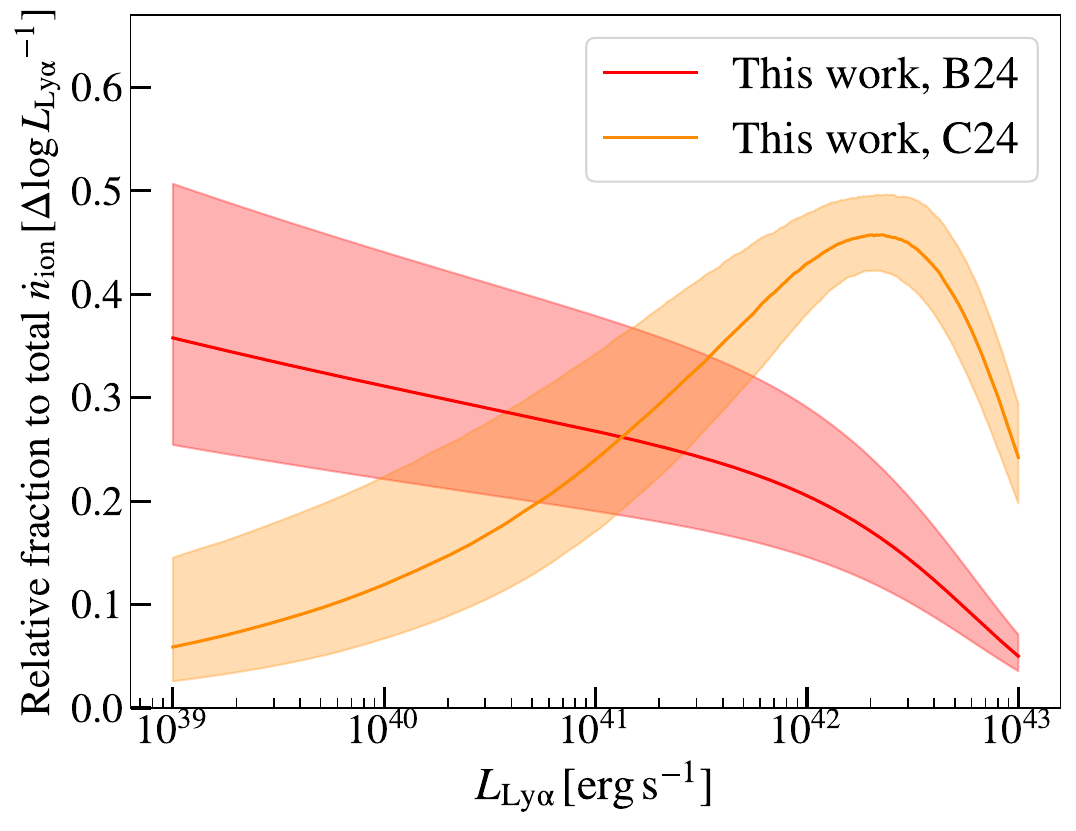}
\caption{The relative contribution of LAEs at each \Lya luminosity to the total \nion\ estimated using the \fescc\ based on the model of \citetalias{Begley+24} (red-shaded region) and \citetalias{Choustikov+24c} (orange-shaded region). }
    \label{fig:nion_cul}
\end{figure}

Whether the high contribution to the ionizing photon budget is due to \Lya-bright LAEs or \Lya-faint LAEs depends on the \fesca-\fescc conversion. 
Fig.~\ref{fig:nion_cul} shows the contribution of LAEs at each \Lya luminosity to the total \nion.
In the \citetalias{Begley+24} model, low \Lya luminosity LAEs are the dominant sources of ionizing photons. 
In contrast, the \citetalias{Choustikov+24c} model indicates that LAEs with \Lya luminosities around $10^{42}\,\mathrm{erg\,s^{-1}}$  are the most dominant contributors.

A potential consideration for this method is the sample completeness with respect to \fesca.
LAE with lower \fesca are inherently more difficult to detect by their \Lya emission, particularly at lower \Lya luminosities.
The result for the \citetalias{Choustikov+24c} model is likely robust against \fesca incompleteness, as these bright contributors are expected to have high completeness in our sample. 
The \citetalias{Begley+24} model, being independent of \fesca in the calculation of \nion, is inherently unaffected by such completeness issues.

\subsection{Implication to reionization history} \label{sec:reionization_history}

Based on \nion\ predicted from each model shown in Fig.~\ref{fig:nion}, we derive the redshift evolution of the volume-averaged IGM neutral hydrogen fraction, $x_{\mathrm{HI}}$.
This calculation is performed by solving the following differential equation for the IGM ionized fraction, $Q_{\mathrm{HII}}$:
\begin{equation} \label{eq:x_HI}
    \dot{Q}_{\mathrm{HII}} = \frac{\dot{n}_{\mathrm{ion}}}{\langle n_{\mathrm{H}} \rangle} - \frac{Q_{\mathrm{HII}}}{t_{\mathrm{rec}}},
\end{equation}
where $\langle n_{\mathrm{H}} \rangle$ is the comoving hydrogen density, and $t_{\mathrm{rec}}$ is the recombination time of ionized hydrogen. 
In these calculations, we adopt the same parameter values used in \citet{Naidu+20}, assuming a clumping factor of $C=3$. 

\begin{figure*}
    \centering
    \includegraphics[width=\textwidth]{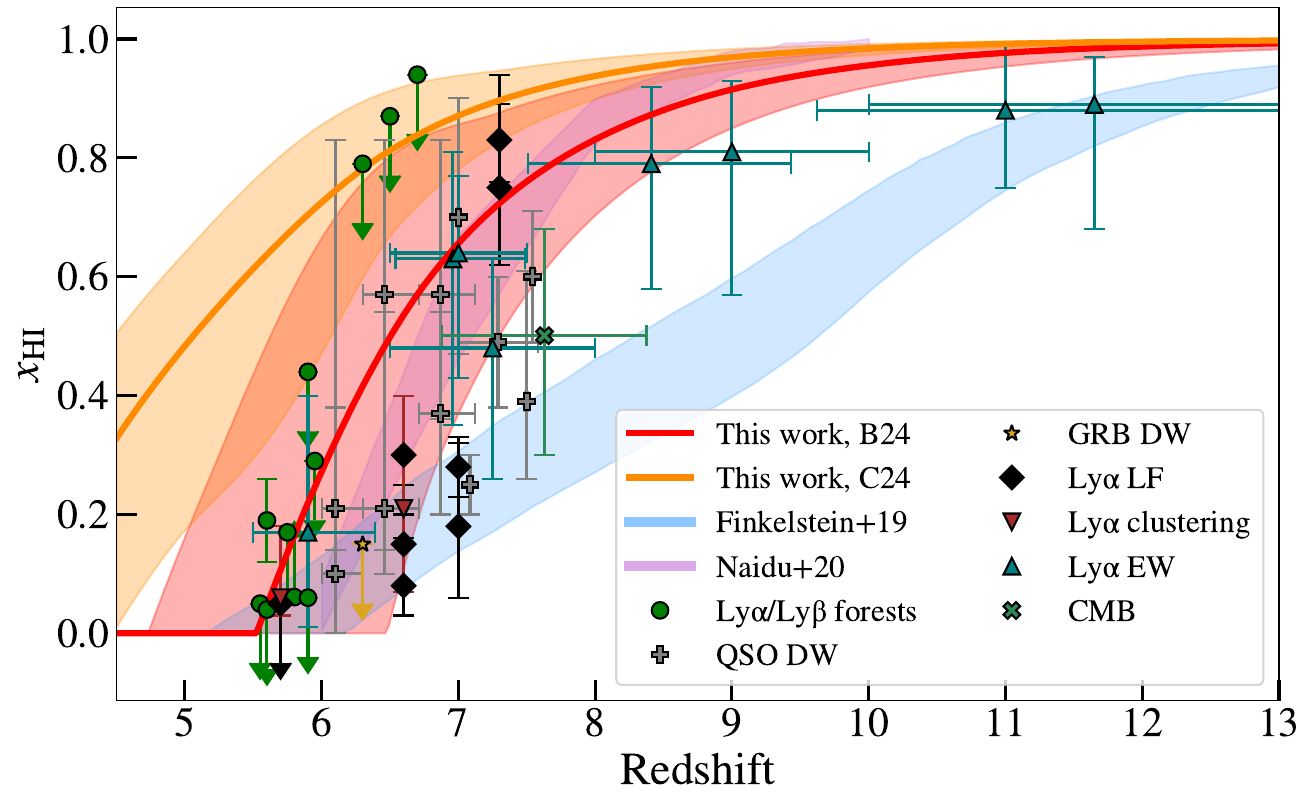}
    \caption{Redshift evolution of $x_{\mathrm{HI}}$. The red and orange lines show the evolution derived using the \fesca and \fescc relations based on the models of \citetalias{Begley+24} and \citetalias{Choustikov+24c}, respectively. For comparison, we also plot two $x_{\mathrm{HI}}$ evolution models (the blue line; \citealt{Naidu+20}, and the purple line;  \citealt{Finkelstein+19}). In addition, previous observational estimates are also shown, derived from \Lya and Ly$\mathrm{\beta}$ forests dark fraction \citep{McGreer+15,Zhu+22,Jin+23,Zhu+24_xhi,Spina+24}, \Lya damping wings of quasars \citep{Duro+20,Wang+20,Yang+20,Greig+22,Duro+24}, \Lya damping wings of GRB \citep{Fausey+25}, \Lya LF \citep{Morales+21,Ning+22,Umeda+25a},  LAE clustering \citep{Umeda+25a}, EW$_{\mathrm{Ly\alpha}}$ distribution \citep{Tang+24,Jones+25,Kageura+25}, 
    %\Lya damping wings of star-forming galaxy \citep{Hsiao+24,Umeda+25b}, 
    and CMB observation \citep{Planck+20}.}
    \label{fig:x_HI}
\end{figure*}
Fig.~\ref{fig:x_HI} shows the redshift evolution of $x_{\mathrm{HI}}$ obtained by solving equation~\eqref{eq:x_HI}. 
The $x_{\mathrm{HI}}$ evolution inferred from the \nion\ based on the \citetalias{Begley+24} model is shown by the red line, while the result from the \citetalias{Choustikov+24c} model is shown in orange.
The \citetalias{Begley+24} model yields a reionization history that is in good agreement with many observational constraints on $x_{\mathrm{HI}}$ derived independently.
The model shows a rapid reionization between $z=6 \mathchar`- 8$, achieving almost full reionization by $z\sim6$, driven solely by the contribution of LAEs.
A similar conclusion is drawn by \citet{Matthee+22}, where reionization is also completed by LAEs alone, with a similarly rapid evolution of the IGM neutral fraction at $z=6\text{-}8$. 
It should be noted that \citet{Matthee+22} focus only on very luminous LAEs as primary ionizers, whereas the \citetalias{Begley+24} model highlights the significant role of \Lya faint galaxies in driving reionization, as shown in Fig.~\ref{fig:nion_cul}.
This result is particularly intriguing as it suggests that LAEs with high ionizing photon production rates and potentially high \fescc can be considered as the major drivers of reionization.

In contrast, the \citetalias{Choustikov+24c} model fails to complete reionization until $z\sim4$, which is in tension with observational evidence suggesting that reionization is already complete by $z = 5$ at the latest (e.g. \citealt{Bosman+22,Gaikwad+23}).
This implies that LAEs alone do not produce a sufficient number of ionizing photons in the \citetalias{Choustikov+24c} model. 
Although we are using this model assuming a constant $Z=0.03\,Z_{\odot}$, even if a different constant metallicity were assumed, this result would not be changed.
However, if there is a metallicity dependence in any physical quantity of LAEs, the relationship between \fesca and \fescc would change, and this discrepancy may be improved.

Furthermore, we consider the impact of different assumptions in our SED fitting (Section~\ref{SED_fitting}), such as the inclusion of binary stars, a top-heavy IMF, and/or a higher high-mass limit. 
These models generally yield a higher production efficiency of ionizing photons, which means a lower SFR is required to match the observed flux \citep{Eldridge+17}. 
As a result, \fesca estimated from equation~\eqref{eq:f_esc} and \eqref{eq:SFR_Ha} increases.
For example, based on the BPASS model \citep{Eldridge+17}, the SFR can be decreased by approximately 0.1 dex at a metallicity of $Z=0.2\,Z_{\odot}$. 
This would lead to a 0.1 dex increase in our \fesca estimate. 
We recalculated the evolution of the $x_{\mathrm{HI}}$ in the case and found that reionization would complete at $z\sim6.1$ with the \citetalias{Begley+24} model, while it would still not complete by $z=5$ with the \citetalias{Choustikov+24c} model.
While this suggests that our main conclusions would not be significantly changed, we note that the results may vary depending on the metallicity and IMF.

Recent observations with JWST \citep{Ning+23,Llerena+24,Mascia+25} have shown that galaxies with lower stellar mass, higher sSFR, more compact morphology, and stronger \Lya emission tend to exhibit higher $\xi_{\mathrm{ion}}$. 
These trends are consistent with the properties of LAEs discussed in Sections~\ref{sec:LAE_LBG} and~\ref{sec:LAE_size}, suggesting that LAEs are galaxies with high ionizing photon production efficiencies \citep{Simmonds+23}. 
In addition, theoretical studies such as \citet{choustikov+24a} suggest that galaxies with high sSFRs, young stellar ages ($3.5 \mathchar`-10$ Myr), and low neutral gas content are more likely to have high \fescc.
LAEs, with their young stellar populations and high sSFR, 
are natural candidates for LyC leakers.
Such an example has actually been found by JWST \citep{Zhiyuan+25}.
This scenario indicates a strong correlation between the \Lya and LyC escape fractions.
These findings support the emerging picture in which LAEs play a central role in cosmic reionization.
The result of this study suggests that ionizing photons from LAEs alone are sufficient to explain the photon budget for reionization, but do not take into account contributions from other ionizing sources, such as AGNs. 
Some studies (e.g. \citealt{Onoue+17, Jiang+22}) suggest that the contribution of quasars and AGN is insufficient for cosmic reionization, while other studies suggest that this is not the case (e.g. \citealt{Grazian+24, Madau+24}).
If these contributions can be accurately measured, a more precise understanding of the photon budget and ionizing history would be achieved.

\section{Summary}

We present a comprehensive analysis of 127 LAEs at $z=2.2\mathchar`-6.6$ by combining Subaru/HSC narrowband imaging with deep HST optical and JWST near‐ and mid‐infrared photometry. 
Through their stellar population revealed by the SED fitting and size measurements, 
we investigate how \fesca of LAEs depend on their age and assess their contribution to cosmic reionization. 
Our main results are as follows:
\begin{enumerate}
    \item The LAE sample divides into young LAEs ($\mathrm{age}<100$ Myr) and old LAEs ($\mathrm{age}>100$ Myr). Young LAEs are dominant (66\%), but a significant fraction of old LAEs exists at every redshift. We find that old LAEs statistically have larger stellar masses and lower sSFRs, while no significant difference is observed in their dust content.
    \item We compare young and old LAEs with LBGs matched by stellar mass. The comparison reveals that young LAEs exhibit starburst-like sSFRs and lower dust content, suggesting that young LAEs are not simply the low-mass end of the LBG population, but may represent a distinct population characterised by active star formation and reduced dust content even among low-mass galaxies. In contrast, old LAEs resemble LBGs in terms of star formation but are less dusty than LBGs. 
    \item  The rest-optical size of LAEs at $z\lesssim3.3$ is more compact than SFGs at the same mass, while at $z\gtrsim4.9$, their sizes converge to those of the SFGs. 
    This supports the interpretation that compact, intensely star‐forming LAEs become increasingly representative of the high-$z$ galaxy population.
    \item We measure \fesca from the observed \Lya flux corrected for IGM absorption, and the inferred \Ha flux derived from the SED fitting. The \fesca declines with stellar mass, but the dependence bifurcates with age. Young LAEs show a steep anticorrelation, whereas old LAEs show higher \fesca compared to young LAEs at the same mass,  and a much weaker mass dependence. 
    \item In young LAEs, the increase in their dust content and stellar mass 
    makes it increasingly difficult for \Lya photons to escape. 
    In contrast, considering that old LAEs are more compact than SFGs at the same stellar mass, it is suggested that in old LAEs, efficient outflows may create low-density channels of \Hi\, and dust along certain lines of sight through which \Lya photons can escape. This scenario is consistent with our finding that old LAEs have the same characteristics as LBGs, except for their low dust content. 
    \item The IGM-corrected average \fesca for young LAEs shows no significant evolution over $z=2\mathchar`-7$, remaining nearly constant at $\sim40\%$. 
    The \fesca of LAEs is almost equivalent to the value obtained by dividing the global \fesca of SFGs by the $X_{\mathrm{LAE}}$ at each redshift under consistent $\mathrm{M_{UV}}$ and EW$_{\mathrm{Ly\alpha}}$ criteria.
    Therefore, the rise of the global \fesca with redshift can be explained by the increasing $X_{\mathrm{LAE}}$ among SFGs rather than by intrinsic changes in \fesca.
    \item To assess the contribution of LAEs to the cosmic reionization, we convert the measured \fesca into \fescc using two relationships reported by \citetalias{Begley+24} based on observational data and \citetalias{Choustikov+24c} derived from theoretical modelling. 
    We derive \nion\ by integrating the \Lya luminosity density with \xiionfesc. For the \citetalias{Begley+24} model, this \xiionfesc is treated as a constant, while for the \citetalias{Choustikov+24c} model, it is fitted as a function of \Lya luminosity for young LAEs. As a result, we obtain $\dot{n}_{\mathrm{ion}}= 10^{50.86\pm0.15} \mathrm{s^{-1}\,Mpc^{-3}}$ for the \citetalias{Begley+24} model and $\dot{n}_{\mathrm{ion}}= 10^{50.50\pm0.09} \mathrm{s^{-1}\,Mpc^{-3}}$ for the \citetalias{Choustikov+24c} model. While \nion remains constant at $z<6$, it decreases for $z>6$, corresponding to the decline in $\phi^*$ of the Ly$\alpha$ LF, which is assumed to be proportional to the decline in that of the UV LF. The lack of evolution of \nion at $z<6$ is consistent with independent previous observational constraints. We find that, under both the \citetalias{Begley+24} and \citetalias{Choustikov+24c} models, LAEs alone can supply the ionizing photon budget required for fully ionizing the universe at $z\sim6$. These models differ in the \Lya luminosity range that dominate the contribution to reionization: the \citetalias{Begley+24} model reflects a scenario in which \Lya faint galaxies are the primary contributors, whereas the \citetalias{Choustikov+24c} model predicts that \Lya brighter galaxies, with \Lya luminosities around $10^{42}\,\mathrm{erg\,s^{-1}}$, play the dominant role. 
    \item When the \citetalias{Begley+24} model is adopted, the inferred $x_{\mathrm{HI}}$ history is consistent with late and rapid reionization, completing at $z\sim6$. 
    This is consistent with independent observational constraints.
    Although the \citetalias{Choustikov+24c} model indicates that LAEs alone cannot supply the ionizing photons required for reionization, the result highlights that, despite the dependence on the conversion model, LAEs are non-negligible contributors to the ionizing photon budget during the reionization epoch.
    
\end{enumerate}

In this study, we investigate the age dependence and evolution of \fesca, and discuss the \Lya photon escape mechanisms and their impact on cosmic reionization.
This study is made possible by improving the accuracy of the SED fitting using deep imaging data across a wide wavelength range obtained by JWST. 
A method for estimating intrinsic \Ha luminosities based on broad multi-wavelength photometry without spectroscopic data is demonstrated to be highly valuable to understand the mechanisms behind \Lya escape statistically.
Moreover, spectroscopic observations of the \Lya line profile, which is highly sensitive to ISM conditions such as \Hi\, column density, outflows, and inflows, are crucial for probing the mechanisms of \Lya photon escape directly.
The Subaru Prime Focus Spectrograph (PFS) will enable wide-area spectroscopic surveys, allowing a more statistical and detailed understanding of the physical processes governing \Lya escape.

In evaluating the contribution of LAEs to reionization, the following three factors addressed in this study could be further improved: 
(1) the correlation between the \Lya and LyC escape fractions, (2) the relationship between \Lya luminosity and \xiionfesc, and (3) the evolution of the \Lya LF at $z>6$.
Regarding (1), upcoming low-redshift LyC leaker surveys 
combined with multivariate analyses are expected to provide substantial improvements in the estimate of \fescc. 
For (2), leveraging combined observations from JWST and VLT/MUSE in gravitational lensing fields will allow investigations of the \Lya luminosity and \xiionfesc relation in broad ranges in both luminosities and redshifts. 
As for (3), direct observational constraint on the intrinsic \Lya LF at $z>6$ is extremely challenging due to strong IGM absorption,
and innovative approaches, such as inferring \Lya luminosities from alternative diagnostics rather than \Lya itself (e.g. \citealt{Yoshioka+25}), should be necessary.

\section*{Acknowledgements}
We appreciate the anonymous reviewer for constructive comments and suggestions.
We thank Takehiro Yoshioka, Hiroki Hoshi, Kazuhiro Shimasaku, and Masami Ouchi for fruitful discussions and valuable comments that greatly improved this manuscript. 
NK was supported by the Japan Society for the Promotion of Science (JSPS) through Grant-in-
Aid for Scientific Research 25H00663, 25K01038, 25K01044.
SK was supported by the JSPS through Grant-in-
Aid for Scientific Research 24KJ0058, 24K17101. 
YT was supported by Forefront Physics and Mathematics Program to Drive Transformation (FoPM), a World-leading Innovative Graduate Study (WINGS) Program, the University of Tokyo, and JSPS KAKENHI Grant Number 23KJ0726.

The data products presented herein were retrieved from the Dawn JWST Archive (DJA). DJA is an initiative of the Cosmic Dawn Center (DAWN), which is funded by the Danish National Research Foundation under grant DNRF140.
The Hyper Suprime-Cam (HSC) collaboration includes the astronomical communities of Japan and Taiwan, and Princeton University.  The HSC instrumentation and software were developed by the National Astronomical Observatory of Japan (NAOJ), the Kavli Institute for the Physics and Mathematics of the Universe (Kavli IPMU), the University of Tokyo, the High Energy Accelerator Research Organization (KEK), the Academia Sinica Institute for Astronomy and Astrophysics in Taiwan (ASIAA), and Princeton University.  Funding was contributed by the FIRST program from the Japanese Cabinet Office, the Ministry of Education, Culture, Sports, Science and Technology (MEXT), the Japan Society for the Promotion of Science (JSPS), Japan Science and Technology Agency  (JST), the Toray Science  Foundation, NAOJ, Kavli IPMU, KEK, ASIAA, and Princeton University.

This paper is based on data collected at the Subaru Telescope and retrieved from the HSC data archive system, which is operated by Subaru Telescope and Astronomy Data Center (ADC) at NAOJ. Data analysis was in part carried out with the cooperation of Center for Computational Astrophysics (CfCA) at NAOJ.  We are honored and grateful for the opportunity of observing the Universe from Maunakea, which has the cultural, historical and natural significance in Hawaii.

This paper makes use of software developed for Vera C. Rubin Observatory. We thank the Rubin Observatory for making their code available as free software at \url{http://pipelines.lsst.io/}.

The Pan-STARRS1 Surveys (PS1) and the PS1 public science archive have been made possible through contributions by the Institute for Astronomy, the University of Hawaii, the Pan-STARRS Project Office, the Max Planck Society and its participating institutes, the Max Planck Institute for Astronomy, Heidelberg, and the Max Planck Institute for Extraterrestrial Physics, Garching, The Johns Hopkins University, Durham University, the University of Edinburgh, the Queen’s University Belfast, the Harvard-Smithsonian Center for Astrophysics, the Las Cumbres Observatory Global Telescope Network Incorporated, the National Central University of Taiwan, the Space Telescope Science Institute, the National Aeronautics and Space Administration under grant No. NNX08AR22G issued through the Planetary Science Division of the NASA Science Mission Directorate, the National Science Foundation grant No. AST-1238877, the University of Maryland, Eotvos Lorand University (ELTE), the Los Alamos National Laboratory, and the Gordon and Betty Moore Foundation.

%%%%%%%%%%%%%%%%%%%%%%%%%%%%%%%%%%%%%%%%%%%%%%%%%%
\section*{Data Availability}
The LAE catalogues and NB images used in this study are obtained from \citet{Kikuta+23}. The JWST photometric data and images are publicly available through the DAWN JWST Archive (DJA) at \href{https://dawn-cph.github.io/dja/imaging/v7/}{https://dawn-cph.github.io/dja/imaging/v7/}.



\bibliographystyle{mnras}
\bibliography{ref} % if your bibtex file is called example.bib

\begin{thebibliography}{}
\makeatletter
\relax
\def\mn@urlcharsother{\let\do\@makeother \do\$\do\&\do\#\do\^\do\_\do\%\do\~}
\def\mn@doi{\begingroup\mn@urlcharsother \@ifnextchar [ {\mn@doi@} {\mn@doi@[]}}
\def\mn@doi@[#1]#2{\def\@tempa{#1}\ifx\@tempa\@empty \href {http://dx.doi.org/#2} {doi:#2}\else \href {http://dx.doi.org/#2} {#1}\fi \endgroup}
\def\mn@eprint#1#2{\mn@eprint@#1:#2::\@nil}
\def\mn@eprint@arXiv#1{\href {http://arxiv.org/abs/#1} {{\tt arXiv:#1}}}
\def\mn@eprint@dblp#1{\href {http://dblp.uni-trier.de/rec/bibtex/#1.xml} {dblp:#1}}
\def\mn@eprint@#1:#2:#3:#4\@nil{\def\@tempa {#1}\def\@tempb {#2}\def\@tempc {#3}\ifx \@tempc \@empty \let \@tempc \@tempb \let \@tempb \@tempa \fi \ifx \@tempb \@empty \def\@tempb {arXiv}\fi \@ifundefined {mn@eprint@\@tempb}{\@tempb:\@tempc}{\expandafter \expandafter \csname mn@eprint@\@tempb\endcsname \expandafter{\@tempc}}}

\bibitem[\protect\citeauthoryear{{Aihara} et~al.,}{{Aihara} et~al.}{2018a}]{Aihara+18a}
{Aihara} H.,  et~al., 2018a, \mn@doi [\pasj] {10.1093/pasj/psx066}, \href {https://ui.adsabs.harvard.edu/abs/2018PASJ...70S...4A} {70, S4}

\bibitem[\protect\citeauthoryear{{Aihara} et~al.,}{{Aihara} et~al.}{2018b}]{Aihara+18b}
{Aihara} H.,  et~al., 2018b, \mn@doi [\pasj] {10.1093/pasj/psx081}, \href {https://ui.adsabs.harvard.edu/abs/2018PASJ...70S...8A} {70, S8}

\bibitem[\protect\citeauthoryear{{Aihara} et~al.,}{{Aihara} et~al.}{2019}]{Aihara+19}
{Aihara} H.,  et~al., 2019, \mn@doi [\pasj] {10.1093/pasj/psz103}, \href {https://ui.adsabs.harvard.edu/abs/2019PASJ...71..114A} {71, 114}

\bibitem[\protect\citeauthoryear{{Aihara} et~al.,}{{Aihara} et~al.}{2022}]{Aihara+22}
{Aihara} H.,  et~al., 2022, \mn@doi [\pasj] {10.1093/pasj/psab122}, \href {https://ui.adsabs.harvard.edu/abs/2022PASJ...74..247A} {74, 247}

\bibitem[\protect\citeauthoryear{{Allen} et~al.,}{{Allen} et~al.}{2025}]{Allen+24}
{Allen} N.,  et~al., 2025, \mn@doi [\aap] {10.1051/0004-6361/202452690}, \href {https://ui.adsabs.harvard.edu/abs/2025A&A...698A..30A} {698, A30}

\bibitem[\protect\citeauthoryear{{Arrabal Haro} et~al.,}{{Arrabal Haro} et~al.}{2020}]{Haro+20}
{Arrabal Haro} P.,  et~al., 2020, \mn@doi [\mnras] {10.1093/mnras/staa1196}, \href {https://ui.adsabs.harvard.edu/abs/2020MNRAS.495.1807A} {495, 1807}

\bibitem[\protect\citeauthoryear{{Becker} \& {Bolton}}{{Becker} \& {Bolton}}{2013}]{Becker+13}
{Becker} G.~D.,  {Bolton} J.~S.,  2013, \mn@doi [\mnras] {10.1093/mnras/stt1610}, \href {https://ui.adsabs.harvard.edu/abs/2013MNRAS.436.1023B} {436, 1023}

\bibitem[\protect\citeauthoryear{{Begley} et~al.,}{{Begley} et~al.}{2024}]{Begley+24}
{Begley} R.,  et~al., 2024, \mn@doi [\mnras] {10.1093/mnras/stad3417}, \href {https://ui.adsabs.harvard.edu/abs/2024MNRAS.527.4040B} {527, 4040}

\bibitem[\protect\citeauthoryear{{Bertin} \& {Arnouts}}{{Bertin} \& {Arnouts}}{1996}]{sextractor}
{Bertin} E.,  {Arnouts} S.,  1996, \mn@doi [\aaps] {10.1051/aas:1996164}, \href {https://ui.adsabs.harvard.edu/abs/1996A&AS..117..393B} {117, 393}

\bibitem[\protect\citeauthoryear{{Bhagwat}, {Napolitano}, {Pentericci}, {Ciardi}  \& {Costa}}{{Bhagwat} et~al.}{2025}]{Bhagwat+24}
{Bhagwat} A.,  {Napolitano} L.,  {Pentericci} L.,  {Ciardi} B.,   {Costa} T.,  2025, \mn@doi [\mnras] {10.1093/mnras/staf1121}, \href {https://ui.adsabs.harvard.edu/abs/2025MNRAS.tmp.1083B} {}

\bibitem[\protect\citeauthoryear{{Boquien}, {Burgarella}, {Roehlly}, {Buat}, {Ciesla}, {Corre}, {Inoue}  \& {Salas}}{{Boquien} et~al.}{2019}]{Boquein+19}
{Boquien} M.,  {Burgarella} D.,  {Roehlly} Y.,  {Buat} V.,  {Ciesla} L.,  {Corre} D.,  {Inoue} A.~K.,   {Salas} H.,  2019, \mn@doi [\aap] {10.1051/0004-6361/201834156}, \href {https://ui.adsabs.harvard.edu/abs/2019A&A...622A.103B} {622, A103}

\bibitem[\protect\citeauthoryear{{Bosman} et~al.,}{{Bosman} et~al.}{2022}]{Bosman+22}
{Bosman} S. E.~I.,  et~al., 2022, \mn@doi [\mnras] {10.1093/mnras/stac1046}, \href {https://ui.adsabs.harvard.edu/abs/2022MNRAS.514...55B} {514, 55}

\bibitem[\protect\citeauthoryear{Bradley et~al.,}{Bradley et~al.}{2024}]{photutils}
Bradley L.,  et~al., 2024, astropy/photutils: 2.0.2, \mn@doi{10.5281/zenodo.13989456}, \url {https://doi.org/10.5281/zenodo.13989456}

\bibitem[\protect\citeauthoryear{{Brammer}}{{Brammer}}{2023a}]{Brammer+23}
{Brammer} G.,  2023a, {grizli}, \mn@doi{10.5281/zenodo.8370018}

\bibitem[\protect\citeauthoryear{{Brammer}}{{Brammer}}{2023b}]{msaexp}
{Brammer} G.,  2023b, {msaexp: NIRSpec analyis tools}, \mn@doi{10.5281/zenodo.8319596}

\bibitem[\protect\citeauthoryear{{Brammer}, {van Dokkum}  \& {Coppi}}{{Brammer} et~al.}{2008}]{EAZY}
{Brammer} G.~B.,  {van Dokkum} P.~G.,   {Coppi} P.,  2008, \mn@doi [\apj] {10.1086/591786}, \href {https://ui.adsabs.harvard.edu/abs/2008ApJ...686.1503B} {686, 1503}

\bibitem[\protect\citeauthoryear{{Bruzual} \& {Charlot}}{{Bruzual} \& {Charlot}}{2003}]{BruChar+03}
{Bruzual} G.,  {Charlot} S.,  2003, \mn@doi [\mnras] {10.1046/j.1365-8711.2003.06897.x}, \href {https://ui.adsabs.harvard.edu/abs/2003MNRAS.344.1000B} {344, 1000}

\bibitem[\protect\citeauthoryear{{Calzetti}, {Armus}, {Bohlin}, {Kinney}, {Koornneef}  \& {Storchi-Bergmann}}{{Calzetti} et~al.}{2000}]{Calzetti+00}
{Calzetti} D.,  {Armus} L.,  {Bohlin} R.~C.,  {Kinney} A.~L.,  {Koornneef} J.,   {Storchi-Bergmann} T.,  2000, \mn@doi [\apj] {10.1086/308692}, \href {https://ui.adsabs.harvard.edu/abs/2000ApJ...533..682C} {533, 682}

\bibitem[\protect\citeauthoryear{{Cantalupo}, {Lilly}  \& {Haehnelt}}{{Cantalupo} et~al.}{2012}]{Cantalupo+12}
{Cantalupo} S.,  {Lilly} S.~J.,   {Haehnelt} M.~G.,  2012, \mn@doi [\mnras] {10.1111/j.1365-2966.2012.21529.x}, \href {https://ui.adsabs.harvard.edu/abs/2012MNRAS.425.1992C} {425, 1992}

\bibitem[\protect\citeauthoryear{{Cappelluti} et~al.,}{{Cappelluti} et~al.}{2009}]{XMM-cosmos}
{Cappelluti} N.,  et~al., 2009, \mn@doi [\aap] {10.1051/0004-6361/200810794}, \href {https://ui.adsabs.harvard.edu/abs/2009A&A...497..635C} {497, 635}

\bibitem[\protect\citeauthoryear{{Chabrier}}{{Chabrier}}{2003}]{Chabrier+03}
{Chabrier} G.,  2003, \mn@doi [\pasp] {10.1086/376392}, \href {https://ui.adsabs.harvard.edu/abs/2003PASP..115..763C} {115, 763}

\bibitem[\protect\citeauthoryear{{Choustikov} et~al.,}{{Choustikov} et~al.}{2024a}]{choustikov+24a}
{Choustikov} N.,  et~al., 2024a, \mn@doi [\mnras] {10.1093/mnras/stae776}, \href {https://ui.adsabs.harvard.edu/abs/2024MNRAS.529.3751C} {529, 3751}

\bibitem[\protect\citeauthoryear{{Choustikov} et~al.,}{{Choustikov} et~al.}{2024b}]{Choustikov+24c}
{Choustikov} N.,  et~al., 2024b, \mn@doi [\mnras] {10.1093/mnras/stae1586}, \href {https://ui.adsabs.harvard.edu/abs/2024MNRAS.532.2463C} {532, 2463}

\bibitem[\protect\citeauthoryear{{Civano} et~al.,}{{Civano} et~al.}{2016}]{chandra-cosmos}
{Civano} F.,  et~al., 2016, \mn@doi [\apj] {10.3847/0004-637X/819/1/62}, \href {https://ui.adsabs.harvard.edu/abs/2016ApJ...819...62C} {819, 62}

\bibitem[\protect\citeauthoryear{{Clarke}, {Shapley}, {Sanders}, {Topping}, {Brammer}, {Bento}, {Reddy}  \& {Kehoe}}{{Clarke} et~al.}{2024}]{Clarke+24}
{Clarke} L.,  {Shapley} A.~E.,  {Sanders} R.~L.,  {Topping} M.~W.,  {Brammer} G.~B.,  {Bento} T.,  {Reddy} N.~A.,   {Kehoe} E.,  2024, \mn@doi [\apj] {10.3847/1538-4357/ad8ba4}, \href {https://ui.adsabs.harvard.edu/abs/2024ApJ...977..133C} {977, 133}

\bibitem[\protect\citeauthoryear{{Davies} et~al.,}{{Davies} et~al.}{2019}]{Davies+19}
{Davies} R.~L.,  et~al., 2019, \mn@doi [\apj] {10.3847/1538-4357/ab06f1}, \href {https://ui.adsabs.harvard.edu/abs/2019ApJ...873..122D} {873, 122}

\bibitem[\protect\citeauthoryear{{Davies} et~al.,}{{Davies} et~al.}{2024}]{Davies+24}
{Davies} F.~B.,  et~al., 2024, \mn@doi [\apj] {10.3847/1538-4357/ad1d5d}, \href {https://ui.adsabs.harvard.edu/abs/2024ApJ...965..134D} {965, 134}

\bibitem[\protect\citeauthoryear{{Delvecchio} et~al.,}{{Delvecchio} et~al.}{2017}]{VLA-cosmos}
{Delvecchio} I.,  et~al., 2017, \mn@doi [\aap] {10.1051/0004-6361/201629367}, \href {https://ui.adsabs.harvard.edu/abs/2017A&A...602A...3D} {602, A3}

\bibitem[\protect\citeauthoryear{{Dijkstra}, {Gronke}  \& {Venkatesan}}{{Dijkstra} et~al.}{2016}]{Dijkstra+16}
{Dijkstra} M.,  {Gronke} M.,   {Venkatesan} A.,  2016, \mn@doi [\apj] {10.3847/0004-637X/828/2/71}, \href {https://ui.adsabs.harvard.edu/abs/2016ApJ...828...71D} {828, 71}

\bibitem[\protect\citeauthoryear{{Dunlop} et~al.,}{{Dunlop} et~al.}{2021}]{Dunlop+21}
{Dunlop} J.~S.,  et~al., 2021, {PRIMER: Public Release IMaging for Extragalactic Research}, JWST Proposal. Cycle 1, ID. \#1837

\bibitem[\protect\citeauthoryear{{Eldridge}, {Stanway}, {Xiao}, {McClelland}, {Taylor}, {Ng}, {Greis}  \& {Bray}}{{Eldridge} et~al.}{2017}]{Eldridge+17}
{Eldridge} J.~J.,  {Stanway} E.~R.,  {Xiao} L.,  {McClelland} L.~A.~S.,  {Taylor} G.,  {Ng} M.,  {Greis} S.~M.~L.,   {Bray} J.~C.,  2017, \mn@doi [\pasa] {10.1017/pasa.2017.51}, \href {https://ui.adsabs.harvard.edu/abs/2017PASA...34...58E} {34, e058}

\bibitem[\protect\citeauthoryear{{Fausey} et~al.,}{{Fausey} et~al.}{2025}]{Fausey+25}
{Fausey} H.~M.,  et~al., 2025, \mn@doi [\mnras] {10.1093/mnras/stae2757}, \href {https://ui.adsabs.harvard.edu/abs/2025MNRAS.536.2839F} {536, 2839}

\bibitem[\protect\citeauthoryear{{Finkelstein}, {Rhoads}, {Malhotra}  \& {Grogin}}{{Finkelstein} et~al.}{2009}]{Finkelstein+09}
{Finkelstein} S.~L.,  {Rhoads} J.~E.,  {Malhotra} S.,   {Grogin} N.,  2009, \mn@doi [\apj] {10.1088/0004-637X/691/1/465}, \href {https://ui.adsabs.harvard.edu/abs/2009ApJ...691..465F} {691, 465}

\bibitem[\protect\citeauthoryear{{Finkelstein} et~al.,}{{Finkelstein} et~al.}{2015}]{Finkelstein+15}
{Finkelstein} S.~L.,  et~al., 2015, \mn@doi [\apj] {10.1088/0004-637X/810/1/71}, \href {https://ui.adsabs.harvard.edu/abs/2015ApJ...810...71F} {810, 71}

\bibitem[\protect\citeauthoryear{{Finkelstein} et~al.,}{{Finkelstein} et~al.}{2019}]{Finkelstein+19}
{Finkelstein} S.~L.,  et~al., 2019, \mn@doi [\apj] {10.3847/1538-4357/ab1ea8}, \href {https://ui.adsabs.harvard.edu/abs/2019ApJ...879...36F} {879, 36}

\bibitem[\protect\citeauthoryear{{Firestone} et~al.,}{{Firestone} et~al.}{2025}]{Firestone+25}
{Firestone} N.~M.,  et~al., 2025, \mn@doi [\apjl] {10.3847/2041-8213/adbf8c}, \href {https://ui.adsabs.harvard.edu/abs/2025ApJ...986L...8F} {986, L8}

\bibitem[\protect\citeauthoryear{{Flury} et~al.,}{{Flury} et~al.}{2022}]{Flury+22}
{Flury} S.~R.,  et~al., 2022, \mn@doi [\apj] {10.3847/1538-4357/ac61e4}, \href {https://ui.adsabs.harvard.edu/abs/2022ApJ...930..126F} {930, 126}

\bibitem[\protect\citeauthoryear{{Fruchter} \& {Hook}}{{Fruchter} \& {Hook}}{2002}]{drizzle}
{Fruchter} A.~S.,  {Hook} R.~N.,  2002, \mn@doi [\pasp] {10.1086/338393}, \href {https://ui.adsabs.harvard.edu/abs/2002PASP..114..144F} {114, 144}

\bibitem[\protect\citeauthoryear{{Gaikwad} et~al.,}{{Gaikwad} et~al.}{2023}]{Gaikwad+23}
{Gaikwad} P.,  et~al., 2023, \mn@doi [\mnras] {10.1093/mnras/stad2566}, \href {https://ui.adsabs.harvard.edu/abs/2023MNRAS.525.4093G} {525, 4093}

\bibitem[\protect\citeauthoryear{{Gawiser} et~al.,}{{Gawiser} et~al.}{2006}]{Gawiser+06}
{Gawiser} E.,  et~al., 2006, \mn@doi [\apjl] {10.1086/504467}, \href {https://ui.adsabs.harvard.edu/abs/2006ApJ...642L..13G} {642, L13}

\bibitem[\protect\citeauthoryear{{Gazagnes}, {Chisholm}, {Schaerer}, {Verhamme}  \& {Izotov}}{{Gazagnes} et~al.}{2020}]{Gazagnes+20}
{Gazagnes} S.,  {Chisholm} J.,  {Schaerer} D.,  {Verhamme} A.,   {Izotov} Y.,  2020, \mn@doi [\aap] {10.1051/0004-6361/202038096}, \href {https://ui.adsabs.harvard.edu/abs/2020A&A...639A..85G} {639, A85}

\bibitem[\protect\citeauthoryear{{Goovaerts} et~al.,}{{Goovaerts} et~al.}{2023}]{Goovaerts+23}
{Goovaerts} I.,  et~al., 2023, \mn@doi [\aap] {10.1051/0004-6361/202347110}, \href {https://ui.adsabs.harvard.edu/abs/2023A&A...678A.174G} {678, A174}

\bibitem[\protect\citeauthoryear{{Goovaerts}, {Thai}, {Pello}, {Tuan-Anh}, {Laporte}, {Matthee}, {Nanayakkara}  \& {Pharo}}{{Goovaerts} et~al.}{2024}]{Goovaerts+24b}
{Goovaerts} I.,  {Thai} T.~T.,  {Pello} R.,  {Tuan-Anh} P.,  {Laporte} N.,  {Matthee} J.,  {Nanayakkara} T.,   {Pharo} J.,  2024, \mn@doi [\aap] {10.1051/0004-6361/202451432}, \href {https://ui.adsabs.harvard.edu/abs/2024A&A...690A.302G} {690, A302}

\bibitem[\protect\citeauthoryear{{Grazian} et~al.,}{{Grazian} et~al.}{2024}]{Grazian+24}
{Grazian} A.,  et~al., 2024, \mn@doi [\apj] {10.3847/1538-4357/ad6980}, \href {https://ui.adsabs.harvard.edu/abs/2024ApJ...974...84G} {974, 84}

\bibitem[\protect\citeauthoryear{{Greig}, {Mesinger}, {Davies}, {Wang}, {Yang}  \& {Hennawi}}{{Greig} et~al.}{2022}]{Greig+22}
{Greig} B.,  {Mesinger} A.,  {Davies} F.~B.,  {Wang} F.,  {Yang} J.,   {Hennawi} J.~F.,  2022, \mn@doi [\mnras] {10.1093/mnras/stac825}, \href {https://ui.adsabs.harvard.edu/abs/2022MNRAS.512.5390G} {512, 5390}

\bibitem[\protect\citeauthoryear{{Harikane} et~al.,}{{Harikane} et~al.}{2025}]{Harikane+25}
{Harikane} Y.,  et~al., 2025, \mn@doi [\apj] {10.3847/1538-4357/ad9b2c}, \href {https://ui.adsabs.harvard.edu/abs/2025ApJ...980..138H} {980, 138}

\bibitem[\protect\citeauthoryear{{Hayes}, {Schaerer}, {{\"O}stlin}, {Mas-Hesse}, {Atek}  \& {Kunth}}{{Hayes} et~al.}{2011}]{Hayes+11}
{Hayes} M.,  {Schaerer} D.,  {{\"O}stlin} G.,  {Mas-Hesse} J.~M.,  {Atek} H.,   {Kunth} D.,  2011, \mn@doi [\apj] {10.1088/0004-637X/730/1/8}, \href {https://ui.adsabs.harvard.edu/abs/2011ApJ...730....8H} {730, 8}

\bibitem[\protect\citeauthoryear{{Hayes}, {Runnholm}, {Gronke}  \& {Scarlata}}{{Hayes} et~al.}{2021}]{Hayes+21}
{Hayes} M.~J.,  {Runnholm} A.,  {Gronke} M.,   {Scarlata} C.,  2021, \mn@doi [\apj] {10.3847/1538-4357/abd246}, \href {https://ui.adsabs.harvard.edu/abs/2021ApJ...908...36H} {908, 36}

\bibitem[\protect\citeauthoryear{{Herenz} et~al.,}{{Herenz} et~al.}{2019}]{Herenz+19}
{Herenz} E.~C.,  et~al., 2019, \mn@doi [\aap] {10.1051/0004-6361/201834164}, \href {https://ui.adsabs.harvard.edu/abs/2019A&A...621A.107H} {621, A107}

\bibitem[\protect\citeauthoryear{{Herenz} et~al.,}{{Herenz} et~al.}{2025}]{Herenz+25}
{Herenz} E.~C.,  et~al., 2025, \mn@doi [\aap] {10.1051/0004-6361/202451012}, \href {https://ui.adsabs.harvard.edu/abs/2025A&A...693A.252H} {693, A252}

\bibitem[\protect\citeauthoryear{{Hummer} \& {Storey}}{{Hummer} \& {Storey}}{1987}]{Hummer}
{Hummer} D.~G.,  {Storey} P.~J.,  1987, \mn@doi [\mnras] {10.1093/mnras/224.3.801}, \href {https://ui.adsabs.harvard.edu/abs/1987MNRAS.224..801H} {224, 801}

\bibitem[\protect\citeauthoryear{{Iani} et~al.,}{{Iani} et~al.}{2024}]{Iani+24}
{Iani} E.,  et~al., 2024, \mn@doi [\apj] {10.3847/1538-4357/ad15f6}, \href {https://ui.adsabs.harvard.edu/abs/2024ApJ...963...97I} {963, 97}

\bibitem[\protect\citeauthoryear{{Inoue} et~al.,}{{Inoue} et~al.}{2020}]{Inoue+20}
{Inoue} A.~K.,  et~al., 2020, \mn@doi [\pasj] {10.1093/pasj/psaa100}, \href {https://ui.adsabs.harvard.edu/abs/2020PASJ...72..101I} {72, 101}

\bibitem[\protect\citeauthoryear{{Ito} et~al.,}{{Ito} et~al.}{2024}]{Ito+24}
{Ito} K.,  et~al., 2024, \mn@doi [\apj] {10.3847/1538-4357/ad2512}, \href {https://ui.adsabs.harvard.edu/abs/2024ApJ...964..192I} {964, 192}

\bibitem[\protect\citeauthoryear{{Izotov}, {Schaerer}, {Worseck}, {Verhamme}, {Guseva}, {Thuan}, {Orlitov{\'a}}  \& {Fricke}}{{Izotov} et~al.}{2020}]{Izotov+20}
{Izotov} Y.~I.,  {Schaerer} D.,  {Worseck} G.,  {Verhamme} A.,  {Guseva} N.~G.,  {Thuan} T.~X.,  {Orlitov{\'a}} I.,   {Fricke} K.~J.,  2020, \mn@doi [\mnras] {10.1093/mnras/stz3041}, \href {https://ui.adsabs.harvard.edu/abs/2020MNRAS.491..468I} {491, 468}

\bibitem[\protect\citeauthoryear{{Izotov}, {Thuan}, {Guseva}, {Schaerer}, {Worseck}  \& {Verhamme}}{{Izotov} et~al.}{2024}]{Izotov+24}
{Izotov} Y.~I.,  {Thuan} T.~X.,  {Guseva} N.~G.,  {Schaerer} D.,  {Worseck} G.,   {Verhamme} A.,  2024, \mn@doi [\mnras] {10.1093/mnras/stad3151}, \href {https://ui.adsabs.harvard.edu/abs/2024MNRAS.527..281I} {527, 281}

\bibitem[\protect\citeauthoryear{{Ji} et~al.,}{{Ji} et~al.}{2025}]{Zhiyuan+25}
{Ji} Z.,  et~al., 2025, \mn@doi [arXiv e-prints] {10.48550/arXiv.2504.01067}, \href {https://ui.adsabs.harvard.edu/abs/2025arXiv250401067J} {p. arXiv:2504.01067}

\bibitem[\protect\citeauthoryear{{Jia} et~al.,}{{Jia} et~al.}{2024}]{Jia+24}
{Jia} C.,  et~al., 2024, \mn@doi [\apj] {10.3847/1538-4357/ad919a}, \href {https://ui.adsabs.harvard.edu/abs/2024ApJ...977..165J} {977, 165}

\bibitem[\protect\citeauthoryear{{Jiang} et~al.,}{{Jiang} et~al.}{2016}]{Jiang+16}
{Jiang} L.,  et~al., 2016, \mn@doi [\apj] {10.3847/0004-637X/816/1/16}, \href {https://ui.adsabs.harvard.edu/abs/2016ApJ...816...16J} {816, 16}

\bibitem[\protect\citeauthoryear{{Jiang} et~al.,}{{Jiang} et~al.}{2022}]{Jiang+22}
{Jiang} L.,  et~al., 2022, \mn@doi [Nature Astronomy] {10.1038/s41550-022-01708-w}, \href {https://ui.adsabs.harvard.edu/abs/2022NatAs...6..850J} {6, 850}

\bibitem[\protect\citeauthoryear{{Jiang} et~al.,}{{Jiang} et~al.}{2024}]{Jiang+24}
{Jiang} H.,  et~al., 2024, \mn@doi [\apj] {10.3847/1538-4357/ad61db}, \href {https://ui.adsabs.harvard.edu/abs/2024ApJ...972..121J} {972, 121}

\bibitem[\protect\citeauthoryear{{Jin} et~al.,}{{Jin} et~al.}{2023}]{Jin+23}
{Jin} X.,  et~al., 2023, \mn@doi [\apj] {10.3847/1538-4357/aca678}, \href {https://ui.adsabs.harvard.edu/abs/2023ApJ...942...59J} {942, 59}

\bibitem[\protect\citeauthoryear{{Jones} et~al.,}{{Jones} et~al.}{2025}]{Jones+25}
{Jones} G.~C.,  et~al., 2025, \mn@doi [\mnras] {10.1093/mnras/stae2670}, \href {https://ui.adsabs.harvard.edu/abs/2025MNRAS.536.2355J} {536, 2355}

\bibitem[\protect\citeauthoryear{{Kageura} et~al.,}{{Kageura} et~al.}{2025}]{Kageura+25}
{Kageura} Y.,  et~al., 2025, \mn@doi [\apjs] {10.3847/1538-4365/adc690}, \href {https://ui.adsabs.harvard.edu/abs/2025ApJS..278...33K} {278, 33}

\bibitem[\protect\citeauthoryear{{Kennicutt}}{{Kennicutt}}{1998}]{Kennicutt+98}
{Kennicutt} Jr. R.~C.,  1998, \mn@doi [\apj] {10.1086/305588}, \href {https://ui.adsabs.harvard.edu/abs/1998ApJ...498..541K} {498, 541}

\bibitem[\protect\citeauthoryear{{Khostovan} et~al.,}{{Khostovan} et~al.}{2019}]{Khostovan+19}
{Khostovan} A.~A.,  et~al., 2019, \mn@doi [\mnras] {10.1093/mnras/stz2149}, \href {https://ui.adsabs.harvard.edu/abs/2019MNRAS.489..555K} {489, 555}

\bibitem[\protect\citeauthoryear{{Kikuta} et~al.,}{{Kikuta} et~al.}{2023a}]{Kikuta+23}
{Kikuta} S.,  et~al., 2023a, \mn@doi [\apjs] {10.3847/1538-4365/ace4cb}, \href {https://ui.adsabs.harvard.edu/abs/2023ApJS..268...24K} {268, 24}

\bibitem[\protect\citeauthoryear{{Kikuta} et~al.,}{{Kikuta} et~al.}{2023b}]{Kikuta+23_halo}
{Kikuta} S.,  et~al., 2023b, \mn@doi [\apj] {10.3847/1538-4357/acbf30}, \href {https://ui.adsabs.harvard.edu/abs/2023ApJ...947...75K} {947, 75}

\bibitem[\protect\citeauthoryear{{Kim} et~al.,}{{Kim} et~al.}{2025}]{Kim+25}
{Kim} K.,  et~al., 2025, \mn@doi [arXiv e-prints] {10.48550/arXiv.2501.07548}, \href {https://ui.adsabs.harvard.edu/abs/2025arXiv250107548K} {p. arXiv:2501.07548}

\bibitem[\protect\citeauthoryear{{Kocevski} et~al.,}{{Kocevski} et~al.}{2018}]{chandra-uds}
{Kocevski} D.~D.,  et~al., 2018, \mn@doi [\apjs] {10.3847/1538-4365/aab9b4}, \href {https://ui.adsabs.harvard.edu/abs/2018ApJS..236...48K} {236, 48}

\bibitem[\protect\citeauthoryear{{Koekemoer} et~al.,}{{Koekemoer} et~al.}{2011}]{CANDELS}
{Koekemoer} A.~M.,  et~al., 2011, \mn@doi [\apjs] {10.1088/0067-0049/197/2/36}, \href {https://ui.adsabs.harvard.edu/abs/2011ApJS..197...36K} {197, 36}

\bibitem[\protect\citeauthoryear{{Konno}, {Ouchi}, {Nakajima}, {Duval}, {Kusakabe}, {Ono}  \& {Shimasaku}}{{Konno} et~al.}{2016}]{Konno+16}
{Konno} A.,  {Ouchi} M.,  {Nakajima} K.,  {Duval} F.,  {Kusakabe} H.,  {Ono} Y.,   {Shimasaku} K.,  2016, \mn@doi [\apj] {10.3847/0004-637X/823/1/20}, \href {https://ui.adsabs.harvard.edu/abs/2016ApJ...823...20K} {823, 20}

\bibitem[\protect\citeauthoryear{{Kron}}{{Kron}}{1980}]{Kron}
{Kron} R.~G.,  1980, \mn@doi [\apjs] {10.1086/190669}, \href {https://ui.adsabs.harvard.edu/abs/1980ApJS...43..305K} {43, 305}

\bibitem[\protect\citeauthoryear{{Kusakabe} et~al.,}{{Kusakabe} et~al.}{2020}]{Kusakabe+20}
{Kusakabe} H.,  et~al., 2020, \mn@doi [\aap] {10.1051/0004-6361/201937340}, \href {https://ui.adsabs.harvard.edu/abs/2020A&A...638A..12K} {638, A12}

\bibitem[\protect\citeauthoryear{{Lai} et~al.,}{{Lai} et~al.}{2008}]{Lai+08}
{Lai} K.,  et~al., 2008, \mn@doi [\apj] {10.1086/524702}, \href {https://ui.adsabs.harvard.edu/abs/2008ApJ...674...70L} {674, 70}

\bibitem[\protect\citeauthoryear{{Le Reste} et~al.,}{{Le Reste} et~al.}{2025}]{Le_Reste+25}
{Le Reste} A.,  et~al., 2025, \mn@doi [arXiv e-prints] {10.48550/arXiv.2504.07056}, \href {https://ui.adsabs.harvard.edu/abs/2025arXiv250407056L} {p. arXiv:2504.07056}

\bibitem[\protect\citeauthoryear{{Li} et~al.,}{{Li} et~al.}{2024}]{Mingyu+24}
{Li} M.,  et~al., 2024, \mn@doi [\apjs] {10.3847/1538-4365/ad812c}, \href {https://ui.adsabs.harvard.edu/abs/2024ApJS..275...27L} {275, 27}

\bibitem[\protect\citeauthoryear{{Lin} et~al.,}{{Lin} et~al.}{2024}]{Lin+24}
{Lin} X.,  et~al., 2024, \mn@doi [\apjs] {10.3847/1538-4365/ad3e7d}, \href {https://ui.adsabs.harvard.edu/abs/2024ApJS..272...33L} {272, 33}

\bibitem[\protect\citeauthoryear{{Liu}, {Dai}, {Wuyts}, {Huang}  \& {Jiang}}{{Liu} et~al.}{2024}]{Liu+24}
{Liu} Y.,  {Dai} Y.~S.,  {Wuyts} S.,  {Huang} J.-S.,   {Jiang} L.,  2024, \mn@doi [\apj] {10.3847/1538-4357/ad3822}, \href {https://ui.adsabs.harvard.edu/abs/2024ApJ...966..210L} {966, 210}

\bibitem[\protect\citeauthoryear{{Llerena} et~al.,}{{Llerena} et~al.}{2025}]{Llerena+24}
{Llerena} M.,  et~al., 2025, \mn@doi [\aap] {10.1051/0004-6361/202453251}, \href {https://ui.adsabs.harvard.edu/abs/2025A&A...698A.302L} {698, A302}

\bibitem[\protect\citeauthoryear{{Madau}}{{Madau}}{1995}]{Madau+1995}
{Madau} P.,  1995, \mn@doi [\apj] {10.1086/175332}, \href {https://ui.adsabs.harvard.edu/abs/1995ApJ...441...18M} {441, 18}

\bibitem[\protect\citeauthoryear{{Madau} \& {Dickinson}}{{Madau} \& {Dickinson}}{2014}]{Madau+14}
{Madau} P.,  {Dickinson} M.,  2014, \mn@doi [\araa] {10.1146/annurev-astro-081811-125615}, \href {https://ui.adsabs.harvard.edu/abs/2014ARA&A..52..415M} {52, 415}

\bibitem[\protect\citeauthoryear{{Madau}, {Haardt}  \& {Rees}}{{Madau} et~al.}{1999}]{Madau+1999}
{Madau} P.,  {Haardt} F.,   {Rees} M.~J.,  1999, \mn@doi [\apj] {10.1086/306975}, \href {https://ui.adsabs.harvard.edu/abs/1999ApJ...514..648M} {514, 648}

\bibitem[\protect\citeauthoryear{{Madau}, {Giallongo}, {Grazian}  \& {Haardt}}{{Madau} et~al.}{2024}]{Madau+24}
{Madau} P.,  {Giallongo} E.,  {Grazian} A.,   {Haardt} F.,  2024, \mn@doi [\apj] {10.3847/1538-4357/ad5ce8}, \href {https://ui.adsabs.harvard.edu/abs/2024ApJ...971...75M} {971, 75}

\bibitem[\protect\citeauthoryear{{Marchi} et~al.,}{{Marchi} et~al.}{2018}]{Marchi+18}
{Marchi} F.,  et~al., 2018, \mn@doi [\aap] {10.1051/0004-6361/201732133}, \href {https://ui.adsabs.harvard.edu/abs/2018A&A...614A..11M} {614, A11}

\bibitem[\protect\citeauthoryear{{Marchi} et~al.,}{{Marchi} et~al.}{2019}]{Marchi+19}
{Marchi} F.,  et~al., 2019, \mn@doi [\aap] {10.1051/0004-6361/201935495}, \href {https://ui.adsabs.harvard.edu/abs/2019A&A...631A..19M} {631, A19}

\bibitem[\protect\citeauthoryear{{Marino} et~al.,}{{Marino} et~al.}{2018}]{Marino+18}
{Marino} R.~A.,  et~al., 2018, \mn@doi [\apj] {10.3847/1538-4357/aab6aa}, \href {https://ui.adsabs.harvard.edu/abs/2018ApJ...859...53M} {859, 53}

\bibitem[\protect\citeauthoryear{{Mascia} et~al.,}{{Mascia} et~al.}{2025}]{Mascia+25}
{Mascia} S.,  et~al., 2025, arXiv e-prints, \href {https://ui.adsabs.harvard.edu/abs/2025arXiv250108268M} {p. arXiv:2501.08268}

\bibitem[\protect\citeauthoryear{{Matharu} et~al.,}{{Matharu} et~al.}{2024}]{Matharu+24}
{Matharu} J.,  et~al., 2024, \mn@doi [\aap] {10.1051/0004-6361/202450522}, \href {https://ui.adsabs.harvard.edu/abs/2024A&A...690A..64M} {690, A64}

\bibitem[\protect\citeauthoryear{{Matthee}, {Sobral}, {Oteo}, {Best}, {Smail}, {R{\"o}ttgering}  \& {Paulino-Afonso}}{{Matthee} et~al.}{2016}]{Matthee+16}
{Matthee} J.,  {Sobral} D.,  {Oteo} I.,  {Best} P.,  {Smail} I.,  {R{\"o}ttgering} H.,   {Paulino-Afonso} A.,  2016, \mn@doi [\mnras] {10.1093/mnras/stw322}, \href {https://ui.adsabs.harvard.edu/abs/2016MNRAS.458..449M} {458, 449}

\bibitem[\protect\citeauthoryear{{Matthee} et~al.,}{{Matthee} et~al.}{2021}]{Matthee+21}
{Matthee} J.,  et~al., 2021, \mn@doi [\mnras] {10.1093/mnras/stab1304}, \href {https://ui.adsabs.harvard.edu/abs/2021MNRAS.505.1382M} {505, 1382}

\bibitem[\protect\citeauthoryear{{Matthee} et~al.,}{{Matthee} et~al.}{2022}]{Matthee+22}
{Matthee} J.,  et~al., 2022, \mn@doi [\mnras] {10.1093/mnras/stac801}, \href {https://ui.adsabs.harvard.edu/abs/2022MNRAS.512.5960M} {512, 5960}

\bibitem[\protect\citeauthoryear{{McGreer}, {Mesinger}  \& {D'Odorico}}{{McGreer} et~al.}{2015}]{McGreer+15}
{McGreer} I.~D.,  {Mesinger} A.,   {D'Odorico} V.,  2015, \mn@doi [\mnras] {10.1093/mnras/stu2449}, \href {https://ui.adsabs.harvard.edu/abs/2015MNRAS.447..499M} {447, 499}

\bibitem[\protect\citeauthoryear{{Momose} et~al.,}{{Momose} et~al.}{2014}]{Momose+14}
{Momose} R.,  et~al., 2014, \mn@doi [\mnras] {10.1093/mnras/stu825}, \href {https://ui.adsabs.harvard.edu/abs/2014MNRAS.442..110M} {442, 110}

\bibitem[\protect\citeauthoryear{{Morales}, {Mason}, {Bruton}, {Gronke}, {Haardt}  \& {Scarlata}}{{Morales} et~al.}{2021}]{Morales+21}
{Morales} A.~M.,  {Mason} C.~A.,  {Bruton} S.,  {Gronke} M.,  {Haardt} F.,   {Scarlata} C.,  2021, \mn@doi [\apj] {10.3847/1538-4357/ac1104}, \href {https://ui.adsabs.harvard.edu/abs/2021ApJ...919..120M} {919, 120}

\bibitem[\protect\citeauthoryear{{Naidu}, {Tacchella}, {Mason}, {Bose}, {Oesch}  \& {Conroy}}{{Naidu} et~al.}{2020}]{Naidu+20}
{Naidu} R.~P.,  {Tacchella} S.,  {Mason} C.~A.,  {Bose} S.,  {Oesch} P.~A.,   {Conroy} C.,  2020, \mn@doi [\apj] {10.3847/1538-4357/ab7cc9}, \href {https://ui.adsabs.harvard.edu/abs/2020ApJ...892..109N} {892, 109}

\bibitem[\protect\citeauthoryear{{Naidu} et~al.,}{{Naidu} et~al.}{2022}]{Naidu+22}
{Naidu} R.~P.,  et~al., 2022, \mn@doi [\mnras] {10.1093/mnras/stab3601}, \href {https://ui.adsabs.harvard.edu/abs/2022MNRAS.510.4582N} {510, 4582}

\bibitem[\protect\citeauthoryear{{Navarro-Carrera}, {Rinaldi}, {Caputi}, {Iani}, {Kokorev}, {Kerutt}  \& {Cooper}}{{Navarro-Carrera} et~al.}{2024}]{Navarro-Carrera+24}
{Navarro-Carrera} R.,  {Rinaldi} P.,  {Caputi} K.~I.,  {Iani} E.,  {Kokorev} V.,  {Kerutt} J.,   {Cooper} R.,  2024, \mn@doi [arXiv e-prints] {10.48550/arXiv.2410.23249}, \href {https://ui.adsabs.harvard.edu/abs/2024arXiv241023249N} {p. arXiv:2410.23249}

\bibitem[\protect\citeauthoryear{{Nilsson} \& {M{\o}ller}}{{Nilsson} \& {M{\o}ller}}{2009}]{Nilsson+09}
{Nilsson} K.~K.,  {M{\o}ller} P.,  2009, \mn@doi [\aap] {10.1051/0004-6361/200913407}, \href {https://ui.adsabs.harvard.edu/abs/2009A&A...508L..21N} {508, L21}

\bibitem[\protect\citeauthoryear{{Ning}, {Jiang}, {Zheng}  \& {Wu}}{{Ning} et~al.}{2022}]{Ning+22}
{Ning} Y.,  {Jiang} L.,  {Zheng} Z.-Y.,   {Wu} J.,  2022, \mn@doi [\apj] {10.3847/1538-4357/ac4268}, \href {https://ui.adsabs.harvard.edu/abs/2022ApJ...926..230N} {926, 230}

\bibitem[\protect\citeauthoryear{{Ning}, {Cai}, {Jiang}, {Lin}, {Fu}  \& {Spinoso}}{{Ning} et~al.}{2023}]{Ning+23}
{Ning} Y.,  {Cai} Z.,  {Jiang} L.,  {Lin} X.,  {Fu} S.,   {Spinoso} D.,  2023, \mn@doi [\apjl] {10.3847/2041-8213/acb26b}, \href {https://ui.adsabs.harvard.edu/abs/2023ApJ...944L...1N} {944, L1}

\bibitem[\protect\citeauthoryear{{Oke} \& {Gunn}}{{Oke} \& {Gunn}}{1983}]{Oke}
{Oke} J.~B.,  {Gunn} J.~E.,  1983, \mn@doi [\apj] {10.1086/160817}, \href {https://ui.adsabs.harvard.edu/abs/1983ApJ...266..713O} {266, 713}

\bibitem[\protect\citeauthoryear{{Ono} et~al.,}{{Ono} et~al.}{2010}]{Ono+10}
{Ono} Y.,  et~al., 2010, \mn@doi [\mnras] {10.1111/j.1365-2966.2009.16034.x}, \href {https://ui.adsabs.harvard.edu/abs/2010MNRAS.402.1580O} {402, 1580}

\bibitem[\protect\citeauthoryear{{Ono} et~al.,}{{Ono} et~al.}{2021}]{Ono+21}
{Ono} Y.,  et~al., 2021, \mn@doi [\apj] {10.3847/1538-4357/abea15}, \href {https://ui.adsabs.harvard.edu/abs/2021ApJ...911...78O} {911, 78}

\bibitem[\protect\citeauthoryear{{Onoue} et~al.,}{{Onoue} et~al.}{2017}]{Onoue+17}
{Onoue} M.,  et~al., 2017, \mn@doi [\apjl] {10.3847/2041-8213/aa8cc6}, \href {https://ui.adsabs.harvard.edu/abs/2017ApJ...847L..15O} {847, L15}

\bibitem[\protect\citeauthoryear{{Oyarz{\'u}n}, {Blanc}, {Gonz{\'a}lez}, {Mateo}  \& {Bailey}}{{Oyarz{\'u}n} et~al.}{2017}]{Oyarzn+17}
{Oyarz{\'u}n} G.~A.,  {Blanc} G.~A.,  {Gonz{\'a}lez} V.,  {Mateo} M.,   {Bailey} III J.~I.,  2017, \mn@doi [\apj] {10.3847/1538-4357/aa7552}, \href {https://ui.adsabs.harvard.edu/abs/2017ApJ...843..133O} {843, 133}

\bibitem[\protect\citeauthoryear{{Pacifici} et~al.,}{{Pacifici} et~al.}{2016}]{Pacifici+16}
{Pacifici} C.,  et~al., 2016, \mn@doi [\apj] {10.3847/0004-637X/832/1/79}, \href {https://ui.adsabs.harvard.edu/abs/2016ApJ...832...79P} {832, 79}

\bibitem[\protect\citeauthoryear{{Pahl}, {Shapley}, {Steidel}, {Chen}  \& {Reddy}}{{Pahl} et~al.}{2021}]{Pahl+21}
{Pahl} A.~J.,  {Shapley} A.,  {Steidel} C.~C.,  {Chen} Y.,   {Reddy} N.~A.,  2021, \mn@doi [\mnras] {10.1093/mnras/stab1374}, \href {https://ui.adsabs.harvard.edu/abs/2021MNRAS.505.2447P} {505, 2447}

\bibitem[\protect\citeauthoryear{{Partridge} \& {Peebles}}{{Partridge} \& {Peebles}}{1967}]{Partridge+67}
{Partridge} R.~B.,  {Peebles} P.~J.~E.,  1967, \mn@doi [\apj] {10.1086/149079}, \href {https://ui.adsabs.harvard.edu/abs/1967ApJ...147..868P} {147, 868}

\bibitem[\protect\citeauthoryear{{Paulino-Afonso} et~al.,}{{Paulino-Afonso} et~al.}{2018}]{Afonso+18}
{Paulino-Afonso} A.,  et~al., 2018, \mn@doi [\mnras] {10.1093/mnras/sty281}, \href {https://ui.adsabs.harvard.edu/abs/2018MNRAS.476.5479P} {476, 5479}

\bibitem[\protect\citeauthoryear{{Pentericci}, {Grazian}, {Fontana}, {Castellano}, {Giallongo}, {Salimbeni}  \& {Santini}}{{Pentericci} et~al.}{2009}]{Pentericci+09}
{Pentericci} L.,  {Grazian} A.,  {Fontana} A.,  {Castellano} M.,  {Giallongo} E.,  {Salimbeni} S.,   {Santini} P.,  2009, \mn@doi [\aap] {10.1051/0004-6361:200810722}, \href {https://ui.adsabs.harvard.edu/abs/2009A&A...494..553P} {494, 553}

\bibitem[\protect\citeauthoryear{{Planck Collaboration} et~al.,}{{Planck Collaboration} et~al.}{2020}]{Planck+20}
{Planck Collaboration} et~al., 2020, \mn@doi [\aap] {10.1051/0004-6361/201833910}, \href {https://ui.adsabs.harvard.edu/abs/2020A&A...641A...6P} {641, A6}

\bibitem[\protect\citeauthoryear{{Protu{\v{s}}ov{\'a}} et~al.,}{{Protu{\v{s}}ov{\'a}} et~al.}{2024}]{Protu+24}
{Protu{\v{s}}ov{\'a}} K.,  et~al., 2024, \mn@doi [arXiv e-prints] {10.48550/arXiv.2412.12256}, \href {https://ui.adsabs.harvard.edu/abs/2024arXiv241212256P} {p. arXiv:2412.12256}

\bibitem[\protect\citeauthoryear{{Pucha}, {Reddy}, {Dey}, {Juneau}, {Lee}, {Prescott}, {Shivaei}  \& {Hong}}{{Pucha} et~al.}{2022}]{Pucha+22}
{Pucha} R.,  {Reddy} N.~A.,  {Dey} A.,  {Juneau} S.,  {Lee} K.-S.,  {Prescott} M. K.~M.,  {Shivaei} I.,   {Hong} S.,  2022, \mn@doi [\aj] {10.3847/1538-3881/ac83a9}, \href {https://ui.adsabs.harvard.edu/abs/2022AJ....164..159P} {164, 159}

\bibitem[\protect\citeauthoryear{{Rasekh} et~al.,}{{Rasekh} et~al.}{2022}]{Rasekh+22}
{Rasekh} A.,  et~al., 2022, \mn@doi [\aap] {10.1051/0004-6361/202140734}, \href {https://ui.adsabs.harvard.edu/abs/2022A&A...662A..64R} {662, A64}

\bibitem[\protect\citeauthoryear{{Rinaldi} et~al.,}{{Rinaldi} et~al.}{2025}]{Rinaldi+24}
{Rinaldi} P.,  et~al., 2025, \mn@doi [\apj] {10.3847/1538-4357/adb309}, \href {https://ui.adsabs.harvard.edu/abs/2025ApJ...981..161R} {981, 161}

\bibitem[\protect\citeauthoryear{{Roberts-Borsani} et~al.,}{{Roberts-Borsani} et~al.}{2024}]{Roberts-Borsani+24}
{Roberts-Borsani} G.,  et~al., 2024, \mn@doi [\apj] {10.3847/1538-4357/ad85d3}, \href {https://ui.adsabs.harvard.edu/abs/2024ApJ...976..193R} {976, 193}

\bibitem[\protect\citeauthoryear{{Rosani}, {Caminha}, {Caputi}  \& {Deshmukh}}{{Rosani} et~al.}{2020}]{Rosani+20}
{Rosani} G.,  {Caminha} G.~B.,  {Caputi} K.~I.,   {Deshmukh} S.,  2020, \mn@doi [\aap] {10.1051/0004-6361/201935782}, \href {https://ui.adsabs.harvard.edu/abs/2020A&A...633A.159R} {633, A159}

\bibitem[\protect\citeauthoryear{{Saldana-Lopez} et~al.,}{{Saldana-Lopez} et~al.}{2022}]{Saldana+22a}
{Saldana-Lopez} A.,  et~al., 2022, \mn@doi [\aap] {10.1051/0004-6361/202141864}, \href {https://ui.adsabs.harvard.edu/abs/2022A&A...663A..59S} {663, A59}

\bibitem[\protect\citeauthoryear{{Salpeter}}{{Salpeter}}{1955}]{Salpeter}
{Salpeter} E.~E.,  1955, \mn@doi [\apj] {10.1086/145971}, \href {https://ui.adsabs.harvard.edu/abs/1955ApJ...121..161S} {121, 161}

\bibitem[\protect\citeauthoryear{{Santos} et~al.,}{{Santos} et~al.}{2020}]{Santos+20}
{Santos} S.,  et~al., 2020, \mn@doi [\mnras] {10.1093/mnras/staa093}, \href {https://ui.adsabs.harvard.edu/abs/2020MNRAS.493..141S} {493, 141}

\bibitem[\protect\citeauthoryear{{Schechter}}{{Schechter}}{1976}]{Schechter}
{Schechter} P.,  1976, \mn@doi [\apj] {10.1086/154079}, \href {https://ui.adsabs.harvard.edu/abs/1976ApJ...203..297S} {203, 297}

\bibitem[\protect\citeauthoryear{{Shibuya} et~al.,}{{Shibuya} et~al.}{2018}]{Shibuya+18}
{Shibuya} T.,  et~al., 2018, \mn@doi [\pasj] {10.1093/pasj/psx122}, \href {https://ui.adsabs.harvard.edu/abs/2018PASJ...70S..14S} {70, S14}

\bibitem[\protect\citeauthoryear{{Shimizu} \& {Umemura}}{{Shimizu} \& {Umemura}}{2010}]{Shimizu+10}
{Shimizu} I.,  {Umemura} M.,  2010, \mn@doi [\mnras] {10.1111/j.1365-2966.2010.16758.x}, \href {https://ui.adsabs.harvard.edu/abs/2010MNRAS.406..913S} {406, 913}

\bibitem[\protect\citeauthoryear{{Simmonds} et~al.,}{{Simmonds} et~al.}{2023}]{Simmonds+23}
{Simmonds} C.,  et~al., 2023, \mn@doi [\mnras] {10.1093/mnras/stad1749}, \href {https://ui.adsabs.harvard.edu/abs/2023MNRAS.523.5468S} {523, 5468}

\bibitem[\protect\citeauthoryear{{Simpson} et~al.,}{{Simpson} et~al.}{2006}]{VLA-uds}
{Simpson} C.,  et~al., 2006, \mn@doi [\mnras] {10.1111/j.1365-2966.2006.10907.x}, \href {https://ui.adsabs.harvard.edu/abs/2006MNRAS.372..741S} {372, 741}

\bibitem[\protect\citeauthoryear{{Sobral} et~al.,}{{Sobral} et~al.}{2018}]{Sobral+18b}
{Sobral} D.,  et~al., 2018, \mn@doi [\mnras] {10.1093/mnras/sty782}, \href {https://ui.adsabs.harvard.edu/abs/2018MNRAS.477.2817S} {477, 2817}

\bibitem[\protect\citeauthoryear{{Solhaug} et~al.,}{{Solhaug} et~al.}{2025}]{Erik+25}
{Solhaug} E.,  et~al., 2025, \mn@doi [The Open Journal of Astrophysics] {10.33232/001c.134065}, \href {https://ui.adsabs.harvard.edu/abs/2025OJAp....8E..35S} {8, 35}

\bibitem[\protect\citeauthoryear{{Spina}, {Bosman}, {Davies}, {Gaikwad}  \& {Zhu}}{{Spina} et~al.}{2024}]{Spina+24}
{Spina} B.,  {Bosman} S. E.~I.,  {Davies} F.~B.,  {Gaikwad} P.,   {Zhu} Y.,  2024, \mn@doi [\aap] {10.1051/0004-6361/202450798}, \href {https://ui.adsabs.harvard.edu/abs/2024A&A...688L..26S} {688, L26}

\bibitem[\protect\citeauthoryear{{Steidel}, {Bogosavljevi{\'c}}, {Shapley}, {Reddy}, {Rudie}, {Pettini}, {Trainor}  \& {Strom}}{{Steidel} et~al.}{2018}]{Steidel+18}
{Steidel} C.~C.,  {Bogosavljevi{\'c}} M.,  {Shapley} A.~E.,  {Reddy} N.~A.,  {Rudie} G.~C.,  {Pettini} M.,  {Trainor} R.~F.,   {Strom} A.~L.,  2018, \mn@doi [\apj] {10.3847/1538-4357/aaed28}, \href {https://ui.adsabs.harvard.edu/abs/2018ApJ...869..123S} {869, 123}

\bibitem[\protect\citeauthoryear{{Tang}, {Stark}, {Topping}, {Mason}  \& {Ellis}}{{Tang} et~al.}{2024}]{Tang+24}
{Tang} M.,  {Stark} D.~P.,  {Topping} M.~W.,  {Mason} C.,   {Ellis} R.~S.,  2024, \mn@doi [\apj] {10.3847/1538-4357/ad7eb7}, \href {https://ui.adsabs.harvard.edu/abs/2024ApJ...975..208T} {975, 208}

\bibitem[\protect\citeauthoryear{{Thai} et~al.,}{{Thai} et~al.}{2023}]{Thai+23}
{Thai} T.~T.,  et~al., 2023, \mn@doi [\aap] {10.1051/0004-6361/202346716}, \href {https://ui.adsabs.harvard.edu/abs/2023A&A...678A.139T} {678, A139}

\bibitem[\protect\citeauthoryear{{Trainor}, {Steidel}, {Strom}  \& {Rudie}}{{Trainor} et~al.}{2015}]{Trainor+15}
{Trainor} R.~F.,  {Steidel} C.~C.,  {Strom} A.~L.,   {Rudie} G.~C.,  2015, \mn@doi [\apj] {10.1088/0004-637X/809/1/89}, \href {https://ui.adsabs.harvard.edu/abs/2015ApJ...809...89T} {809, 89}

\bibitem[\protect\citeauthoryear{{Trainor}, {Strom}, {Steidel}  \& {Rudie}}{{Trainor} et~al.}{2016}]{Trainor+16}
{Trainor} R.~F.,  {Strom} A.~L.,  {Steidel} C.~C.,   {Rudie} G.~C.,  2016, \mn@doi [\apj] {10.3847/0004-637X/832/2/171}, \href {https://ui.adsabs.harvard.edu/abs/2016ApJ...832..171T} {832, 171}

\bibitem[\protect\citeauthoryear{{Ueda} et~al.,}{{Ueda} et~al.}{2008}]{XMM-uds}
{Ueda} Y.,  et~al., 2008, \mn@doi [\apjs] {10.1086/591083}, \href {https://ui.adsabs.harvard.edu/abs/2008ApJS..179..124U} {179, 124}

\bibitem[\protect\citeauthoryear{{Umeda} et~al.,}{{Umeda} et~al.}{2025}]{Umeda+25a}
{Umeda} H.,  et~al., 2025, \mn@doi [\apjs] {10.3847/1538-4365/adb1c0}, \href {https://ui.adsabs.harvard.edu/abs/2025ApJS..277...37U} {277, 37}

\bibitem[\protect\citeauthoryear{{Valentino} et~al.,}{{Valentino} et~al.}{2023}]{Valentino+23}
{Valentino} F.,  et~al., 2023, \mn@doi [\apj] {10.3847/1538-4357/acbefa}, \href {https://ui.adsabs.harvard.edu/abs/2023ApJ...947...20V} {947, 20}

\bibitem[\protect\citeauthoryear{{Verhamme}, {Schaerer}  \& {Maselli}}{{Verhamme} et~al.}{2006}]{Verhamme+06}
{Verhamme} A.,  {Schaerer} D.,   {Maselli} A.,  2006, \mn@doi [\aap] {10.1051/0004-6361:20065554}, \href {https://ui.adsabs.harvard.edu/abs/2006A&A...460..397V} {460, 397}

\bibitem[\protect\citeauthoryear{{Verhamme}, {Orlitov{\'a}}, {Schaerer}  \& {Hayes}}{{Verhamme} et~al.}{2015}]{Verhamme+15}
{Verhamme} A.,  {Orlitov{\'a}} I.,  {Schaerer} D.,   {Hayes} M.,  2015, \mn@doi [\aap] {10.1051/0004-6361/201423978}, \href {https://ui.adsabs.harvard.edu/abs/2015A&A...578A...7V} {578, A7}

\bibitem[\protect\citeauthoryear{{Verhamme}, {Orlitov{\'a}}, {Schaerer}, {Izotov}, {Worseck}, {Thuan}  \& {Guseva}}{{Verhamme} et~al.}{2017}]{Verhamme+17}
{Verhamme} A.,  {Orlitov{\'a}} I.,  {Schaerer} D.,  {Izotov} Y.,  {Worseck} G.,  {Thuan} T.~X.,   {Guseva} N.,  2017, \mn@doi [\aap] {10.1051/0004-6361/201629264}, \href {https://ui.adsabs.harvard.edu/abs/2017A&A...597A..13V} {597, A13}

\bibitem[\protect\citeauthoryear{{Wang} et~al.,}{{Wang} et~al.}{2020}]{Wang+20}
{Wang} F.,  et~al., 2020, \mn@doi [\apj] {10.3847/1538-4357/ab8c45}, \href {https://ui.adsabs.harvard.edu/abs/2020ApJ...896...23W} {896, 23}

\bibitem[\protect\citeauthoryear{{Weiss} et~al.,}{{Weiss} et~al.}{2021}]{Weiss+21}
{Weiss} L.~H.,  et~al., 2021, \mn@doi [\apj] {10.3847/1538-4357/abedb9}, \href {https://ui.adsabs.harvard.edu/abs/2021ApJ...912..100W} {912, 100}

\bibitem[\protect\citeauthoryear{{Witstok} et~al.,}{{Witstok} et~al.}{2024}]{Witstok+24}
{Witstok} J.,  et~al., 2024, \mn@doi [arXiv e-prints] {10.48550/arXiv.2408.16608}, \href {https://ui.adsabs.harvard.edu/abs/2024arXiv240816608W} {p. arXiv:2408.16608}

\bibitem[\protect\citeauthoryear{{Xue} et~al.,}{{Xue} et~al.}{2017}]{Xue+17}
{Xue} R.,  et~al., 2017, \mn@doi [\apj] {10.3847/1538-4357/837/2/172}, \href {https://ui.adsabs.harvard.edu/abs/2017ApJ...837..172X} {837, 172}

\bibitem[\protect\citeauthoryear{{Yang} et~al.,}{{Yang} et~al.}{2020}]{Yang+20}
{Yang} J.,  et~al., 2020, \mn@doi [\apjl] {10.3847/2041-8213/ab9c26}, \href {https://ui.adsabs.harvard.edu/abs/2020ApJ...897L..14Y} {897, L14}

\bibitem[\protect\citeauthoryear{{Yoshioka} et~al.,}{{Yoshioka} et~al.}{2025}]{Yoshioka+25}
{Yoshioka} T.,  et~al., 2025, \mn@doi [\mnras] {10.1093/mnras/stae2796}, \href {https://ui.adsabs.harvard.edu/abs/2025MNRAS.536.3386Y} {536, 3386}

\bibitem[\protect\citeauthoryear{{Zhu} et~al.,}{{Zhu} et~al.}{2022}]{Zhu+22}
{Zhu} Y.,  et~al., 2022, \mn@doi [\apj] {10.3847/1538-4357/ac6e60}, \href {https://ui.adsabs.harvard.edu/abs/2022ApJ...932...76Z} {932, 76}

\bibitem[\protect\citeauthoryear{{Zhu} et~al.,}{{Zhu} et~al.}{2024}]{Zhu+24_xhi}
{Zhu} Y.,  et~al., 2024, \mn@doi [\mnras] {10.1093/mnrasl/slae061}, \href {https://ui.adsabs.harvard.edu/abs/2024MNRAS.533L..49Z} {533, L49}

\bibitem[\protect\citeauthoryear{{Zhu}, {Zheng}, {Yuan}, {Jiang}  \& {Lin}}{{Zhu} et~al.}{2025}]{Zhu+24}
{Zhu} S.,  {Zheng} Z.-Y.,  {Yuan} F.-T.,  {Jiang} C.,   {Lin} R.,  2025, \mn@doi [\apjl] {10.3847/2041-8213/adc125}, \href {https://ui.adsabs.harvard.edu/abs/2025ApJ...982L..58Z} {982, L58}

\bibitem[\protect\citeauthoryear{{{\v{D}}urov{\v{c}}{\'\i}kov{\'a}}, {Katz}, {Bosman}, {Davies}, {Devriendt}  \& {Slyz}}{{{\v{D}}urov{\v{c}}{\'\i}kov{\'a}} et~al.}{2020}]{Duro+20}
{{\v{D}}urov{\v{c}}{\'\i}kov{\'a}} D.,  {Katz} H.,  {Bosman} S. E.~I.,  {Davies} F.~B.,  {Devriendt} J.,   {Slyz} A.,  2020, \mn@doi [\mnras] {10.1093/mnras/staa505}, \href {https://ui.adsabs.harvard.edu/abs/2020MNRAS.493.4256D} {493, 4256}

\bibitem[\protect\citeauthoryear{{{\v{D}}urov{\v{c}}{\'\i}kov{\'a}} et~al.,}{{{\v{D}}urov{\v{c}}{\'\i}kov{\'a}} et~al.}{2024}]{Duro+24}
{{\v{D}}urov{\v{c}}{\'\i}kov{\'a}} D.,  et~al., 2024, \mn@doi [\apj] {10.3847/1538-4357/ad4888}, \href {https://ui.adsabs.harvard.edu/abs/2024ApJ...969..162D} {969, 162}

\makeatother
\end{thebibliography}

%\begin{thebibliography}{99}
%\end{thebibliography}

%%%%%%%%%%%%%%%%%%%%%%%%%%%%%%%%%%%%%%%%%%%%%%%%%%

%%%%%%%%%%%%%%%%% APPENDICES %%%%%%%%%%%%%%%%%%%%%

\appendix

\section{robustness of H$\alpha$ luminosity calculation} \label{sec:Ha_check}
We estimate the \fesca of LAEs by converting the dust-corrected SFRs obtained from the SED fitting into the \Ha luminosities using equation~\eqref{eq:SFR_Ha}, since spectroscopic measurements of Balmer emission lines such as \Ha are unavailable. 
This method, however, may introduce systematic uncertainties due to assumptions in the SED fitting. 
To assess the robustness of this approach, we follow a method similar to that of \citet{Begley+24} and independently estimate the \Ha luminosity from the BB excess due to the \Ha line, combined with continuum estimates from the SED fitting.
For LAEs at $z=\,$2.2, 3.3, 4.9, and 5.7, the \Ha line appears at wavelengths of approximately 2.09, 2.84, 3.87, and 4.41 $\mathrm{\mu m}$, respectively. 
We therefore use the F200W, F277W, F356W, and F444W filters to estimate \Ha luminosities in each case. At $z=\,$6.6, no BB filter fully covers the expected \Ha wavelength ($\sim$4.97 $\mathrm{\mu m}$), so this redshift sample is excluded from this analysis.

We modify equation~\eqref{eq:shibuya} by replacing the NB parameters with those of the BB filter to estimate the \Ha flux.
Unlike the calculation for \Lya luminosity, no IGM attenuation correction is applied to either the continuum or the emission line. 
Additionally, while NB filters permit the assumption of a flat continuum within the transmission curve, in the BB case we adopt a simplifying assumption that the continuum flux density is constant at rest-frame 6563\,\AA\, across the filter transmission.
The estimated \Ha luminosities naturally include contributions from nearby emission lines such as [NII] and [SII] ([OIII] and H$\mathrm{\beta}$ lines are not included, as they fall outside the wavelength coverage of the selected filters). 
To remove these contributions, we assume sub-solar metallicity and apply a correction factor of 0.82, as adopted in \citet{Navarro-Carrera+24}. 
Finally, we apply a dust attenuation correction using the E(B-V) value derived from the SED fitting and the \citet{Calzetti+00} extinction law.

\begin{figure}
    \centering
    \includegraphics[width=\columnwidth]{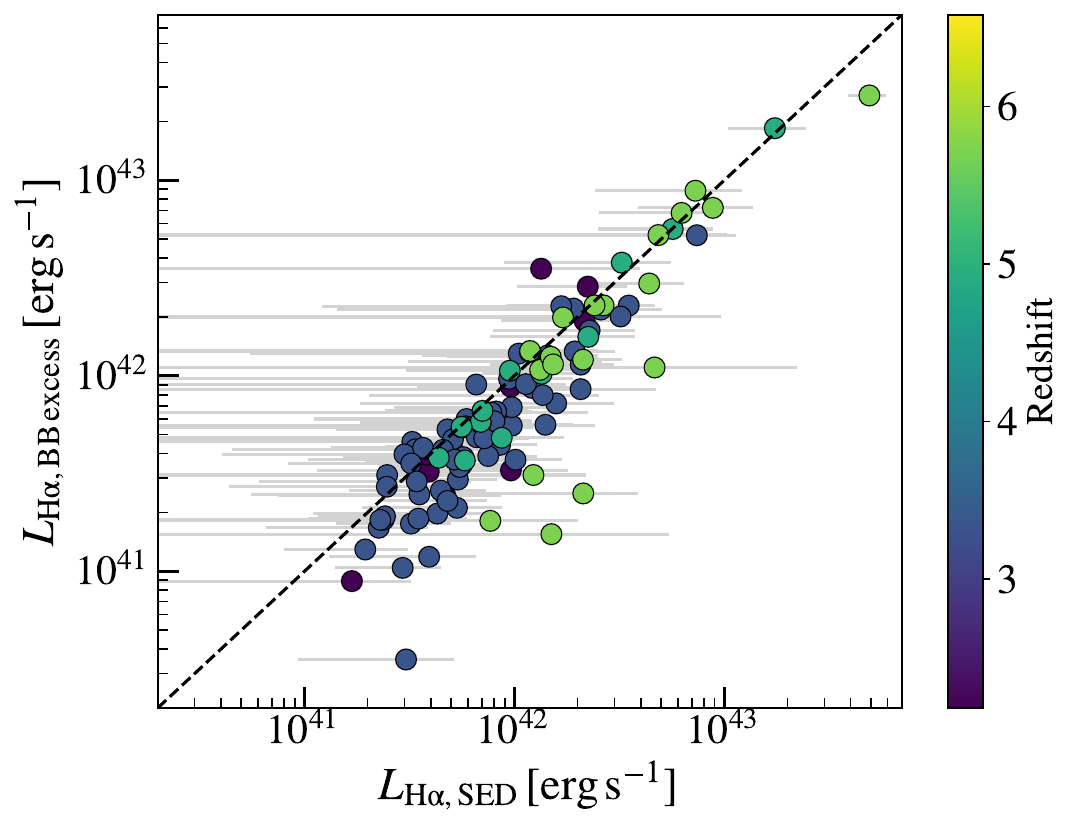}
    \caption{Comparison between \Ha luminosities derived from SED-based SFRs and those estimated from BB photometric excesses, for LAEs at $z=2.2,3.3,4.9,5.7$. Each point represents an individual LAE, colour-coded by redshift. The black dashed line indicates the one-to-one relation.}
    \label{fig:Ha_check}
\end{figure}

In Fig.~\ref{fig:Ha_check}, we compare the \Ha luminosities obtained by this BB excess method with those derived from the SED-based SFRs using equation~\eqref{eq:SFR_Ha}. 
We find overall agreement, although there is some scatter, which is likely due to uncertainties in the dust correction and metallicity assumptions. 
At higher \Ha luminosities ($L_{\mathrm{H\alpha}}>10^{42}\,\mathrm{erg/s}$
), the agreement is excellent, while at lower luminosities the SED-based estimates tend to be slightly lower. % ($\sim 0.2\,\mathrm{dex}$). 
This offset ($\sim0.2$ dex in $L_{\mathrm{H\alpha}}$) may reflect a systematic bias, but could also arise from underestimated \Ha fluxes in the BB excess method due to low S/N in the faint regime.
One might be concerned that this comparison is not fully independent, as the same BB photometry, including the \Ha excess, also contributes to the SED-based SFR estimates. 
However, we confirm that removing the BB filter containing the \Ha line from the SED fitting does not significantly affect the derived SFR values.
These results suggest that the influence of the \Ha excess on the SED fitting is minimal, and support the robustness of the \Ha luminosities estimated in this paper.

\section{Spatial extent of Ly$\alpha$ emission} \label{app:halo}
\Lya photons emitted from LAEs are known to exhibit more extended spatial profiles compared to the stellar component of their host galaxies due to resonant scattering in the surrounding circum-galactic medium (CGM).
In Section~\ref{sec:LAE_size}, we show that old LAEs have larger stellar sizes than young LAEs, but are still more compact than typical SFGs, suggesting that old LAEs may represent a particularly compact subset of the SFG population.
In this appendix, we investigate whether a similar difference is present in the spatial extent of \Lya emission between young and old LAEs.

To examine the systematic difference in the \Lya spatial extent between old and young LAEs, we perform image stacking for each population separately. 
The analysis is conducted only at $z=2.2$ and $z=3.3$, where we have sufficient numbers of both young and old LAEs: at $z=2.2$, the sample comprises 13 young and 7 old LAEs, while at $z=3.3$, it includes 46 young and 23 old LAEs.
However, as will be discussed later, the two old LAEs at $z=2.2$ are likely to possess exceptionally extended \Lya haloes. 
Therefore, we exclude them from the stacking analysis and instead analyse their \Lya profiles individually.
To construct \Lya images, we follow the method of \citet{Momose+14}, subtracting the BB image (UV image) from the NB image. 
For the UV continuum images to be subtracted, we use the HSC \textit{g}-band image for $z=2.2$, and the HSC \textit{r}-band images for $z=3.3$, which have the same pixel scale as the NB images.
As shown in Table~\ref{tab:limitmag}, the PSF sizes of the BB and NB images differ. 
To account for this, we first match the PSF FWHM to the image with the larger PSF, applying a corresponding Gaussian kernel.
Next, we extract $10\arcsec \times 10\arcsec$ cutouts centred on the coordinates of each LAE from both the BB and NB images. 
We subtract the BB image from the NB image, assuming that the UV continuum is flat, to create \Lya images. 
For \Lya images, we apply 3$\sigma$ clipping and perform mean stacking. 
We calculate the surface brightness (SB) profiles of the stacked images. 
To focus on the extent of the profiles rather than the absolute SB, we normalize the SB at the centre to one. 
To estimate the SB errors, we extract the same size cutouts free from any sources (sky images) from NB and BB images and stack them 1,000 times for various combinations, matching the number of images in the stack. 
The SB measurement error is evaluated from the standard deviation of the SB in the stacked sky images at each radius.
\begin{figure}
    \centering
    \includegraphics[width=\columnwidth]{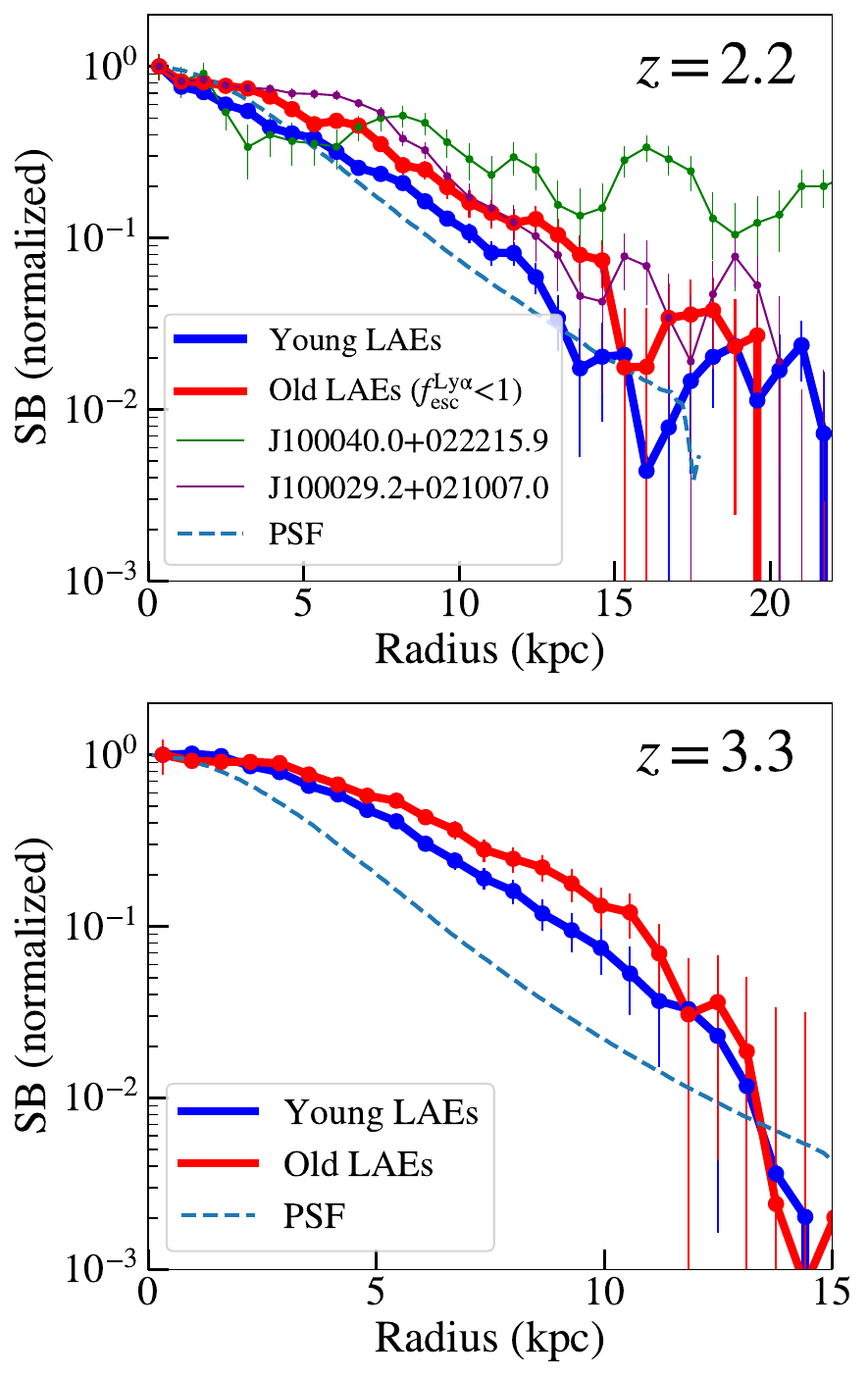}
    \caption{Comparison of the mean SB radial profiles of LAEs in \Lya images. The top panel shows the comparison between old LAEs (red) and young LAEs (blue) at $z=2.2$, and the bottom panel presents the same comparison at $z=3.3$. SB profiles of two LAEs at $z=2.2$ that exhibit extended \Lya halo are also shown in the top panel (J100040.0+022215.9:green line, J100029.2+021007.0:purple line). All SB profiles are normalised to unity at $r=0$. The light blue dashed lines indicate the PSF profiles corresponding to the larger FWHM between the NB and BB images at each redshift.}
    \label{fig:Lya_halo}
\end{figure}

Fig.~\ref{fig:Lya_halo} presents the radial SB profiles of stacked \Lya images of old and young LAE subsamples at $z=2.2$ (top panel) and $z=3.3$ (bottom panel). 
Old LAEs exhibit more extended \Lya emission in both redshift panels than young LAEs.
This trend may be interpreted in two possible ways. 
One is a simple scenario in which the more extended \Lya profile of old LAEs directly reflects the larger size of their host galaxies (Fig.~\ref{fig:size_mass}), as discussed in \citet{Rasekh+22}. 
The other potential explanation is the contribution from \Lya halos driven by mechanisms such as outflows, which can scatter \Lya photons (e.g. \citealt{Xue+17}). 
As discussed in Section~\ref{escape_oldLAE}, in old LAEs, \Lya photons may escape through low-density channels in the ISM. 
However, escape from this low-density channel alone would likely result in a relatively compact \Lya spatial extent. 
Therefore, it is plausible that resonant scattering of \Lya photons in the CGM, enhanced by outflows, contributes significantly to the extended \Lya SB profile observed in old LAEs.

This scenario may be particularly relevant for J100040.0+022215.9 and J100029.2+021007.0 shown in Fig.~\ref{fig:Lya_halo} at $z=2.2$, where such a mechanism could be acting more prominently.
These two LAEs exhibit more extended \Lya SB profiles than the other old LAEs.
As indicated by the diamond symbols in Fig.~\ref{fig:f_esc_mass}, both show unusually high \fesca ($>1$), which is unlikely by definition and deviates from the old LAE sequence.
The background flux of J100040.0+022215.9, however, is locally high, making it difficult to distinguish whether the LAE has a real extended structure or whether it is noise. 
On the other hand, the J100029.2+021007.0 exhibits an extended ($\sim8$\,kpc) profile compared to those of young and old LAEs with \fesca$<1$. 
Although it has an average stellar mass as an old LAE ($\sim10^{8.7}\,\mathrm{M_{\odot}}$), its effective radius, as measured from the UV and optical continuum (Section~\ref{sec:LAE_size}), is relatively small, 0.47\,kpc in the UV and 0.39\,kpc in the optical wavelength. 
These characteristics place J100029.2+021007.0 among the most compact objects in the old LAE population, and make it an LAE with a particularly dominant and extended \Lya halo.
Therefore, in this old LAE, it is plausible that a strong outflow drives both its high \Lya escape and significantly extended \Lya halo.
However, confirming the hypothesis would require spectroscopic observation of the \Lya emission line. 
It should be noted that the \Lya emissions of J100029.2+021007.0 and J100040.0+022215.9 do not extend as large as those of typical \Lya blobs, based on the criteria defined by \citet{Mingyu+24}.
To conclusively determine why these two LAEs exhibit anomalously high \fesca, deeper imaging and spectroscopic data are necessary. 
Nevertheless, it is important to consider the potential influence of \Lya halos in driving \fesca$>1$, because \Lya photons escaping directly from the central galaxy coexist with those originating from within the halo on different timescales.

%%%%%%%%%%%%%%%%%%%%%%%%%%%%%%%%%%%%%%%%%%%%%%%%%%

% Don't change these lines
\bsp	% typesetting comment
\label{lastpage}
\end{document}